\documentclass[12pt]{article}
\usepackage{amsmath}
\usepackage{jcappub}
\usepackage{pifont}
\usepackage{braket}
\usepackage{multirow}
\usepackage{cancel}
\newcommand{\spdot}{^{\!.}}

\title{Cosmic Energy Density:  Particles, Fields and the Vacuum}
\author[a]{Cristian Armendariz-Picon}
\emailAdd{carmendarizpicon@stlawu.edu}
\author[b]{and Alberto Diez-Tejedor}
\emailAdd{alberto.diez@fisica.ugto.mx}
\affiliation[a]{Physics Department,\\St.~Lawrence University, \\23 Romoda Dr., Canton, NY 13617, USA.}                                            
\affiliation[b]{Departamento de F\'isica,\\ Divisi\'on de Ciencias e Ingenier\'ias,\\
Campus Le\'on, Universidad de Guanajuato, Le\'on 37150, M\'exico.}

\makeatletter
\def\@fpheader{\relax} 
\makeatother
                                
\abstract{We revisit the cosmic evolution of the energy density of a quantized free scalar field and assess under what conditions the particle  production and classical field  approximations reproduce its correct value.  Because the unrenormalized energy-momentum tensor diverges in the ultraviolet, it is necessary to frame our discussion within an appropriate regularization and renormalization scheme.   Pauli-Villars avoids  some  of the drawbacks of  adiabatic subtraction and dimensional regularization and is particularly convenient in this context. In some cases, we can predict the evolution of the energy density irrespectively of the quantum state of the field modes.   To further illustrate our results we focus however on the {\it in} vacuum, the preferred quantum state  singled out by inflation, and explore to what extent the latter determines the subsequent evolution of the energy density regardless of the unknown details of reheating. We contrast this discussion with  examples of transitions to radiation domination that avoid some of the problems of the one commonly studied in the literature, and point out some instances in which the particle production or the classical field approximations lead to the incorrect energy density. Along the way, we also elaborate on the connection of our analysis to dynamical dark energy models and axion-like dark matter candidates.}

\begin{document}
 \vspace{-5cm}
 
\maketitle
 
\section{Introduction}

A variation of the old particle/wave duality that permeated the origins of quantum mechanics is still present in many of the current analyses of the evolution of the universe and its content.  Whereas we typically assume that the  theories that describe nature are quantum \emph{field} theories, we often frame our study of cosmic phenomena in terms of \emph{particles}.  This is what happens, for instance, with dark matter, which is usually presumed to consist of  nonrelativistic particles \cite{Szalay:1976ef,Lee:1977ua,Pagels:1981ke,Turner:1986tb}. In addition, even when we recognize that  the fundamental constituents of the universe are \emph{quantum} fields, we typically ignore their quantization and treat them \emph{classically} instead. The paradigmatic example here is that of dynamical dark energy, which is commonly modeled as a classical scalar \cite{Wetterich:1987fm,Peebles:1987ek,Caldwell:1997ii}. In other cases, such as some axion models, one often switches back and forth between the particle and classical field descriptions, many a time implicitly and without further discussion of the validity of either picture \cite{Turner:1986tb}.

Yet in the  semiclassical gravity approximation \cite{Birrell:1982ix, Parker:2009uva, Mukhanov:2007zz},  which we take as our starting point and do not question,   the Einstein equations are sourced by  the expectation of the energy-momentum tensor.  Hence,  cosmic expansion and structure formation are then basically determined  by the  expectation of the energy and pressure of matter and characterized by field correlations. In this context, the description of the cosmic matter content in terms of particles or  classical fields can therefore be only an approximation.   Certainly  there are cases  when, say,  a particle interpretation is convenient, such as when one sets up  underground dark matter direct detection experiments, but these cases appear to be rather limited. 

In this manuscript we explore the limits in which a quantized field can be approximated either by an ensemble of particles or by a classical excitation. We limit our analysis to an homogeneous and isotropic universe, where the expectation of the energy density and pressure must be spatially constant by symmetry.   Furthermore, to simplify the presentation we restrict our attention to a free scalar, although our formalism can be extended to fields of higher spin  or with additional interactions \cite{Edgar}. In order to frame our discussion in a specific context we focus on an aspect of cosmic evolution in which particles and classical fields, justifiably or not, appear to play an important role: the transition from inflation to the radiation era at the onset of the standard hot Big Bang mode (for a study with a similar motivation in a different context, see reference \cite{Alcubierre:2022rgp}.) According to the conventional wisdom, departures from adiabaticity in the early universe cause the gravitational production of particles out of the vacuum, which leave an imprint on cosmic evolution through their energy density and pressure.  Alternatively, when field fluctuations of a light scalar leave the horizon during inflation, they freeze and   ``classicalize." The continuous accumulation of these superhorizon modes is then argued to result in an effectively homogeneous classical field of appropriate amplitude that may latter dominate the universe, for instance, as a form of dark energy \cite{Starobinsky:1986fx}. In both cases, the  production of particles or classical  excitations is ``gravitational" because the corresponding quantum fields only interact gravitationally.

Transitions from inflation  may also have important phenomenological consequences. At this point, absent  a direct detection of dark matter,  an hypothesis that appears to be gaining ground is that dark matter only couples to gravitation \cite{Peebles:1999fz}. If that were the case, there ought to be a mechanism to produce the relic  density required by the analysis of cosmological observations. Gravitational  production at the end of inflation  is one possibility \cite{Turner:1987vd,Chung:1998zb,Hashiba:2018tbu,Cembranos:2019qlm,Garcia:2021iag}. Similarly, there appears to be a tension between the need to keep the couplings of the inflaton to matter small, in order to protect the flatness of its potential, and that of making them large enough  in order for the inflaton to efficiently decay and reheat the universe \cite{Armendariz-Picon:2020tkc}. If, instead,  radiation could be produced gravitationally at the end of inflation \cite{Ford:1986sy}, the inflaton would not need to couple to matter directly, and  the problems related to the flatness of its potential might thus be ameliorated.

From a more theoretical perspective, our objective here is not only to bypass the particle or classical field descriptions by directly computing  the expectation of the energy density upon cosmic transitions like the above, but also to establish under which general conditions the particle or classical field interpretations yield a quantitatively correct estimate of this density.  The literature on gravitational particle production, particularly in the context of cosmology, is  vast. Fortunately, a significant fraction of its foundations  is  reviewed in the now standard monographs \cite{Birrell:1982ix, Parker:2009uva, Mukhanov:2007zz}. Yet surprisingly, as far as the energy density is concerned, none of these references discusses the precise relation between the field  and particle production formalisms, nor how  renormalization enters the latter, and the resulting confusion has diffused into the literature on cosmic transitions and beyond.  This is a gap that this manuscript also aims to close.  As we shall see, our analysis uncovers serious shortcomings in some of the conventional implementations of  particle production. At long wavelengths,  previous approaches have missed or ignored  important contributions to the energy-momentum tensor, whereas at short wavelengths  ultraviolet divergent \emph{renormalized} energy densities  render some transitions unphysical.  

Our analysis also overlaps with works that directly consider  the renormalized energy density of a scalar after a  transition from inflation, such as those in references \cite{Glavan:2013mra,Aoki:2014ita,Aoki:2014dqa}.  On top of a different focus and scope, our treatment deviates form these in several important aspects. Whereas  previous work appears to take highly idealized cosmic transitions rather literally, we emphasize here their model-independent features, and how these survive in more realistic cases with an expanded parameter space.    In order to obtain sensible results that we can connect mode-by-mode to the total renormalized energy density we rely  on Pauli-Villars regularization \cite{Pauli:1949zm}, which offers notable advantages when compared with the previously employed adiabatic scheme and dimensional regularization \cite{Weinberg:2010wq}. In addition, because Pauli-Villars preserves diffeomorphism invariance, the expectation of the energy-momentum tensor is covariantly conserved. In the homogeneous and isotropic background that we consider the pressure of the scalar is hence uniquely determined by the energy density through the continuity equation, and we may just concentrate on the latter \cite{Armendariz-Picon:2020tkc}.

On the other hand, the literature on the validity of the classical  field approximation  is rather sparse, although similar questions have been raised in  connection to axions \cite{Erken:2011dz,Sikivie:2016enz,Hertzberg:2016tal} and axion-like particles \cite{Aguirre:2015mva}, where self-interactions play an important role. Most research on the topic, though,  has  focused  on the statistical properties exhibited by  cosmological perturbations generated during inflation, after a purported ``quantum to classical" transition \cite{Grishchuk:1990bj,Albrecht:1992kf, Polarski:1995jg,Kiefer:1998qe} (see \cite{Berjon:2020vdv, Agullo:2022ttg} and references therein for different perspectives.) Here we are specifically concerned instead with a somewhat more limited question, namely, whether in the context of semiclassical gravity the expectation of the energy density  can be approximated by that of a classical scalar.  Such an approximation is often either simply taken for granted or invoked without further justification. 

As we elaborate  in detail below,  the modes that dominate  the energy density broadly determine the behavior of a quantum field. In general,  high frequency modes allow a particle interpretation, be it as radiation (if they are relativistic) or as  dust (if they are nonrelativistic). On the other hand, nonrelativistic modes also admit a classical field interpretation, whether as a dark energy component (if they are low frequency) or as dust (if they are   high frequency.) Hence, although  the particle and classical field interpretations may look very different or even incompatible,  at a fundamental level they only constitute different regimes of the same entity: a quantum field. An interesting case where the two regimes are realized together is that of high frequency nonrelativistic modes, which admit a simultaneous interpretation in terms of particles and classical fields. This is why even when dark matter is commonly accepted to consist on nonrelativistic particles, it can sometimes be described as a classical field, as in some of the axion-like models that have recently gained renewed interest \cite{Suarez:2013iw,Marsh:2015xka,Hui:2016ltb}. Dynamical dark energy, on the contrary, arises only from nonrelativistic low frequency modes and is restricted to a field interpretation. These and similar questions will be discussed at length  in the main text.

The table of contents reproduces the outline of the paper, and we recommend the reader take a look at the conclusion section to identify its most significant results.  Regarding the conventions that we follow, we use the mostly plus metric signature $(-,+,+,+)$ and work with natural units where $\hbar=c=1$. 
 
\section{Formalism}
\label{sec:Cosmological Transitions}

Our main target is the evolution of the energy density of a free, real scalar field $\phi$ coupled to gravity,
\begin{equation}
S_\phi=-\frac{1}{2}\int d^4 x  \sqrt{-g} \left(\partial_\mu  \phi \, \partial^\mu\phi+m^2 \phi^2 +\xi\,  R \, \phi^2\right).
\end{equation}
Gauge fields and massless fermions are conformally invariant, and do not react to changes in the expansion history. To avoid similar conclusions in the massless case $m=0$, we   assume that the scalar $\phi$ is  minimally coupled to gravity ($\xi=0$), as opposed to conformally coupled ($\xi=1/6$). Even though we set $\xi=0$, our results should remain qualitatively the same for nonzero $\xi$ as long as it is not too close to the conformal value. 

We  also assume that the spacetime metric is that of a spatially flat  homogeneous and isotropic universe,
\begin{equation}
	ds^2=a^2(\eta) \left(-d\eta^2+d\vec{x}\cdot d\vec{x}\right),
\end{equation}
where $a$ is the scale factor and $\eta$ labels conformal time. In these coordinates, the comoving Hubble parameter is $\mathcal{H}\equiv\dot{a}/a$, where  a dot denotes a derivative with respect to conformal time, and the ``physical" Hubble constant  reads $H=\mathcal{H}/a$. Similarly, while $m$ refers to the actual mass of the field, it will often be convenient to consider its comoving mass $m a$.  

We are particularly interested in tracking changes in the scalar energy density as the universe transitions from an \emph{in} region during which the universe inflates, to an \emph{out} region during which the expansion is decelerating, say, during radiation domination. At this point \emph{in} and \emph{out} are to be regarded as labels for the two different epochs, and no connection with the particle production formalism is implied or required. In the $in$ region we regard the scalar $\phi$ as a test field, which allows us to control the ``initial" conditions for the field fluctuations. In the $out$ region  our results often do not depend on the nature of cosmic expansion, and therefore apply even if the scalar energy density eventually comes to dominate the cosmic budget. 

\subsection{Mode Functions}

In order to quantize the scalar field, we expand the field operator into plane waves as usual, 
\begin{equation}\label{eq:mode expansion}
	\hat \phi=\frac{1}{\sqrt{V}} \frac{1}{a}\sum_{\vec{k}} \left[\hat a_{\vec{k}} \, \chi_k(\eta) e^{i \vec{k}\cdot \vec{x}}+
		\hat a^\dag_{\vec{k}} \, \chi^*_k(\eta) e^{-i \vec{k}\cdot \vec{x}}\right].
\end{equation}
In this expression  $V$ is the (monentarily finite) comoving volume of the universe. The   creation operators $\hat a_{\vec{k}}^\dag$ can be interpreted as creating ``particles" of comoving momentum $\vec{k}$, and constitute the only instance in which the particle notion shall sneak into our analysis. In fact, these states are actually eigenvectors of the momentum operator, and do not really represent localized particles.

 The  mode functions $\chi_k$ in the expansion (\ref{eq:mode expansion}) satisfy the mode equation
\begin{equation}\label{eq:mode equation}
	\ddot{\chi}_k+\tilde\omega_k^2 \,\chi_k=0, \quad \text{where}\quad
	\tilde\omega_k^2\equiv \omega_k^2-\frac{\ddot{a}}{a} \quad
	\text{and} \quad 
	\omega_k^2\equiv k^2+m^2 a^2.
\end{equation}
The quantity $\tilde{\omega}_k$ corresponds to the oscillation frequency of the mode functions $\chi_k$, which is not the same as that of the actual modes of the scalar field, $ \omega_k$, since we have rescaled the latter by $1/a$.  The canonical commutation relations  imply  the normalization condition 
\begin{equation}\label{eq:Wronskian}
\chi_k \dot{\chi}_k^*-\dot{\chi}_k\chi^*_k =i.
\end{equation}
In particular, because of the  homogeneity and isotropy of the metric, we may assume that the mode functions only depend on $k\equiv |\vec{k}|$. 

Since the different modes of a free  field decouple, they can be treated separately.  Amongst them, the zero mode $\vec{k}=0$ needs special consideration, not just because symmetry  allows it  to have a nonzero expectation, but also because in some cases there is no  preferred  state for that mode.  States typically considered in the literature are classical-like, with  a nonzero expectation that evolves like a homogeneous classical field. This property of the zero mode will be clarified in  subsequent subsections. 

In general, there are no exact analytical solutions to the mode equation (\ref{eq:mode equation}), so we shall  rely on approximate solutions instead. In the next two subsections we present high and low frequency expansions that will be useful in the remainder of this work. 

\subsubsection{High Frequencies}
\label{sec:High Frequencies}

At high frequencies $\omega_k$ we can obtain approximate solutions of the mode equation (\ref{eq:mode equation}) by adopting an ``adiabatic" expansion in the number of time derivatives,
\begin{equation}\label{eq:out adiabatic}
	 \chi_k^{\mathrm{ad}(n)}(\eta)=  \frac{1}{\sqrt{2 W^{(n)}_k(\eta)}}
	\exp\left(-i \int^\eta W^{(n)}_k(\tilde{\eta})\,  d\tilde{\eta}\right).
\end{equation}
Note that we have not specified the lower limit of integration, which simply shifts the phase of the mode functions by a constant. In integrals like that of equation (\ref{eq:out adiabatic}) we shall omit the lower limit of integration when it is  physically irrelevant.

The adiabatic mode functions (\ref{eq:out adiabatic}) are    $n$-derivative  approximations  to the actual solutions of the mode equation. In particular, the  $W^{(n)}_k$ themselves  are  $n$-derivative approximate solutions  of the differential equation
\begin{equation}
 	W^2_k=\omega_k^2 -\frac{\ddot{a}}{a}-\frac{1}{2}\left(\frac{\ddot{W}_k}{W_k}-\frac{3}{2}\frac{\dot{W}_k^2}{W_k^2}\right),
\end{equation}
which follows from  the mode equation (\ref{eq:mode equation}) upon substitution of the ansatz  (\ref{eq:out adiabatic}). The nature of the approximation thus guarantees that $n$ is a positive even number.  Up to four derivatives, the $W_k^{(n)}$ are 
\begin{subequations}\label{eq:W}
\begin{equation}
	W_k^{(0)}=\omega_k,
	\quad
	W_k^{(2)}=W_k^{(0)}+{} ^{(2)}{}W_k,
	\quad
	W_k^{(4)}=W_k^{(2)}+{}^{(4)}{}W_k, 
\end{equation}
where
\begin{eqnarray}
	{}^{(2)}{}W_k&=&\frac{3}{8}\frac{\dot\omega^2_k}{\omega_k^3}-\frac{\ddot{\omega}_k}{4\omega_k^2}-\frac{1}{2\omega_k}\frac{\ddot{a}}{a},
	\\
	{}^{(4)}{}W_k&=&-\frac{297}{128}\frac{\dot{\omega}_k^4}{\omega_k^7}
	+\frac{1}{4}\frac{\dot{a}^2 \, \ddot{a}}{a^3\, \omega_k^3}
	+\frac{5}{8}\frac{\dot{a}\,\ddot{a}\,\dot{\omega}_k}{a^2\,\omega_k^4}
	+\frac{19}{16}\frac{\ddot{a}\,\dot{\omega}^2_k}{a\,\omega_k^5}
	-\frac{\ddot{a}^2}{4 a^2 \,\omega_k^3} 
	+\frac{99}{32} \frac{\dot{\omega}_k^2 \, \ddot{\omega}_k}{\omega_k^6}
	 \\
	 &-&\frac{3}{8}\frac{\ddot{a}\, \ddot{\omega}_k}{a\, \omega^4_k}
	-\frac{13}{32}\frac{\ddot{\omega}_k^2}{\omega_k^5}
	-\frac{\dot{a} \,\dddot{a}}{4a^2 \,\omega_k^3}
	-\frac{5}{8}\frac{\dddot{a}\,\dot{\omega}_k }{a \, \omega_k^4}
	-\frac{5}{8}\frac{\dot{\omega}_k\, \dddot{\omega}_k}{\omega_k^5}
	+\frac{1}{8}\frac{a^{(4)}}{a \, \omega_k^3}
	+\frac{1}{16}\frac{\omega_k^{(4)}}{\omega_k^4}. \nonumber
\end{eqnarray}
\end{subequations}
As we discuss further in Appendix \ref{sec:Adiabaticity During Radiation Domination}, it follows from this expansion that the adiabatic approximation generically  applies when $\omega_k\gg \mathcal{H}$, that is, for sufficiently short wavelength modes or massive fields.  But, strictly speaking,  the  ``adiabatic regime" holds whenever  equation (\ref{eq:out adiabatic}) is a valid approximation to the solution of the mode equation, no matter what the value of the  frequency $\omega_k$ is.  In section~\ref{sec:Intermediate Regime} we discuss an example in which modes are in the adiabatic regime  even though their  frequencies satisfy $\omega_k=k\ll \mathcal{H}$.   In general, however,  once the mode frequency becomes small, $\omega_k\ll\mathcal{H}$, the mode function $\chi_k$ stops oscillating and the adiabatic approximation (\ref{eq:out adiabatic}) breaks down. This is precisely where the following low frequency expansion begins to apply.

\subsubsection{Low Frequencies}
\label{sec:Low Frequencies}

When the  mode frequency $\omega_k$ is sufficiently small, we shall rely on the solution to the mode equation with $k=0$ and $m=0$ as lowest order approximation.   For any scale factor $a(\eta)$, the complex mode function
\begin{equation}\label{eq:chi bar infrared}
	 \chi^{m=0}_0(\eta) = -\frac{a}{\sqrt{2M}}\left(i +M\, b\right),
	 \quad \text{where} \quad
	b\equiv \int^{\eta}\frac{d\tilde{\eta}}{a^2} ,
\end{equation}
 solves equation (\ref{eq:mode equation}) when $\omega_k=0$, and also satisfies the normalization condition (\ref{eq:Wronskian}). Here, $M$ is an arbitrary real and positive mass scale that shall drop out of our final expressions. 

If $\omega_k$ is nonzero, equation (\ref{eq:chi bar infrared})  is clearly not a solution of the mode equation. Instead, it can be regarded as the lowest order solution of the mode equation in the limit of small $\omega_k$, with  a correction $\Delta\chi_k\equiv \chi_k- \chi^{m=0}_0$ that is implicitly determined by 
\begin{equation}\label{eq:scattering}
	\Delta \chi_k=-\int^\eta d\tilde{\eta}\,\,
	 G(\eta;\tilde{\eta}) 
	 \,\omega_k^2(\tilde{\eta}) \chi_k(\tilde{\eta}),
\end{equation}
where $G(\eta;\tilde{\eta})$ is the retarded Green's function of the zero-frequency equation, which can be readily constructed as a linear combination of the two solutions $a$ and $ab$ we just identified above.   
Equation (\ref{eq:mode equation}) and its solution $\chi^{m=0}_0+\Delta\chi_k$ are analogous to those encountered in a scattering problem in nonrelativistic quantum mechanics, in which   $\omega^2_k$ plays the role of the potential, $\chi^{m=0}_0$ is the incoming wave function and $\Delta\chi_k$ the scattered one.  Just as in the scattering problem, we can obtain an explicit solution of the mode equation by recursively expanding (\ref{eq:scattering}) in powers of $\omega_k^2$. The $n$-th order in such an expansion thus contains $n$ powers of $\omega_k$ and is related to the next one by
\begin{equation}\label{eq:chi0 correction}
	 \chi^{\mathrm{low}(n+2)}_k (\eta) =\chi_k^{\mathrm{low}(0)}(\eta) -\int^{\eta} d\tilde{\eta} \, \left[a(\tilde{\eta})\, a(\eta)b(\eta)
		-a(\eta)\,  a(\tilde{\eta}) b(\tilde{\eta})
	\right]  \omega_k^2(\tilde{\eta}) 
 	\chi_k^{\mathrm{low}(n)}(\tilde{\eta}),
\end{equation}
where $\chi_k^{\mathrm{low}(0)} \equiv  \chi_0^{m=0}$, $n$ is an even natural number,   and we have used the explicit form of the  Green's function.  In the Born approximation, one keeps just the  leading order correction, $n=2$. As we discuss further in appendix \ref{sec:Validity of the Low Frequency Approximation}, it follows from this expansion that the low frequency approximation generically  applies when $\omega_k \ll \mathcal{H}$, that is, for sufficiently long wavelength modes and light fields.

\begin{figure}[t!]
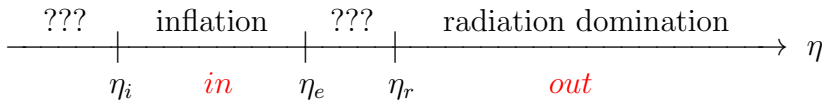


\hspace{2.8cm}??? \hspace{0.55cm} inflation \hspace{0.55cm}  ???  \hspace{0.55cm} radiation domination

\vspace{-.7cm}
\[ \xrightarrow{\hspace*{10.2cm}} \; \eta \]

\vspace{-.745cm}
\hspace{3.6cm} $|$ \hspace{2.1cm} $|$ \hspace{0.8cm} $|\quad$

\hspace{3.54cm} $\eta_i$\hspace{0.8cm}  {\color{red}$in$} \hspace{0.6cm}  $\eta_e$ \hspace{.55cm} $\eta_r$
\hspace{1.5cm} {\color{red}$out$}\\ 
  \caption{Different epochs of  early cosmic evolution. What happened before inflation and between inflation and radiation domination is not fully understood. In a sharp transition $\eta_e=\eta_r$ and there is a discontinuity in the second derivative of the scale factor. In a smooth transition $\eta_e<\eta_r$ and the second derivative of the scale factor is continuous at $\eta=\eta_e$,  though there is still a discontinuity in the higher order derivatives. In the chaotic transition  $\eta_e<\eta_r$ and all the derivatives of the scale factor remain continuous.}
\label{fig:cosmic evolution}
\end{figure}

\subsection{Cosmological Epochs}
\label{sec:Cosmological Epochs}

In order to single out the ``initial" state of the quantum field, we shall consider a spacetime that proceeds from an initial inflationary {\it in} region of accelerated expansion, to a subsequent decelerated {\it out} region. The $out$ region is initially radiation dominated, subsequently undergoes a period of matter domination and currently experiences  a new stage of cosmic acceleration.  The transition between the $in$ and the $out$ region is mediated by  reheating, which is highly model-dependent \cite{Allahverdi:2010xz}.   In this section we present the details of the two asymptotic regimes, as well as some models for the transition between them. Figure \ref{fig:cosmic evolution} sketches a timeline of our model universe.

\subsubsection{$In$ Region}
\label{sec:In Region}

The mode expansion of the field  (\ref{eq:mode expansion}) allows us to introduce the notion of  the number of quanta in each momentum mode. Yet the notion of vacuum, as well as the associated particle number,  relies on the particular choice of  mode functions, which are not uniquely determined  by the dynamical equation (\ref{eq:mode equation}) and the normalization condition (\ref{eq:Wronskian}). We are primarily interested here in transitions from an $in$ region in which a preferred notion of vacuum exists. This is the case if for any fixed value of $k$ the mode equation admits solutions that  in the asymptotic past match the zeroth order ``positive frequency" adiabatic solutions we identified above,
\begin{equation}\label{eq:in}
	\chi^\mathrm{in}_k\to \frac{1}{\sqrt{2 \omega}_k}\exp\left(-i \int^\eta \omega_k(\tilde{\eta})\,  d\tilde{\eta}\right).
\end{equation}
For any field mass, this condition is met in any universe that underwent an early period of inflation, not necessarily of the de Sitter type.  Then  $\mathcal{H}$ tends to zero as $\eta\to -\infty$, and the mode equation admits solutions  that approach (\ref{eq:in}) in this limit. Up to an irrelevant phase, such initial conditions single out the $in$ mode functions $\chi_k^{\mathrm{in}}$, which select a preferred notion of vacuum.  Since as $k\to \infty$ the adiabatic approximation  always remains valid, it also follows then that the mode functions approach (\ref{eq:in}) in the ultraviolet at any moment of cosmic history. 

Strictly speaking,  though,   inflation is not expected to be past eternal  \cite{Borde:2001nh}. When we set initial conditions for the mode functions with equation (\ref{eq:in}) we act as if the $in$ region extended all the way to the asymptotic past. But this is just a device to single out the quantum state of the adiabatic modes at the beginning of inflation, and it does not really presuppose that inflation extends indefinitely into the past. In practice, when the field is massless or light, the finite duration of inflation does require the introduction of an infrared cutoff $\Lambda_\mathrm{IR}\sim \mathcal{H}_i,$ where $\mathcal{H}_i$ is the comoving Hubble constant when inflation starts, since in that case there is no preferred quantum state for  superhorizon modes at the beginning of inflation (if the field is heavy, we can set $\Lambda_\mathrm{IR}=0$.)  In general, we refer to ``light fields" as those with a mass that is much smaller than the Hubble parameter. The latter evolves in time, so a light field in the early universe can become heavy at sufficiently late times. In some cases, however, as in this paragraph, we refer to light fields as those that were light  during inflation. The meaning should be clear from the context.

\begin{figure}[t!]
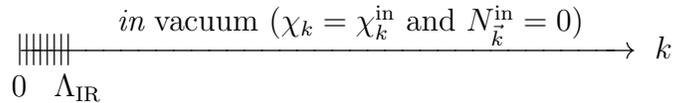

 
\hspace{4.7cm}{\it in} vacuum ($\chi_k=\chi_k^{\mathrm{in}}$ and $N_{\vec{k}}^{\mathrm{in}}=0$)
 
\vspace{-.65cm}
\[ \xrightarrow{\hspace*{8cm}} \; k \]
 
\vspace{-.75cm}
\hspace{3.35cm}$|$ \hspace{-.3cm} $|$ \hspace{-.3cm} $|$ \hspace{-.3cm} $|$ \hspace{-.3cm} $|$ \hspace{-.3cm} $|$ \hspace{-.3cm} $|$ \hspace{-.3cm} $|$
 
\hspace{3.3cm}0 \hspace{0.08cm} $\Lambda_{\mathrm{IR}}$
 
  \caption{Inflation sets a preferred state for the quantum field, the {\it in} vacuum. However, if the field is massless or light, the state of the superhorizon modes at the beginning of inflation, including the zero mode $\vec{k} = 0$, is not determined by inflation and remains unknown to us. This introduces an effective infrared cutoff at $\Lambda_{\textrm{IR}}\sim\mathcal{H}_i$ that we can set to zero if the field is heavy.}
\label{fig:in vacuum}
\end{figure}

When the mode functions in the mode expansion (\ref{eq:mode expansion}) satisfy the $in$ initial conditions (\ref{eq:in}), the corresponding ladder operators  $\hat a^\mathrm{in}_{\vec{k}}$ and $\hat a^\mathrm{in}_{\vec{k}}{}^\dag$ can be identified with those of the $in$ states.   We define then the particle number operator associated with the mode $\vec{k}$ by $\hat N^\mathrm{in}_{\vec{k}}\equiv \hat a^\mathrm{in}_{\vec{k}}{}^\dag \hat a^\mathrm{in}_{\vec{k}}$. The ``$in$ vacuum," $\hat a^\mathrm{in}_{\vec{k}}\ket{0_\mathrm{in}}=0$, has no $in$ quanta, $\hat N^\mathrm{in}_{\vec{k}}\ket{0_\mathrm{in}}=0$, while states with $\hat N^\mathrm{in}_{\vec{k}}\ket{\psi}=N^\mathrm{in}_{\vec{k}}\ket{\psi}$ can be thought of as containing a definite number of quanta $N^\mathrm{in}_{\vec{k}}$.  This is how inflation allows us to identify a preferred  quantum state for all modes with $k>\Lambda_\mathrm{IR}$, the $in$ vacuum. Figure~\ref{fig:in vacuum} represents schematically this situation.

According to the inflationary paradigm, the early universe actually went through a stage of accelerated expansion during which the scale factor grew almost as in de Sitter spacetime. Although de Sitter is often a good approximation for inflation, its group of symmetries is larger than that of a generic (inflationary) Friedman-Robertson-Walker spacetime. We shall thus consider a  generic power-law expansion, which allows us to work with  a one parameter group of inflationary spacetimes that includes  de Sitter,
\begin{equation} \label{eq:a inflation}
	a=a_0 \left( \frac{\eta}{\eta_0}\right)^p,
\end{equation} 
where $p\leq-1$ and $\eta_0<0$.  The effective equation of state in such a universe is ${w=(2-p)/3p}$, which ranges from that of a cosmological constant at $p=-1$ to that of  curvature at $p=-\infty$.  Also note that there is a direct connection between $p$ and the slow roll parameter $\epsilon_1\equiv -\dot{H}/(aH^2)=(1+p)/p$, which is clearly constant during power-law inflation (higher-order slow-roll parameters hence vanish.)   In equation (\ref{eq:a inflation}) $\eta_0$ is an arbitrary time during inflation that we shall choose by convenience, and may take different values in different contexts. At time $\eta_e\geq\eta_0$ inflation ends, and equation (\ref{eq:a inflation}) ceases to describe  the evolution of the scale factor.

\subsubsection{$Out$ Region}
\label{sec:Out Region}

The phenomenological success of the hot Big Bang model implies  that inflation must have been  followed by a radiation dominated epoch, albeit not  necessarily right away. Hence, for phenomenological reasons we shall consider an $out$ region that contains a radiation dominated universe  starting at $\eta=\eta_r\geq\eta_e$, as shown in figure \ref{fig:cosmic evolution}. This choice is not only realistic, but  also streamlines the analysis considerably, because in a radiation dominated universe $\ddot{a}/{a}=0$, which  simplifies the mode equation (\ref{eq:mode equation})  significantly. 

In order to determine the mode functions in the $out$ region, we  need to find the solution of equation (\ref{eq:mode equation}) that  evolved through the transition from the one in the $in$ region. For that purpose, it is often convenient to work with a set of solutions of (\ref{eq:mode equation}) that do not necessarily satisfy the required initial conditions set by the preceding inflationary period.    Let then $\chi^\mathrm{in}_k$ be the solution of (\ref{eq:mode equation})   with the appropriate $in$ boundary conditions, and let $\chi_k$ be an arbitrary solution of equation (\ref{eq:mode equation}). We shall refer to the $\chi_k$ generically as the {\it out} mode functions, although they are not  necessarily related to the {\it out} region. Assuming that   $\chi_k$ and $\chi^*_k$ are linearly independent, we can express the solution $\chi^\mathrm{in}_k$  as a linear combination of the two,
\begin{equation}\label{eq:B transform}
	\chi^\mathrm{in}_k(\eta)=\alpha_k \chi_k(\eta)+\beta_k\chi_k^*(\eta).
\end{equation}
This equation has the form of a Bogolubov transformation.  At this point the nature of the $\chi_k$ is irrelevant. We  just demand  that they solve  (\ref{eq:mode equation}) and that they satisfy the normalization condition (\ref{eq:Wronskian}).  The Bogolubov coefficients $\alpha_k$ and $\beta_k$ are  then constant and, because of equation (\ref{eq:Wronskian}), they satisfy
\begin{equation}\label{eq:constraint}
	|\alpha_k|^2-|\beta_k|^2=1.
\end{equation}
Combining equation (\ref{eq:B transform}) and its first derivative to solve for $\alpha_k $ and $\beta_k$, and making use of the normalization of the mode functions (\ref{eq:Wronskian}), we arrive at
\begin{subequations}\label{eq:Bogolubov coeff}
\begin{eqnarray}
	i \alpha_k=&\chi^\mathrm{in}_k(\eta)  \, \dot{\chi}_k^*(\eta)-\dot{\chi}^\mathrm{in}_k (\eta) \, \chi_k^*(\eta), \\
	i \beta_k=&-\chi^\mathrm{in}_k(\eta) \, \dot{\chi}_k(\eta)+\dot{\chi}^\mathrm{in}_k(\eta)\,\chi_k(\eta).
\end{eqnarray}
\end{subequations}
 The Bogolubov coefficients  (\ref{eq:Bogolubov coeff}) are  readily seen to satisfy equation (\ref{eq:constraint}). If, say, $\chi_k=\chi^\mathrm{in}_k$, these equations imply that $\alpha_k=1$ and $\beta_k=0$, as expected. 
 
Substitution of equation (\ref{eq:B transform}) into  (\ref{eq:mode expansion}) returns an analogous mode expansion, simply with all the mode functions and ladder operators replaced by their $out$ counterparts, provided that we identify 
\begin{equation}
	\hat a_{\vec{k}}\equiv \alpha_k \,  \hat a_{\vec{k}}^\mathrm{in}+ \beta_k^* \, \hat a_{-\vec{k}}^\mathrm{in}\!\!{}^\dag .
\end{equation}
The two mode expansions allow us then to define the number of $in$ and $out$ particles in a given mode $\vec{k}$ as the expectation of the number operators $\hat N^\mathrm{in}_{\vec{k}}\equiv \hat a^\mathrm{in}_{\vec{k}}{}^\dag \hat a^\mathrm{in}_{\vec{k}}$ and $\hat N_{\vec{k}}\equiv \hat a^\dag_{\vec{k}}  \hat a_{\vec{k}}$, respectively. By definition the $in$ vacuum $|0_\mathrm{in}\rangle$ contains no $in$ particles, $\hat a^\mathrm{in}_{\vec{k}}|0_\mathrm{in}\rangle=0$, and  the $out$ vacuum $| 0\rangle$ contains no $out$ particles, $\hat a_{\vec{k}}|0\rangle=0$. But when $\beta_k$ is nonzero, the $in$ vacuum  does contain $out$ particles, $\langle 0_\mathrm{in} | \hat N_{\vec{k}}|0_\mathrm{in}\rangle=|\beta_k|^2$, although it is not an eigenvector of the number operator,    $\hat N_{\vec{k}}|0_{\mathrm{in}}\rangle = |\beta_k|^2|0_{\mathrm{in}}\rangle + \alpha_k^*\beta_k^*|1_{-\vec{k}}^{\mathrm{in}},1_{\vec{k}}^{\mathrm{in}}\rangle$.
As a matter of fact, the $in$ vacuum of a single mode $\vec{k}$ can be regarded as a two-mode  state with an indefinite number of entangled $out$ particles of momentum $\vec{k}$ and $-\vec{k}$.  The limit in which the   magnitudes of the Bogolubov coefficients are large, as we shall encounter later on, has peculiar statistical properties; it  corresponds to what is known as a highly ``squeezed state," which has been argued to display classical behavior \cite{Grishchuk:1990bj,Albrecht:1992kf,Polarski:1995jg,Kiefer:1998qe}. This possible (and somewhat controversial) feature is not relevant for our purposes, though.
 
In general, however, the Bogolubov coefficients do not have an independent meaning by themselves, since they are inherently linked to the arbitrary choice of mode functions $\chi_k$. Only in the adiabatic regime, where from now on we shall assume the latter  approach the positive frequency approximations (\ref{eq:out adiabatic}), do they acquire a relatively context-independent significance. In analogy with section \ref{sec:In Region}, we can therefore claim that in this case there exists a preferred choice of mode functions. Whenever applicable, we shall refer to such set of $\chi_k$ as the ``adiabatic {\it out} mode functions," and to the vacuum state that they define as the ``{\it out} adiabatic vacuum." This will be relevant when we discuss the particle production formalism in section \ref{sec:Particle Production}.

In standard references, such as \cite{Birrell:1982ix}, the adiabatic vacuum corresponds to the \emph{actual} solutions of the mode equation that  match  (\ref{eq:out adiabatic}) and its time derivative at an arbitrary time $\eta_0$. This leads to a two-parameter class of  ``adiabatic vacua" that depends on the adiabatic order $n$ and the matching time $\eta_0$. Our definition is essentially the same, though we are not explicit in our choice of $n$ and $\eta_0$. In most cases, we are just interested in the $in$ vacuum, and the choice of $out$ mode basis functions is inconsequential, or simply dictated by the order of the adiabatic approximation we work at. Only when we introduce the renormalized energy density of the $out$ vacuum it is important to choose $n\geq 4$. In that case, our results are independent of the choice of $\eta_0$, up to subdominant terms of sixth adiabatic order.  

\subsubsection{Transitions}
\label{sec:Transitions} 

We have now established the nature of the $in$ and $out$ regions, but  still need to specify how the transition between the two occurs. Because the transition itself  clearly  depends on the unknown details of reheating, we shall only make rather generic assumptions about its properties. We do assume that the energy-momentum tensor  remains finite, which implies the continuity of $a$ and $\dot{a}$ as  a consequence of the Darmois-Israel junction conditions \cite{Israel:1966rt}, although we also expect the scale factor to be infinitely differentiable in any realistic transition.  In addition, we postulate that cosmic expansion  is such that ${\mathcal{H}_r\leq\mathcal{H}\leq\mathcal{H}_e}$ during reheating. 

In a realistic transition in which the scale factor and its derivative  evolve smoothly, but the equation of state parameter or its derivatives experience a relatively sudden change, we expect an inverse time $s$ to capture the sharpness of the transition. If the transition is ``abrupt," $\mathcal{H}_e\ll s$, modes with $\mathcal{H}_e\ll \omega_k \ll  s$ should  experience departures from adiabaticity that lead to significant particle production. Ultraviolet modes with $\mathcal{H}_e \ll  s \ll \omega_k$, however, will remain adiabatic throughout the transition, and particle production in this range will be negligible. On the other hand in a ``gradual" transition, the scale $s$ obeys $s\leq \mathcal{H}_e$, so we  expect departures from adiabaticity mostly at most around $\omega_k=\mathcal{H}_e$.  When specific examples of these two classes are necessary, the following three transitions to radiation domination shall prove to be useful.

\paragraph{Sharp Transition.} The simplest example of an abrupt transition  consists in taking the limit in which the duration of the transition  goes to zero,  $\eta_r\to \eta_e$, by just  matching the scale factor and its first derivative at the  time of the transition, $\eta_e<0$, as it is mostly done in the literature on the topic \cite{Ford:1986sy,Glavan:2013mra,Aoki:2014ita,Aoki:2014dqa,Yajnik:1990un,Damour:1995pd}. Proceeding that way, we   find
\begin{equation}\label{eq:disc transition}
	a=a_e+\dot{a}_e  \, (\eta-\eta_e) \quad \quad (\eta>\eta_e),
\end{equation}
where $a_e$ and $\dot{a}_e$ respectively are the value of the scale factor and its first derivative at the end of inflation.   This corresponds to a sharp change in the equation of state from $-1\leq w<-1/3$ for $\eta<\eta_e$, to $w=1/3$ for $\eta>\eta_e$, while the energy density remains continuous at $\eta=\eta_e$. One can look at such a transition as a toy model for an abrupt transition, where the  discontinuity in the second derivative of the scale factor is just an idealization of a sudden change within the actual  continuous and differentiable evolution of the scale factor. In the previous language it correspond to  a limit in which the parameter $s$ is sent to infinity. 

As we shall see, the main consequence of such an idealization is an unphysical behavior of the scalar energy density at ultrahigh frequencies. This is not a problem per se, as long as long as the contribution of those frequencies remains subdominant. The problem here is that the \emph{renormalized} energy density diverges in the ultraviolet. Thus, strictly speaking, such a transition is not simply unrealistic; it is also unphysical. We certainly expect a sharp change in the equation of state parameter at the end of inflation to yield  substantial particle production, but we need a  characterization of this process  that avoids a jump in the second derivative of the scale factor.

\paragraph{Smooth Transition.} We thus prefer to consider a scale factor with continuous second time derivatives as an example of an abrupt transition. We achieve this by choosing
\begin{equation}\label{eq:cont transition}
	a=c\left[K_0(e^{-r\eta})+d\, I_0(e^{-r\eta})\right] \quad \quad (\eta>\eta_e),
\end{equation} 
where $c$ and $d$ are dimensionless coefficients, $r$ is a positive constant with dimensions of inverse time, and  $I_\nu$ and $K_\nu$ respectively are the first and second kind modified Bessel functions of order $\nu$.  The main  advantage of equation (\ref{eq:cont transition}) is that
\begin{equation}\label{eq:addoa smooth}
	\frac{\ddot{a}}{a}=r^2 e^{-2r\eta},
\end{equation}
which  simplifies the mode equation after the end of inflation and also rapidly approaches zero, as in a radiation dominated universe. The latter can be readily appreciated by expanding the scale factor in equation (\ref{eq:cont transition}) for small arguments of the Bessel functions,
\begin{equation}\label{eq:a smooth limit}
	a\to c\left(r\, \eta+d+\log 2-\gamma\right)  \quad \quad( r\eta\to \infty),
\end{equation}
where $\gamma$ is the Euler-Mascheroni constant. 

Imposing the continuity of $\ddot{a}/a$ at $\eta_e$  fixes the value of the otherwise arbitrary $r$,
\begin{equation}\label{eq:t0}
	r \eta_e=-W_0\left(\sqrt{p(p-1)}\right),
\end{equation}
where $W_0$ is the principal branch of the Lambert function. In the de Sitter limit, for instance,  $r\eta_e=-W_0(\sqrt{2})\approx -0.7$.   In that particular case $r$ is of the order of the comoving Hubble constant at time $\eta_e$, but in general $r$ can be thought of  as 
a parameter analogous to $s$ above.   The  constants $c$ and $d$ in equation (\ref{eq:cont transition}) are  determined by the continuity of the scale factor and its first derivative at the transition,
\begin{equation}\label{eq:b and c}
	c=\frac{a_e}{r\eta_e}\left[p I_0-\sqrt{p(p-1)}I_1\right],
	\quad
	d=-\frac{ p K_0+\sqrt{p(p-1)} K_1}{p  I_0-\sqrt{p(p-1)} I_1},  
\end{equation}
where the Bessel functions are evaluated at $z=e^{-r\eta_e}$ and we have used their Wronskian to simplify $c$. With these choices of $r$, $c$ and $d$ the energy density and the equation of state remain continuous at $\eta_e$. 

By construction, both  $\dot{a}(\eta)$ and  $\ddot{a}(\eta)$ remain positive after the transition, so the scale factor and its derivative grow monotonically for any value of $p\leq -1$.  An additional advantage of such a smooth transition is that  the time between the end of inflation and the onset of radiation domination is different from zero, as expected in any realistic cosmology.  In order to determine the time $\eta_r$ after which it is safe to assume that the universe is radiation dominated, we shall impose  $\sqrt{\ddot{a}/(a\mathcal{H}^2)}\leq 10^{-1}$. This condition for instance implies that the comoving Hubble scale at time $\eta_r$ is $\mathcal{H}_r\approx 0.3\, \mathcal{H}_e$ after a transition from de Sitter. 

As we shall see, even though the  ultraviolet divergence that was present in the sharp transition is absent here,  the jump in the third and higher derivatives of the scale factor still leads to  substantial particle production in the ultraviolet. But, once again, the spectral density  at ultrahigh frequencies cannot be taken  literally. 

\paragraph{Chaotic Transition.}

Finally, to illustrate some of our results, we shall also consider an example  based on  chaotic inflation with potential \cite{Linde:1983gd}
 \begin{equation}\label{eq:chaotic}
 	V(\varphi)=\frac{\lambda}{4}\varphi^4.
 \end{equation}
Although this inflationary potential was ruled out long ago by observations of the cosmic microwave background \cite{Akrami:2018odb}, it does provide a simple model to study  a gradual transition to a radiation dominated universe in which all the derivatives of the scale factor remain continuous and $\mathcal{H}_e$ is the only scale in the problem.   Indeed, after the end of inflation, around 
${H_e\approx 2 \sqrt{\lambda} M_P}$,
 the oscillations of the inflaton about the minimum of a quartic potential lead to an inflaton energy density that behaves on average like radiation \cite{Greene:1997fu}.  At the time at which a radiation dominated  universe becomes a good approximation for the scale factor evolution the Hubble parameter is $H_r\approx 0.1 \sqrt{\lambda}M_P$ and the universe has expanded by a factor $a_r/a_e\approx 4$; hence, $\mathcal{H}_r\approx 0.2\mathcal{H}_e$ in this case. In these formulae, and throughout the text, $M_P$ denotes the reduced Planck mass $M_P\equiv (8\pi G)^{-1/2}$.

As opposed to what we have assumed in the foregoing, in this model the equation of state during inflation does not remain constant. By matching the equation of state of the inflaton to the exponent in equation (\ref{eq:a inflation}) we find that
\begin{equation}\label{eq:p slow roll}
	p\equiv-1+\Delta p, \quad \Delta p=-\left(\log \frac{a_e}{a}+\frac{1}{3}\right)^{-1},
\end{equation}
 where we have used the slow-roll approximation and  how $\phi$ depends on  the scale factor \cite{Linde:2005ht}.  Hence, the value of $p$ deviates significantly from the de Sitter value $p=-1$ as the end of inflation nears. This is relevant because, as we shall see,  the energy density of the scalar field depends on the value of $p$ during inflation. 
 
As we shall show below, the main difference with  the two previous (abrupt)  transitions is that in the chaotic case the production of particles is highly suppressed for modes with frequencies $\mathcal{H}_e\ll \omega_k$. This is what we expect to happen in a realistic transition that is  gradual.  
 
 \paragraph{Bogolubov Coefficients.}
 
 In order to determine the form of the $in$ mode functions after a jump in the derivatives of the scale factor,  we need to find the solution of equation (\ref{eq:mode equation}) that matches the one in the {\it in} region at the transition time $\eta=\eta_e$. To do so, we shall use an arbitrary set of mode functions $\chi_k$ that solve (\ref{eq:mode equation})  after the transition.  Because the mode equation is second order, imposing continuity of the solution and its derivative at the future boundary of the $in$ region we arrive at the same expressions for the Bogolubov coefficients as in equations (\ref{eq:B transform}) and (\ref{eq:Bogolubov coeff}), with the arbitrary $\eta$ in (\ref{eq:Bogolubov coeff}) now replaced by the matching time $\eta_e$.

Provided that the mode function $\chi^\mathrm{in}_k$ remains  in the adiabatic regime throughout the $in$ region, and under the assumption that  the mode function $\chi_k$ is well approximated by the adiabatic  expression  (\ref{eq:out adiabatic}),  we can already estimate  the Bogolubov coefficients after such a transition (see also \cite{Glavan:2013mra}), 
\begin{subequations}\label{eq:B UV}
\begin{eqnarray}
\displaystyle
	\alpha^\mathrm{ad}_k&\approx&1+i\frac{({}^{(2)}W^+_k/\omega_k)^\cdot-({}^{(2)}W^-_k/\omega_k)^\cdot}{4\omega_k}
	+\frac{({}^{(2)}W^+_k{}-{}^{(2)}W^-_k)^2}{8\omega_k^2}+\cdots,
	\\
	\beta^\mathrm{ad}_k&\approx&\frac{{}^{(2)}W^+_k-{}^{(2)}W^-_k}{2\omega_k}
	-i\frac{({}^{(2)}W^+_k/\omega_k)^\cdot-({}^{(2)}W^-_k/\omega_k)^\cdot}{4\omega_k}
	-\cdots, 
\end{eqnarray}
\end{subequations}
where we have used the continuity of the scale factor and its first derivative at $\eta_e$, and the plus and minus superscripts denote the limits in which $\eta$ approaches $\eta_e$ from above and below, respectively.   The superscripts on the Bogolubov coefficients $\alpha_k $ and $\beta_k$ denote the basis of mode functions that  we employ in equation (\ref{eq:Bogolubov coeff}), which in this case are the adiabatic $out$ mode functions that we introduced in section \ref{sec:Out Region}. Even though the situation in which the adiabatic approximation works well before and after the transition time $\eta_e$ is somewhat idealized, it is expected to appropriately  describe the ultraviolet modes and all the modes of a heavy field. Note that the different terms are organized by growing number of time derivatives.  These formulas establish a link between the smoothness of the transition  and the  behavior of the Bogolubov coefficients in the adiabatic regime. If the transition remains differentiable (in the sense above), the $out$ adiabatic mode functions can be chosen to be the same as those in the $in$ region, so the coefficients $\beta^\mathrm{ad}_k$ vanish, as  noticed after equations (\ref{eq:Bogolubov coeff}). As we have previously argued, the modulus square $|\beta^\mathrm{ad}_k|^2$ represents the expected  number of $out$ particles in that mode, and particle production hence requires departures from adiabaticity.  

As we shall show, the spectral density in the ultraviolet depends on the sharpness of the transition, and,  particularly, on which derivative of the scale factor experiences a sudden change.  On the other hand, on causality grounds one  expects that the long wavelength modes at the end of inflation are unaltered by the transition. Consequently, as far as the scalar energy density is concerned, the exact nature of the transition is relatively unimportant in this regime. In  subsequent sections we  discuss  this universality in more detail.

\subsection{Energy Density}
\label{sec:Energy Density}

 As we mentioned earlier, we are concerned here with the evolution of the energy density of the   scalar $\hat{\phi}$ as the universe transitions from the $in$ to the $out$ region. Just like the energy density of a classical field is determined by the time-time component of its energy-momentum tensor,  in the semiclassical approximation the  energy density of the  quantum field is related to the expectation of the same component, $\rho\equiv a^2 \braket{\hat T^{00}}$.  More rigorously, this identification follows, for instance, from the appropriate quantum-corrected equation of motion of the metric \cite{Armendariz-Picon:2020tkc}, where the expectation value of the energy-momentum tensor sources the semiclassical Einstein equations. 
 
  To further explore the energy density of the scalar, it is convenient to replace the mode sum in the expectation of the energy-momentum tensor by an integral, by taking  
the continuum limit $V\to\infty$. In that case, as in the analysis of Bose-Einstein condensation, the density of momentum states vanishes at $k=0$ and the  contribution of the zero mode $\rho_0$ needs to be treated separately, 
\begin{subequations}
\begin{equation}\label{eq:rho 0+neq0}
 \rho = \rho_0 + \int_0^{\Lambda} \frac{dk}{k}\frac{d\rho}{d\log k},
\end{equation}
where
 \begin{equation}\label{eq:rho0}
\rho_0=  \frac{1}{2a^2V}\left\{\left(N_0+\frac{1}{2}\right)\left[\left|\left(\frac{\chi_0}{a}\right)\spdot\right|^2+m^2\left|\chi_0\right|^2\right]+
 L_0\left[{\left(\frac{\chi_0}{a}\right)\spdot}^2+m^2 \chi_0 ^2\right]+\mathrm{c.c.}\right\},
\end{equation}
and
\begin{equation}\label{eq:spectral generic}
	\frac{d\rho}{d\log k}\equiv \frac{k^3}{4\pi^2a^2}
	 \left\{
	 \left(N_k+\frac{1}{2}\right)\left[\left|\left(\frac{\chi_k}{a}\right)\spdot\right|^2+\omega_k^2 \left|\frac{\chi_k}{a} \right|^2\right] 
	+L_k\left[{\left(\frac{\chi_k}{a}\right)\spdot}^2+\omega_k^2 \, \left(\frac{\chi_k}{a}\right)^2\right]+ \mathrm{c.c.}
	\right\}.
\end{equation}
\end{subequations}
Equation (\ref{eq:spectral generic}) is a ``spectral energy density'' (per logarithmic interval), which can be interpreted as the energy density of a given mode $k\neq 0$. In these expressions the mode functions $\chi_k$ are arbitrary and ``c.c." denotes complex conjugation of all the terms inside the curly brackets, including those that are manifestly real. We have also implicitly defined ${\langle \hat a^{\dagger}_{\vec{k}} \hat a_{\vec{k}'}\rangle \equiv N_k\, \delta_{\vec{k},\vec{k}'}}$   and ${\langle \hat a_{\vec{k}} \hat a_{\vec{k}'}\rangle \equiv L_{k} \, \delta_{\vec{k},-\vec{k}'}}$, where the structure of the expectations is determined by the homogeneity and isotropy of the quantum state, with $N_k$ real and positive, and $L_k$ complex.\footnote{If the state 
is invariant under translations by $\vec{\Delta}$, and the latter are represented by the unitary operator $\hat U$, then 
$\langle \hat a_{\vec{k}} \hat a_{\vec{k}'}\rangle=\langle \hat U^\dag \hat a_{\vec{k}}\hat U \, \hat U^\dag \hat a _{\vec{k}'}\hat U\rangle=e^{-i(\vec{k}+\vec{k}')\cdot \vec{\Delta}}\langle \hat a_{\vec{k}} \hat a_{\vec{k}'}\rangle$, 
which implies that the expectation is proportional to $\delta_{\vec{k},-\vec{k}'}$. Here, we have used the action of $\hat U$ on the annihilation operator, 
${\hat U^\dag \hat a_{\vec{k}} \hat U=e^{-i\vec{k}\cdot\vec{\Delta}}\hat a_{\vec{k}}}$. A similar argument with rotations indicates that the expectation can only depend on the magnitude of the wave 
vectors.}  

Motivated by our discussion in section \ref{sec:In Region} (see also figure \ref{fig:in vacuum}),  we  will divide the integral in~(\ref{eq:rho 0+neq0}) in two pieces
\begin{equation}\label{eq:rhoug}
 	 \rho_{<\Lambda_{\mathrm{IR}}} \equiv \int_0^{\Lambda_\mathrm{IR}}\frac{dk}{k}  \frac{d\rho}{d\log k},
	 \quad
 	 \rho_{>\Lambda_{\mathrm{IR}}} \equiv \int_{\Lambda_\mathrm{IR}}^\Lambda\frac{dk}{k}  \frac{d\rho}{d\log k},
 \end{equation}
where $\Lambda_{\textrm{IR}}$ plays the role of an infrared cutoff in future analyses. 
As we mentioned earlier, if the field is massless or light, the cutoff is expected to be of the order of the comoving Hubble radius at the beginning of inflation, $\Lambda_{\textrm{IR}}\sim\mathcal{H}_i$, while if the field is heavy we can simply set  $\Lambda_{\textrm{IR}}=0$. In the first case, $\rho_{<\Lambda_\mathrm{IR}}$ contains the contribution of those 
modes in the interval $0< k < \Lambda_\mathrm{IR}$ that were already outside the horizon at the beginning of inflation, and whose state hence remains undetermined.  Modes in the interval $\Lambda_\mathrm{IR}<k<\infty$, on the contrary, do have a preferred state, though we have limited their contribution up to those below an ultraviolet cutoff $\Lambda$ that 
we have introduced for later convenience. Note that if inflation is responsible for the origin of the cosmic structure, then $\Lambda_{\mathrm{IR}} < \mathcal{H}$
at any time $\eta_i < \eta <\eta_{\mathrm{today}}$.

 \subsubsection{Modes $0\leq k < \Lambda_\mathrm{IR}$}
\label{sec:Beyond the Infrared Cutoff}

Let us consider first the contribution to the energy density of modes below the infrared cutoff. To proceed, it is convenient to differentiate between the  homogeneous zero mode, $\vec{k}=0$,  and the continuum of nonzero  modes  at $0<k<\Lambda_{\mathrm{IR}}$.

\paragraph{Zero Mode.} As the volume of the universe approaches infinity the energy density of the zero mode (\ref{eq:rho0}) goes to zero, unless the latter finds itself in a ``macroscopic" excitation that results in  a nonzero limit of $\rho_0$. In that case its energy density  (\ref{eq:rho0}) appears to match  that of a \textit{complex} classical homogeneous scalar
\begin{equation}\label{eq:phi cl}
	\phi_\mathrm{cl}=\frac{1}{a}
	\left( A_0 \chi_0+B_0 \chi_0^* \right),
\end{equation}
whose energy density $\rho_{\mathrm{cl}}=(|\dot{\phi}_{\mathrm{cl}}|^2+m^2a^2|\phi_{\mathrm{cl}}|^2)/(2a^2)$  is given by  
\begin{equation}\label{eq:rho phi cl}
\rho_\mathrm{cl}=  \frac{1}{2a^2}\left\{\frac{|A_0|^2+|B_0|^2}{2}\left[\left|\left(\frac{\chi_0}{a}\right)\spdot\right|^2+m^2|\chi_0|^2\right]+
 A_0 B_0^*\left[{\left(\frac{\chi_0}{a}\right)\spdot}^2+m^2\chi_0^2\right]+\mathrm{c.c.}\right\}.
\end{equation}
Such a classical description  works provided that we are able to identify 
\begin{equation}\label{eq:identification}
	\frac{|A_0|^2+|B_0|^2}{2} = \frac{1}{V}\left( N_0+\frac{1}{2}\right), \quad
	A_0 B_0^*=\frac{L_0}{V}.
\end{equation}
These two equations  admit a solution for $A_0$ and $B_0$ whenever $|L_0|\leq N_0\left[1+1/(2N_0)\right]$. 
On the other hand, the Cauchy-Schwarz inequality, applied to the vectors $\hat a_0|\psi\rangle$  and $\hat a_0^\dag |\psi\rangle$,   implies ${|L_0|\leq N_0\sqrt{1+1/N_0}}.$ Therefore, all quantum states allow a classical characterization, in  the sense that there exists a classical field configuration with the same energy density. 
Yet only when  $|L_0|= N_0+1/2$ it is possible to choose  ${B_0=A_0^*}$, which is the condition necessary for the classical field (\ref{eq:phi cl}) to be real. This is the case, for example, if the zero mode finds itself in a coherent state $|A_0\rangle$, with ${\hat a_0|A_0\rangle= \sqrt{V} A_0|A_0\rangle}$.  In general, there is no preferred choice of $\chi_0$. Any  solution of the mode equation (\ref{eq:mode equation}) with $k=0$ that satisfies the normalization condition (\ref{eq:Wronskian}) is equally valid, so the numbers $N_0$ and $L_0$ (and $A_0$ and $B_0$) do not possess any particular meaning by themselves.   

If the field is massless, the mode function of the zero mode $\chi_0$ can be  chosen to be that in  equation~(\ref{eq:chi bar infrared}). In this case $\rho_0$ is determined by the kinetic energy density, which behaves like that of a stiff fluid with equation of state $w=1$,
\begin{equation}\label{eq:stiff zero mode}
 \rho_0 = \frac{M}{4V}\left[\left(N^0_0+\frac{1}{2}\right)+L^0_0+\textrm{c.c.}\right]\frac{1}{a^6}.
\end{equation}
 Therefore, the evolution of the energy density is dictated by the  solution proportional to $a b$, rather than the one that grows with the scale factor, which drops out of the time derivative of the field.  The superscripts in $N^0_0$  and $L_0^0$ emphasize that these are the expectation values  of the ladder operator bilinears linked to  the mode functions (\ref{eq:chi bar infrared}).

If the mass is different from zero, on the other hand,  there are  two different regimes. As long as the field remains light, $m\ll H$, making use of the low frequency expansion we can still  approximate $\chi_0= \chi_0^{\textrm{low}} \approx \chi_0^{m=0}$ in equation~(\ref{eq:rho0}), but in addition to (\ref{eq:stiff zero mode}) there is a nonvanishing contribution to the energy density stemming from the mass term. As a consequence, on top of  the one that mimics a stiff fluid, there  appears a  ``frozen field'' contribution that behaves like a cosmological constant and dominates the energy density once the contribution in (\ref{eq:stiff zero mode}) has redshifted away (we have neglected the contribution of $ab$ to the potential energy,  which decays in an expanding universe with $-1\leq w<1$.) After some time, when the field becomes heavy, $H \ll m$, the classical solution (\ref{eq:phi cl}) starts to oscillate around the minimum of the potential and the low frequency expansion~(\ref{eq:chi bar infrared}) breaks down. In this regime we may use the adiabatic approximation, equation~(\ref{eq:out adiabatic}), to describe the field evolution. Substituting the latter into equation~(\ref{eq:rho0}), we obtain that at zeroth order in the adiabatic approximation the oscillating zero mode behaves like a pressureless fluid, whose overall density is proportional to the value of $N^\mathrm{ad}_0$, where, again, the superscript indicates that we have chosen the mode functions (\ref{eq:out adiabatic}) in the field expansion (\ref{eq:mode expansion}). As we discuss in  section \ref{sec:Particle Production}, this is precisely the behavior  expected from the energy density of  $N^\mathrm{ad}_0$  particles with zero spatial momentum.  As the volume  $V$ approaches infinity,  their number $N^\mathrm{ad}_0$ must grow with $V$ in order for their energy density to approach a non-zero limit. At this point, note that only in the context of the preferred choice of adiabatic $out$  mode functions (\ref{eq:out adiabatic}) does the standard identification of $N_0$ with a particle number and macroscopic excitations with highly populated states really make sense. At any rate, since there is no natural way to specify the state of this mode,  we shall not dwell on its contribution any further. 

\paragraph{Continuum.} A similar problem afflicts the  energy density $\rho_{<\Lambda_\mathrm{IR}}$, which captures the contribution of  modes whose state we also ignore.  In the massless case,  as long as $\Lambda_\mathrm{IR}<\mathcal{H}$, we can set  $\chi_k=\chi^\mathrm{low}_k\approx \chi_0^{m=0}$, as discussed in section \ref{sec:Low Frequencies}. Substituting then the known form of $\chi_0^{m=0}$ into (\ref{eq:rhoug}) we find that 
\begin{eqnarray}\label{eq:rho below IR}
 \rho_{<\Lambda_{\mathrm{IR}}} &\approx& \frac{M}{4}\left[\int_0^{\Lambda_{\mathrm{IR}}}
 	\frac{dk}{k}
 	\frac{k^3}{2\pi^2} \left(N^\mathrm{low}_k+\frac{1}{2}\right) 
	+\int_0^{\Lambda_{\mathrm{IR}}}
 \frac{dk}{k}\frac{k^3}{2\pi^2}
 L^\mathrm{low}_k+\mathrm{c.c.}\right]\frac{1}{a^6}\nonumber\\
 && + \frac{1}{4M} \left[\int_0^{\Lambda_{\mathrm{IR}}}
 \frac{dk}{k}\frac{k^5}{2\pi^2}
 \left(N^\mathrm{low}_k+\frac{1}{2}\right)
	- \int_0^{\Lambda_{\mathrm{IR}}}
 \frac{dk}{k}\frac{k^5}{2\pi^2}
 L^\mathrm{low}_k+\mathrm{c.c.}\right]\frac{1}{a^2},
\end{eqnarray}
where in the second line we have again neglected the contribution of $ab$. The first line of equation (\ref{eq:rho below IR}), which contains the time derivatives of the field, behaves like that of a fluid with equation of state $w=1$, cf. (\ref{eq:stiff zero mode}) above.  As with the zero mode, it  can be cast as the energy density of the homogeneous classical scalar  (\ref{eq:rho phi cl}), provided that we solve equations (\ref{eq:identification}) with $N_0$ and $L_0$ now defined by  
\begin{subequations}\label{eq.n_0.int}
\begin{eqnarray}
	\frac{1}{V}\left(N_0+\frac{1}{2}\right) &\equiv &
 	\frac{1}{2\pi^2}\int_0^{\Lambda_{\mathrm{IR}}}\frac{dk}{k}k^3\left(N_k+\frac{1}{2}\right), 
	\\
	 \frac{L_0}{V} & \equiv &  \frac{1}{2\pi^2}\int_0^{\Lambda_{\mathrm{IR}}}\frac{dk}{k}k^3 \,L_k,
\end{eqnarray}
\end{subequations}
where in these expressions we have omitted the superscripts because they are valid for other choices of mode functions. The second line contribution in (\ref{eq:rho below IR})   scales like spatial curvature and cannot be interpreted as the energy density of a homogeneous scalar; it arises from the field gradients, which are absent if the scalar is homogeneous.  At late times the contribution of the curvature term dominates over that of the stiff fluid, so  the identification of modes $k\neq 0$ with a classical field  is not attainable if the scalar is massless.

In the massive case we need to distinguish between two possible limits.  When $\Lambda_\mathrm{IR}<ma$ all the relevant modes  are nonrelativistic, so we can  approximate the dispersion relation in equation (\ref{eq:spectral generic}) by ${\omega_k\approx ma}$. This allows us to cast the energy density as that of the zero mode  (\ref{eq:rho0}), provided that we make the identifications in (\ref{eq.n_0.int}).  In that case, the density evolution matches the dynamics of a massive classical scalar field  in an expanding universe above:  When $ma\ll \mathcal{H}$, the mode functions are well approximated by equation (\ref{eq:chi bar infrared}),  so $\rho_{<\Lambda_\mathrm{IR}}$ effectively behaves like a cosmological constant, whereas when $\mathcal{H}\ll ma$, the mode functions are well approximated by the oscillatory (\ref{eq:out adiabatic}),  and the energy density scales like that of nonrelativistic matter.  Since the comoving mass grows monotonically, we expect $\Lambda_\mathrm{IR}<ma$ to hold at sufficiently late times, and this is the relevant limit then. 

If the mass satisfies $ma< \Lambda_\mathrm{IR}$, modes in the interval $0<k<ma$ are nonrelativistic, so their spectral density behaves like that of a massive case we just discussed. Similarly, modes in the interval $ma<k<\Lambda_\mathrm{IR}$ are relavistic, and their spectral density behave like that of a massless field.  But since the boundary between the two regimes at $k=ma$ changes with time, we cannot in general make definite predictions about the time evolution of the integrated spectral densities. Hence,  we shall not study this case explicitly here, though the methods we have discussed so far, along with those we present below, could be similarly deployed, and in any case the inequality $ma<\Lambda_{\mathrm{IR}}$ will be violated at sufficiently late times.  It is nevertheless quite remarkable that, even though we ignore the state of modes with $k< \Lambda_\mathrm{IR}$, we can still make quite definite predictions about the behavior of their energy density.   

\subsubsection{Modes $\Lambda_\mathrm{IR}<k<\infty$}

Modes in the range $\Lambda_\mathrm{IR}<k$ find themselves effectively in Minkowski space at the beginning of inflation, where a preferred choice of state exists: the $in$ vacuum.  If, as opposed to a general state, the field is in the $in$ vacuum, $\chi_k=\chi^\mathrm{in}_k$ and $N^\mathrm{in}_k= L^\mathrm{in}_k= 0$,  the energy density $\rho_{>\Lambda_\mathrm{IR}}$ simplifies to
 \begin{subequations}\label{eq:rho ren}
 \begin{equation}\label{eq:rho}
 	\rho^\mathrm{in}_{>\Lambda_\mathrm{IR}}=\int_{\Lambda_\mathrm{IR}}^\Lambda \! \frac{dk}{k}\, \frac{d\rho_\mathrm{in}}{d\log k}, 
	\quad \text{where} \quad 
	\frac{d\rho_\mathrm{in}}{d\log k}\equiv\frac{k^3}{4\pi^2 a^2}   
	\left[\left|\left(\frac{\chi^\mathrm{in}_k}{a}\right)\spdot\right|^2 
	+\omega_k^2\, \left|\frac{\chi^\mathrm{in}_k}{a}\right|^2\right].
 \end{equation}
From now on, unless stated otherwise, the energy density $\rho$ and  the spectral density $d\rho/d\log k$ will be those of modes above the infrared cutoff in the  $in$ vacuum, as in equation (\ref{eq:rho}), but for notational simplicity, we shall omit the labels ``$>\Lambda_\mathrm{IR}$'' and ``$in$'' from  our expressions.

As it stands, the energy density (\ref{eq:rho}) diverges in the ultraviolet, when ${\Lambda\to \infty}$. This follows from equation (\ref{eq:in}), which implies that at large $k$ the leading term in the spectral density is proportional to $k^4$. Pictorially, this divergence arises from a Feynman loop diagram in which a particle is created and annihilated at the same spacetime location.  As described, for instance, in reference \cite{Armendariz-Picon:2020tkc}, it is possible to regularize this  quantity while preserving diffeomorphism invariance by introducing a set of Pauli-Villars regulator fields. The contribution of these regulator fields and the counterterms lead to the renormalized energy density 
\begin{equation}\label{eq:rho ren def}
	\rho_\mathrm{ren}=\rho-\rho_\mathrm{sub},
\end{equation}
where the subtraction terms are
\begin{align}\label{eq:rho sub}
	\rho_\mathrm{sub}&=\frac{1}{2\pi^2}  \Bigg\{
	\frac{\Lambda^4}{8a^4}
	+\frac{m^2 \Lambda^2}{8a^2}
	-\left[\delta\Lambda^f-\frac{m^4}{64}\left(1-2\log \frac{4\Lambda^2}{a^2 \mu^2}\right)\right]
	\\
	 &+\frac{\Lambda^2\mathcal{H}^2}{8a^4}+\left[3(\delta M_P^2)^f-\frac{m^2 }{6}\left(1-\frac{3}{8}\log \frac{4\Lambda^2}{a^2 \mu^2}\right)\right]\frac{\mathcal{H}^2}{a^2}	
	 \notag \\
	& - \left[48\delta c^f-\frac{1}{8}\log\frac{4\Lambda^2}{a^2\mu^2}\right]\left(\frac{\mathcal{H}^2 \ddot{a}}{a^5}+\frac{\ddot{a}^2}{4a^6}-\frac{\mathcal{H}\dddot{a}}{2a^5} \right)
	 -\frac{\mathcal{H}^4}{480 a^4}
	-\frac{\mathcal{H}^2 \ddot{a}}{30 a^5}
	-\frac{19\ddot{a}^2}{480a^6}
	+\frac{19\mathcal{H}\dddot{a}}{240 a^5} \Bigg\}. \notag
\end{align}
\end{subequations}
Then, after the subtraction in equation (\ref{eq:rho ren def}), as the cutoff $\Lambda$ is sent to infinity  the renormalized energy density remains finite and cutoff-independent by construction.

In  expression (\ref{eq:rho sub}) $\mu$ is an arbitrary renormalization scale, and $\delta \Lambda^f$, $(\delta M_P^2)^f$ and $\delta c^f$ are the finite  pieces of the counterterms associated with a cosmological constant, the Einstein-Hilbert term and dimension four curvature invariants, respectively.  Changes in the arbitrary renormalization scale $\mu$ effectively amount to changes in the finite values of these constants, which are determined by appropriate renormalization conditions. In that sense, observables that depend on the values of the counterterms are not predictions of the quantum theory.

The subtraction terms in  (\ref{eq:rho sub}) arise from an adiabatic expansion of the vacuum energy of the regulators. Note in particular that the first line contains no derivatives of the scale factor, the second line contains two, and the third line has four. On dimensional grounds, terms with a higher number of time derivatives vanish as the cutoff $\Lambda$ is sent to infinity.  Leaving the counterterms aside, it  is  in fact straight-forward (though somewhat tedious) to check that, when the field is massive,   $\rho_\mathrm{sub}$ is just the integral of the adiabatic expansion to fourth order   of the vacuum integrand in  (\ref{eq:rho}),
\begin{align}\label{eq:spectral sub}
&\frac{d\rho^\mathrm{ad(4)}}{d\log k}= \frac{1}{4\pi^2} \frac{k^3}{a^3}\Bigg\{
	\frac{\omega_k}{a}+\frac{\mathcal{H}^2}{2a^2}\left(\frac{a}{\omega_k}\right)^5 \left[\frac{9 m^4}{4}+ \frac{3 m^2 k^2}{a^2}+\frac{k^4}{a^4}\right]
	\notag \\
	&+\frac{1}{2}\left(\frac{a}{\omega_k}\right)^{11}
	\bigg[m^8\left( -\frac{189 \mathcal{H}^4}{64a^4}+\frac{45\mathcal{H}^2 \ddot{a}}{8 a^5}+\frac{9\ddot{a}^2}{16 a^6}-\frac{9 \mathcal{H}\dddot{a}}{8a^5}\right) 
	\notag \\
	&+\frac{m^6 k^2}{a^2}\left(\frac{9 \mathcal{H}^4}{8a^4}+\frac{111 \mathcal{H}^2 \ddot{a}}{8a^5}
	+\frac{15 \ddot{a}^2}{8 a^6}-\frac{15 \mathcal{H} \dddot{a}}{4 a^5}\right)
	\notag \\
	&+\frac{m^4 k^4}{a^4}\left(\frac{43 \mathcal{H}^4}{16a^4}+\frac{51 \mathcal{H}^2 \ddot{a}}{4a^5}
	+\frac{37\ddot{a}^2}{16 a^6}-\frac{37 \mathcal{H} \dddot{a}}{8 a^5}\right)
	+\frac{m^2k^6}{a^6}\left(\frac{\mathcal{H}^4}{4a^4}+\frac{11\mathcal{H}^2 \ddot{a}}{2 a^5}+\frac{5\ddot{a}^2}{4 a^6}-\frac{5 \mathcal{H} \dddot{a}}{2a^5}\right) 
	\notag \\
	&+\frac{k^8}{a^8}\left(\frac{\mathcal{H}^2 \ddot{a}}{a^5} +\frac{\ddot{a}^2}{4 a^6}
	-\frac{\mathcal{H}\dddot{a}}{2 a^5}\right)
	\bigg]
	\Bigg\}.
\end{align}
Therefore,  our regularization scheme appears to reproduce and justify the often employed adiabatic scheme \cite{Parker:1974qw}, at least in the massive case,  but it goes beyond it because it makes the role of the counterterms explicit and it also explains the origin of the subtraction terms. 

Yet, from the perspective of Pauli-Villars regularization, the subtraction of adiabatic approximations to the spectral density is not fully justified \cite{Weinberg:2010wq}. In Pauli-Villars the masses of the regulators  are assumed to be much larger than any accessible scale $k$, so their contribution to the \emph{spectral} energy density at  long distances is highly suppressed. For this reason, we shall not distinguish between the unrenormalized and renormalized spectral densities, as long as cosmological scales $k$ are concerned. The regularization and renormalization afforded by the Pauli-Villars regulators is only of consequence in  the ultraviolet, and only there does it play a role. Hence, we shall subtract equation (\ref{eq:rho sub}) from the energy density only when the mode integral includes the contributions of the ultraviolet. 

Differences between Pauli-Villars and the adiabatic scheme renormalization are further underscored by  an unphysical infrared singularity that appears in equation (\ref{eq:spectral sub}) in the massless limit. In that case, the adiabatic scheme leads to a renormalized  energy density that diverges in the infrared, even when the unrenormalized spectral  density and its renormalized   Pauli-Villars counterpart are perfectly well-behaved there. In that limit, the adiabatic scheme fails.
 
Renormalizability  also places constraints on the physically allowed states of the field. Since the energy density is rendered finite by the fixed contributions of the subtraction terms, the occupation numbers of the high-momentum modes need to decay sufficiently fast. In particular, in order for the \emph{renormalized} energy density to remain finite,   $N^\mathrm{in}_k$ needs to decay faster than $1/k^4$. Thermal states with $N^\mathrm{in}_k\propto \exp(-\omega_k/T)$ are thus  physically allowed, while states with $N^\mathrm{in}_k\propto 1/k^4$ are not. We are assuming that the occupation numbers only depend on the magnitude of $\vec{k}$ because of isotropy.

\subsubsection{Particle Production}
\label{sec:Particle Production}

Equations (\ref{eq:rho ren})  suffice to compute the energy density of the scalar $\hat\phi$ in the $in$ vacuum at any time in cosmic history. All one needs is an $in$ region to single out the appropriate state of the field. This fixes the initial conditions for the mode functions $\chi^\mathrm{in}_k$ in the asymptotic past, and equation (\ref{eq:mode equation}) then determines their evolution all the way to the asymptotic future. However, the use of $\chi_k=\chi_k^{\mathrm{in}}$ in equation  (\ref{eq:rho}) is only a possible choice, and the same  energy density can be expressed in any basis of mode functions. In that case, under appropriate circumstances, the spectral density admits an interpretation in terms of produced particles, which we explore next. 

\paragraph{Spectral Density.} In order to obtain the spectral density of the $in$ vacuum in terms of the arbitrary mode functions $\chi_k$, it suffices to  plug  equation (\ref{eq:B transform}) into equation (\ref{eq:rho}). Clearly, by construction, the end result does not depend on the nature of the chosen mode functions $\chi_k$, as long  the state of the field remains unaltered. Carrying out the substitution,  we thus find
\begin{equation}\label{eq:rho bar}
\frac{d\rho}{d\log k}=\frac{k^3}{4\pi^2 a^2}  \Bigg\{
\!\!\left(|\beta_k|^2+\frac{1}{2}\right)\left[\left|\left(\frac{\chi_k}{a}\right)\spdot\right|^2 
	+\omega_k^2\,\left|\frac{\chi_k}{a}\right|^2\right]
	+\alpha_k\beta_k^* 
	\bigg[ {\left(\frac{\chi_k}{a}\right)\spdot}^2+\omega_k^2\, \frac{ \chi_k^2}{a^2}\bigg]
	+\mathrm{c.c.}
	\Bigg\},
\end{equation}
where we have used that $|\alpha_k|^2-|\beta_k|^2=1$. Comparing equations (\ref{eq:spectral generic}) and (\ref{eq:rho bar}) reveals that the $in$ vacuum appears to effectively contain $N_k= |\beta_k|^2$  $out$ particles that are not in an eigenvector of the $out$ number operator,     $L_k= \alpha_k\beta^*_k$.   This formal similarity is behind what is referred to as ``particle production."  In this approach, one would associate  the spectral energy density
\begin{equation}\label{eq:rho produced}
\frac{d\rho_\mathrm{p}}{d\log k}\equiv
	\frac{k^3}{4\pi^2 a^2} \left\{
	|\beta_k|^2 \left[\left|\left(\frac{\chi_k}{a}\right)\spdot\right|^2 
	+\omega_k^2\,\left| \frac{\chi_k}{a}\right|^2\right]
	+\alpha_k\beta_k^*
	\bigg[{\left(\frac{\chi_k}{a}\right)\spdot}^2 +\omega_k^2\, \frac{\chi_k^2}{a^2}\bigg]
	+\mathrm{c.c.}\right\}
\end{equation}
to  the   produced particles, and the contribution left over when $\beta_k=0$ with the spectral density of the $out$ vacuum,
\begin{equation}\label{eq.density.out}
 \frac{d\rho_\mathrm{out}}{d\log k} \equiv \frac{k^3}{4\pi^2a^2}\left[\left|\left(\frac{\chi_k}{a}\right)^{\cdot}\right|^2+\omega_k^2\left|\frac{\chi_k}{a}\right|^2\right].
\end{equation}
Hence, one could regard equation (\ref{eq:rho produced})  as the spectral density of the field  (\ref{eq:rho bar}) from which the spectral density of the $out$ vacuum (\ref{eq.density.out}) has been subtracted.  As a matter of fact, however,  the $out$ vacuum plays no role in our analysis, first because  it depends on the arbitrary choice of mode functions $\chi_k$, and second because we assume that the field is in the $in$ vacuum. Furthermore, since we are interested in the gravitational effects of the field, there is no physical basis for the removal of the $out$ vacuum energy density. In the adiabatic scheme, if the field is massive, renormalization amounts to the subtraction of the spectral density in equation (\ref{eq:spectral sub}). But the latter is the spectral density of the  $out$  vacuum only when the $out$ adiabatic vacuum is actually defined,  in the adiabatic regime,  and only up to factors of sixth adiabatic order.

In any case,   we shall not adopt  adiabatic regularization here, and equation (\ref{eq:rho produced}) is not what is  usually associated with the particle production formalism. Rather,  see for instance \cite{Yajnik:1990un}, the spectral density is often approximated by 
\begin{equation}\label{eq:rho pp}
\frac{d\rho_\mathrm{p}}{d\log k}\approx\frac{k^3}{2\pi^2 a^2} 
	|\beta_k|^2 \left[\left|\left(\frac{\chi_k}{a}\right)\spdot\right|^2 
	+\omega_k^2\, \left|\frac{\chi_k}{a}\right|^2\right],
\end{equation}
which neglects the $out$ vacuum contribution and assumes that the $in$ vacuum is an eigenvector of the $out$ number operator, with eigenvalue $|\beta_k|^2$.  Since the normalization condition (\ref{eq:constraint}) implies  $|\alpha_k \beta_k^*| \geq |\beta_k|^2$, equation (\ref{eq:rho pp}) does not necessarily follow from (\ref{eq:rho produced}).   To  explore the potential applicability of equation (\ref{eq:rho pp}) it is useful to consider the spectral density  when  the corresponding modes  are in the adiabatic regime, where we can approximate $\chi_k$ by equation (\ref{eq:out adiabatic}). This does not generically hold, but  applies, for instance, for massive fields at late times or sufficiently large wavenumbers.   In that case, up to an arbitrary phase, the spectral  density (\ref{eq:rho produced}) reduces to
 \begin{align}\label{eq:rho adiabatic}
 	\frac{d\rho_\mathrm{p}}{d\log k}  
	 \approx  \frac{k^3}{2\pi^2 a^4} \Bigg\{  &|\beta^\mathrm{ad}_k|^2    
	\left[\frac{W_k}{2} +\frac{\omega_k^2+\mathcal{H}^2}{2W_k}+\frac{\mathcal{H}\dot{W}_k}{2W_k^2}+\frac{\dot{W}_k^2}{8W_k^3}\right]
	\nonumber \\
	&+|\alpha^\mathrm{ad}_k \beta^\mathrm{ad}_k|  
	\Bigg[\left(-\frac{W_k}{2}+\frac{\omega_k^2+\mathcal{H}^2}{2W_k}+\frac{\mathcal{H}\dot{W}_k}{2W_k^2}+\frac{\dot{W}_k^2}{8W_k^3}\right) \cos \left(\! 2 \!\int^\eta \!\!\! W_k \, d\tilde\eta\right)
	 \nonumber  \\
	 &\quad\quad\quad\quad \quad {}+\left(\mathcal{H}+\frac{\dot{W}_k}{2W_k}\right)\sin \left(\! 2 \!\int ^\eta \!\!\! W_k \, d\tilde\eta\right) 
	\Bigg]
	\Bigg\}. 
\end{align}
In the light of equation (\ref{eq:rho adiabatic}), the approximation of  the particle spectral density (\ref{eq:rho produced}) by (\ref{eq:rho pp}) does receive some support when frequencies are large, $\omega_k\gg \mathcal{H}$. Then, the leading terms on the second line of equation (\ref{eq:rho adiabatic}) are doubly suppressed: First, because as opposed to those on the first line proportional to $W_k\approx\omega_k$,  they are proportional to $\mathcal{H}$, since the terms of order $\omega_k$ cancel, and second, because they rapidly oscillate with time.  In fact, on cosmological timescales of order $\mathcal{H}$, the evolution of the scale factor is only sensitive to the time average of the energy density, which is  strongly suppressed  when $W_k \gg \mathcal{H} $. The strength of the suppression depends on the particular details of the time average, and we shall simply assume for the time being that the average is such that the terms on the second line remain subdominant in the adiabatic expansion. 

But adiabaticity and high frequencies are still  not sufficient to guarantee the validity of the particle production formula (\ref{eq:rho pp}). In the limit $|\beta^\mathrm{ad}_k|\ll1$, when particle production is ``ineffective," the terms on the first line in equation (\ref{eq:rho adiabatic}), of order $|\beta^\mathrm{ad}_k|^2$,   are suppressed with respect to those on the second line, which are of order $|\alpha^\mathrm{ad}_k \beta^\mathrm{ad}_k|\sim |\beta^\mathrm{ad}_k|$.  Therefore, we can only claim that the terms on the first line are necessarily dominant if, in addition, particle production is ``effective," $|\beta^\mathrm{ad}_k|\gtrsim 1$. If  that is the case,  in the mode range in which the three conditions are  simultaneously met, the spectral  density  is well approximated by
 \begin{equation}\label{eq:rho ppf}
 	\frac{d\rho_\mathrm p}{d\log k}\approx  \frac{k^3}{2\pi^2 a^3}|\beta^\mathrm{ad}_k|^2 \,  \frac{\omega_k}{a},
 \end{equation}
which possesses a clear interpretation in terms of particles, once we identify $|\beta^\mathrm{ad}_k|^2$ with the number of created $out$ particles in the  mode $k$. According to this interpretation, $k^3 |\beta^\mathrm{ad}_k|^2/(2\pi^2)$ is then the comoving number density of particles per logarithmic interval of $k$, the factor $1/a^3$ accounts for the physical volume of  our universe, and the factor  $1/a$ accompanying $\omega_k$ captures the redshift of the particle's energy.   Since equations (\ref{eq:rho pp}) and (\ref{eq:rho ppf}) are valid under the same conditions, but the latter is simpler and more intuitive, we are referring to (\ref{eq:rho ppf}) whenever we invoke the ``particle production formalism." For massless particles the dispersion relation is $\omega_k=k$, and the spectral density  (\ref{eq:rho ppf}) scales like radiation. For massive particles $\omega_k\approx ma$ at late times, and the spectral density  would scale like nonrelativistic matter. Those are the two behaviors usually attributed to free particles.    Although the meaning of the Bogolubov coefficients $\alpha_k$ and $\beta_k$ is tied in general to the arbitrary choice of mode functions $\chi_k$ in equation (\ref{eq:B transform}),  in order to arrive at (\ref{eq:rho ppf}) we have employed the adiabatic approximation (\ref{eq:out adiabatic}).  Hence, the $\beta^\mathrm{ad}_k$ in equation (\ref{eq:rho ppf}) are uniquely determined by that choice of mode functions.   Since  (\ref{eq:rho ppf}) neglects terms with one derivative,  it is inconsequential to calculate $\beta^\mathrm{ad}_k$ beyond the zeroth order adiabatic approximation.  

It is also worth pointing out that  equation  (\ref{eq:rho ppf}) fails at small frequencies even when the modes themselves are in the adiabatic regime and particle production is effective, because to justify it  we need  to assume that $\mathcal{H}\ll \omega_k$ (see for instance the second term in the first line of equation (\ref{eq:rho adiabatic}).) Though this condition usually amounts to the validity of the adiabatic regime,  there are cases in which modes are adiabatic even when their frequencies are small; see  appendix \ref{sec:Validity of the Low and High  Frequency Approximations} for details. Conversely, since the validity of the adiabatic approximation demands that ${}^{(n)}W_k\gg {}^{(n+2)}W_k$ for all $n$, it is conceivable for one of these conditions to  be violated even when frequencies are large.

In conclusion, the particle production formula  (\ref{eq:rho ppf}) is well-justified provided that
\begin{enumerate}
 \item[$i$)] the relevant modes are in the adiabatic regime (${}^{(n)}W_k\gg {}^{(n+2)}W_k$),
 \item[$ii)$] the mode frequencies are large ($\omega_k\gg\mathcal{H})$,
 \item[$iii$)] particle production is effective ($|\beta^\mathrm{ad}_k|\gtrsim 1$).
 \end{enumerate}
Even then one should recognize that the approximation  (\ref{eq:rho ppf}) does not extend beyond the leading adiabatic order, since the terms that are neglected on the second line of equation (\ref{eq:rho adiabatic}) are of first  order.  Note that when particle production is ineffective, $|\beta^\mathrm{ad}_k|\ll 1$, the spectral density of the field is dominated by that of the {\it out} vacuum, equation (\ref{eq.density.out}). If, on the contrary, particle production is not only efficient, but also ``significant," $|\beta^\mathrm{ad}_k|\gg 1$, the particle production formula (\ref{eq:rho ppf}) approximates the full spectral density (\ref{eq:rho bar}), which also includes the {\it out} vacuum contribution. We summarize the conditions under which the different particle production equations  correctly approximate the  spectral density (\ref{eq:rho produced}) in table \ref{tab:conditions}.

\begin{table}
\begin{center}
\begin{tabular}{c c c c}
 \hline
 \hline
  Eq. &$\;$ $ {}^{(n)}W_k\gg {}^{(n+2)}W_k$ $\;$ &  $\;$ $\omega_k \gg \mathcal{H}$ $\;$    &   $\;$ $|\beta^\mathrm{ad}_k|\gtrsim 1$ $\;$\\
 \hline
  (\ref{eq:rho pp}) & \ding{51} & \ding{51}    & \ding{51} \\
  (\ref{eq:rho adiabatic}) & \ding{51} & \ding{55}  & \ding{55}\\
  (\ref{eq:rho ppf}) & \ding{51} & \ding{51}  & \ding{51}\\
 \hline
 \hline
\end{tabular}
\end{center}
\caption{Conditions under which the different particle production formulae are valid approximations to the actual spectral density of produced particles   (\ref{eq:rho produced}). A checkmark and a cross denote whether  the corresponding condition is necessary or not, respectively. Note that  $ {}^{(n)}W_k\gg {}^{(n+2)}W_k$ when modes are adiabatic,  $\omega_k \gg \mathcal{H}$ when frequencies are large, and $|\beta^\mathrm{ad}_k| \gtrsim 1$ when particle production is effective. For minimally coupled scalar fields, the condition $\omega_k\gg \mathcal{H}$ generically implies ${}^{(n)}W_k\gg {}^{(n+2)}W_k$.}
\label{tab:conditions}
\end{table}
  
The  behavior  of the adiabatic $out$ mode functions in the ultraviolet  also allows us to determine under what conditions the renormalized energy density after the transition remains  finite. By construction, the spectral energy density of the  $out$ vacuum (\ref{eq.density.out}) leads to an ultraviolet divergent integral that is regulated and renormalized by the subtraction terms in (\ref{eq:rho sub}). Therefore, the spectral density of the produced particles in (\ref{eq:rho produced}) must yield a finite contribution to the energy density. To estimate its behavior in the ultraviolet, we substitute the leading approximation  $W_k\approx \omega_k\approx  k$ into equation (\ref{eq:rho adiabatic}). At leading order we obtain
\begin{equation}\label{eq:rho UV}
	\frac{d\rho_\mathrm{p}^\mathrm{UV}}{d\log k}\approx \frac{1}{2\pi^2 a^4}
	\left[|\beta^\mathrm{ad}_k|^2 k^4+\mathcal{H} |\alpha^\mathrm{ad} _k \beta^\mathrm{ad}_k| k^3 \sin (2k\eta+\varphi)+\cdots\right],
\end{equation}
which implies that $|\beta^\mathrm{ad}_k|$ has to decay faster than $1/k^2$ in order for the integral to remain finite, since $\alpha^\mathrm{ad}_k\approx 1$ in the ultraviolet. This is the same as the restriction on  the number of \textit{in} particles $N^\mathrm{in}_k$ in the ultraviolet that we discussed previously.   Looking back at equations (\ref{eq:B UV}) it means that the second derivative of the scale factor has to be continuous throughout the transition. Otherwise, it is not just that the energy density diverges; the structure of the divergences is incompatible with our regularization scheme. With $|\beta^\mathrm{ad}_k|\sim 1/k^3$, the spectral density (\ref{eq:rho ppf}) is ultraviolet finite.

\paragraph{Energy Density.} Even though the spectral density is particularly convenient to study the contribution of the different modes to the total energy density, it is not an actual observable. The gravitational equations are sourced by the renormalized total energy density, which, once we have removed the subtraction terms,   is given by the integral of the spectral density. 

In order to establish contact with the particle production formalism, it shall prove to be convenient to split the total energy density into that of the modes for which the $out$ adiabatic vacuum is defined, which we shall label by ``$\mathrm{ad}$," and those for which it is  not, which we shall denote by ``$\cancel{\mathrm{ad}}$." The former are precisely those that satisfy condition $i)$ above, whereas the latter  typically include those beyond the horizon when the field is light or massless.   It then follows from equations (\ref{eq:rho ren}), (\ref{eq:rho bar}), (\ref{eq:rho produced}) and (\ref{eq.density.out}) that
\begin{subequations}\label{eq:rho p ren}
 \begin{equation}\label{eq:rho ren ad nad}
  \rho_\mathrm{ren}= \rho_{\cancel{\mathrm{ad}}}+ \rho^\mathrm{p}_\mathrm{ad}
	 +(\rho_\mathrm{ad}^\mathrm{out})_\mathrm{ren},
 \end{equation}
where 
\begin{equation}
 (\rho_\mathrm{ad}^\mathrm{out})_\mathrm{ren}\equiv \rho_\mathrm{ad}^\mathrm{out}-\rho_\mathrm{sub},
\end{equation}
and
\begin{equation}\label{eq:rho ad def}
 	\rho_{\cancel{\mathrm{ad}}}\equiv \int_{\cancel{\mathrm{ad}}} 
 	\frac{dk}{k} \frac{d\rho}{d\log k}, 
 	\quad
	\rho^\mathrm{p}_{\mathrm{ad}}\equiv \int_\mathrm{ad}
	\frac{dk}{k} \frac{d\rho_\mathrm{p}}{d\log k}, 
	\quad
	\rho_\mathrm{ad}^\mathrm{out}\equiv \int_\mathrm{ad} \frac{dk}{k} \frac{d\rho_\mathrm{out}}{d\log k}.
\end{equation}
\end{subequations}
Note that we have split the energy density of the adiabatic modes, which include those in the ultraviolet, into two contributions, one of the ``produced particles,"$\rho^\mathrm{p}_{\textrm{ad}}$, and that of the renormalized $out$ vacuum for those modes, $(\rho_\mathrm{ad}^\mathrm{out})_\mathrm{ren}$.  It is important to stress that the spectral density that enters the energy density $\rho^\mathrm{p}_\mathrm{ad}$ here is the one in (\ref{eq:rho produced}), since only then is the quoted expression for  $\rho_\mathrm{ren}$ in (\ref{eq:rho ren ad nad}) exact.  Because $|\beta^\mathrm{ad}_k|$ has to decay faster than $1/k^2$, the adiabatic energy density $\rho_{\textrm{ad}}^{\textrm{p}}$ is ultraviolet finite,  and we can directly set the cutoff $\Lambda$ to infinity therein. On the other hand, both $\rho_\mathrm{ad}^\mathrm{out}$ and $\rho_\mathrm{sub}$ are ultraviolet divergent, and only their difference,  $(\rho^\mathrm{out}_\mathrm{ad})_\mathrm{ren}$, remains finite as $\Lambda\to \infty.$  The value of $\rho_{\cancel{\mathrm{ad}}}$ does not depend on our choice of mode functions, since it corresponds to the energy density of the non-adiabatic modes in the $in$ vacuum, and neither does the sum $\rho^\mathrm{p}_\mathrm{ad}+(\rho_\mathrm{ad}^\mathrm{out})_\mathrm{ren}$, which is that of the adiabatic modes in the same state.  In the last case the individual $\rho_{\textrm{ad}}^{\textrm{p}}$ and $(\rho_{\textrm{ad}}^{\textrm{out}})_{\textrm{ren}}$ do actually depend on the election of mode functions, but adiabaticity singles out a “preferred” state, the {\it out} adiabatic vacuum.

Just like the {\it in} vacuum,  the {\it out} vacuum is in general only defined for modes shorter than a certain  length.  However, there are some particular situations in which the {\it out} vacuum can be extended to the whole mode range. In particular, the {\it out} adiabatic vacuum is defined for all modes  when the field is heavy, and the resulting energy density is of sixth adiabatic order, as we find in equation (\ref{eq:rho dS massive}). When the field is massless, it is also possible to define this quantity during radiation domination, although in this case its value is of fourth adiabatic order,  see equation (\ref{eq:rho out massless}).

In the end, though, which of the three components in $\rho_\mathrm{ren}$ dominates the energy density depends on the properties of the $in$ vacuum and the evolution of the universe since the beginning of inflation.  But the applicability of the particle production formalism also requires the validity of conditions $ii)$ and $iii)$. Just as we did above, it is hence useful to split the adiabatic modes into those  that additionally satisfy conditions  $ii)$ and $iii)$ from those that do not.  When the former ``particle production modes" dominate the renormalized energy density, it is possible to interpret our results under a new light.   In that case, the spectral density of the dominant modes can be approximated by equation (\ref{eq:rho ppf}), and  only under those circumstances it is then justified to write
\begin{equation}\label{eq:rho ren ppf}
	\rho_\mathrm{ren}\approx \int_\mathrm{pp} \frac{dk}{k} \frac{k^3}{2\pi^2 a^3} |\beta^\mathrm{ad}_k|^2 \,  \frac{\omega_k}{a},
\end{equation}
where ``$\mathrm{pp}$" denotes that the integral only runs over the modes that satisfy conditions  $i)$, $ii)$ and $iii)$.  Although not necessarily so, this approximation   is expected to apply when the adiabatic approximation is violated for a sufficiently large period of time in the early universe, leading to large values of the coefficients $\beta^\mathrm{ad}_k$. 
Equation (\ref{eq:rho ren ppf}) can be interpreted as the integral over phase space of the distribution function $f(k)=|\beta^\mathrm{ad}_k|^2$ associated to an isotropic classical ensemble of particles of physical energy $\omega_k/a$. With the integration range replaced by all modes,  (\ref{eq:rho ren ppf})  is  the blanket ``particle production" equation often used in the literature (see, for instance, \cite{Kofman:1997yn,Chung:1998zb}.)   Alas, since one or several of the conditions stated above typically fails,  the latter  is often \emph{not} a valid approximation to the field energy density, as we shall see below.

Incidentally, by combining  equations (\ref{eq:rho}), (\ref{eq:rho bar}) and (\ref{eq:rho ppf})  we arrive at an alternative characterization of the effective particle number that appears in equation  (\ref{eq:rho ren ppf}),
\begin{equation}\label{eq:invariant}
	 |\beta_k^\mathrm{ad}|^2= \frac{1}{2\omega_k}\left[\left|\left(\chi^\mathrm{in}_k\right)\spdot\right|^2 
	+\omega_k^2\, \left| \chi^\mathrm{in}_k \right|^2\right]-\frac{1}{2}.
\end{equation}
This expression is an adiabatic invariant, that is, it is a constant in the limit of constant scale factor, that happens to agree with the squared magnitude of the coefficient $\beta_k$ we introduced in (\ref{eq:B transform}).  This is why equation (\ref{eq:invariant}) is sometimes used to define the particle number $N_k\equiv |\beta^\mathrm{ad}_k|^2$ in the literature (see, e.g. \cite{Greene:1997fu,Kolb:2023dzp}), though it  is  only useful  when (\ref{eq:rho ren ppf}) is a valid approximation to the particle spectral density $d\rho_\mathrm{p}/d\log k$, namely, when  the three conditions $i), ii)$ and $iii)$ hold.

At this point it becomes clear  that in most cases the particle production formalism is just an approximation at best. As far as the spectral density is concerned, equation (\ref{eq:rho bar}) remains true no matter whether the notion of particle exists, and regardless of  how the mode functions $\chi_k$ are chosen. Furthermore, if we knew the form of the $in$ mode functions throughout cosmic history, $\chi_k^\mathrm{in}$,  there would be no need to go through the process of introducing Bogolubov coefficients and evaluating (\ref{eq:rho bar}) or its approximations,  (\ref{eq:rho produced}) to (\ref{eq:rho ppf}); it would just suffice to evaluate equations (\ref{eq:rho ren}) at any desired time. The particle production formalism is useful in the adiabatic regime and at high frequencies, where we can interpret the field excitations  as actual particles on top of the $out$ adiabatic vacuum.  However, it does not universally apply to all modes of the field, as it is sometimes implicitly assumed in the literature, nor its use is restricted to asymptotically flat spacetimes, as it is often presented  in the standard monographs. 

\subsubsection{Classical Field Description}
\label{sec:Classical Field Description}

It is also interesting to think about whether there exists a regime in which the classical field formalism applies. 
We already noted earlier that when the excitation of the $\vec{k}=0$ mode is macroscopic, its contribution to the energy density can be cast as that of a homogeneous classical field. 
Remarkably, such a description can be extended in some cases to modes with $k\neq 0$, as we did previously in the interval $0<k<\Lambda_{\mathrm{IR}}$. In this subsection we analyze if this description can be extended further to the mode range $k>\Lambda_{\mathrm{IR}}$, where the state of the field is not arbitrary but is  assumed to be  the {\it in} vacuum. 

Consider for that purpose the renormalized energy density that we introduced in equations~(\ref{eq:rho ren}). If we now split $\rho_\mathrm{ren}$ into the contribution of the nonrelativistic and the relativistic modes, we obtain 
\begin{subequations}\label{eq.decompositionrhop2}
 \begin{equation}
  \rho_\mathrm{ren}=  \rho_{<ma}
	 	+\rho^{>ma}_\mathrm{ren},
 \end{equation}
where 
\begin{equation}
 \rho^{>ma}_\mathrm{ren} \equiv \rho_{>ma}-\rho_\mathrm{sub}
\end{equation}
and
\begin{equation}
 \rho_{<ma}\equiv \int_{\Lambda_\mathrm{IR}}^{ma} \frac{dk}{k} \, \frac{d\rho}{d\log k},\quad
	 \rho_{>ma}\equiv \int_{ma}^{\infty} \frac{dk}{k} \, \frac{d\rho}{d\log k}.
\end{equation}
\end{subequations} 
Note that there is no ambiguity in this decomposition, as the spectral density refers to that of the $in$ vacuum, as opposed to that of the particle and {\it out} vacuum in (\ref{eq:rho ad def}).  In (\ref{eq.decompositionrhop2}) we have assumed that $\Lambda_\mathrm{IR}<ma$; we discuss the opposite case below. To proceed, let us  focus on $\rho_{<ma}$ first. Since the corresponding modes are nonrelativistic, their dispersion relation is 
$k$-independent, $\omega_k^2\approx m^2 a^2$, and their mode functions must be of the form $\chi_k^{\mathrm{in}}\approx \alpha_k\chi_0 + \beta_k\chi_0^*$. Substituting the latter 
into the spectral density (\ref{eq:rho}), we find that we can cast the energy density of the nonrelativistic modes as that of the classical homogeneous field (\ref{eq:rho phi cl}), provided we  solve equations (\ref{eq:identification})  with $N_0$ and $L_0$ now set to  
\begin{subequations}\label{eq.identification0mode}
\begin{eqnarray}
	\frac{1}{V}\left(N_0+\frac{1}{2}\right) & \equiv & \frac{1}{2\pi^2}\int_{\Lambda_{\mathrm{IR}}}^{ma}\frac{dk}{k}k^3\left(|\beta_k|^2+\frac{1}{2}\right),
 	 \\
	 \frac{L_0}{V} & \equiv & \frac{1}{2\pi^2}\int_{\Lambda_{\mathrm{IR}}}^{ma}\frac{dk}{k}k^3 \alpha_k\beta_k^*.
\end{eqnarray}
\end{subequations}
Note that with $N_k=|\beta_k|^2$ and $L_k=\alpha_k \beta_k^*$ these equations differ from (\ref{eq.n_0.int}) only in the integration limits. Indeed, since $N_0$ and $L_0$ are supposed to be constants, such an identification only works if the previous mode integrals are time-independent, which is not guaranteed, since the upper limit of integration depends on time. Assuming this is the case, equations (\ref{eq:identification}) always admit a solution, in analogy  with  the discussion that follows  the same equation.  In particular, the classical interpretation holds even when ``particle production" is suppressed, since $|\beta_k|^2\ll 1/2$ implies $|L_0|\ll N_0+1/2$. In this case the classical scalar would be necessarily complex. In the opposite limit of significant ``particle production," $|\beta_k|^2\gg 1$,  the classical interpretation works too, and whether $N_0+1/2\approx |L_0|$ (which leads to a real scalar)  or $N_0+1/2< |L_0|$  (which leads to a complex one) depends on how the phase of the product $\alpha_k \beta_k^*$ varies with $k$. It is important to stress, though,  that this classical field is not the result of a coherent state, but of quantum vacuum fluctuations of modes that are stretched to superhorizon scales during inflation. Indeed,  when the Bogolubov coefficients are large, the $in$ vacuum can be thought of as a highly squeezed state of $out$ modes.

The classical field interpretation is not generally possible in the case of the relativistic modes, $\rho_\mathrm{ren}^{>ma}$, and for that reason it cannot be applied when $ma<\Lambda_\mathrm{IR}$.  Their mode functions  cannot be approximated by that of the  homogeneous mode, and their dispersion relations imply that  gradients typically contribute to the energy density, unlike what happens with a homogeneous scalar.  As we previously argued in section~\ref{sec:Beyond the Infrared Cutoff}, this is the reason why the interpretation in terms of a classical field of modes $k\neq 0$ is not attainable for massless fields. In particular,   the classical field formalism  misses the contribution of the relativistic modes. 

Even when $\Lambda_{\textrm{IR}}<ma$ and the classical contribution $\rho_{<ma}$ is well defined, in general  it is not justified to assume that  $\rho_\mathrm{ren}^{>ma}$ is subdominant. For example, we discuss in section  \ref{sec:Light Fields Inflation} a case in which the energy density of the nonrelativistic modes (what we call $\rho_\mathrm{fIR}$ therein)  is much smaller than that of the relativistic ones, see the discussion under equation (\ref{eq:rho fIR ultra}).  Nevertheless, if for any reason $\rho_{<ma}$ dominates over $\rho_\mathrm{ren}^{>ma}$, we can approximate 
\begin{equation}\label{eq:rho ren class}
 \rho_{\mathrm{ren}} \approx \frac{1}{2a^2}\left(|\dot{\phi}_{\mathrm{cl}}|^2+m^2a^2|\phi_{\mathrm{cl}}|^2\right),
\end{equation}
where $\phi_\mathrm{cl}$ is computed from equations (\ref{eq:phi cl}), (\ref{eq:identification}) and (\ref{eq.identification0mode}). It is under these circumstances that the classical field formalism becomes really useful. A case of particular interest, with applications to dark matter, arises when the spectral density is dominated by adiabatic nonrelativistic modes. In this case it is possible in principle that the classical field interpretation and the particle production  formalism work simultaneously, with the apparently different approximations (\ref{eq:rho ren ppf}) and (\ref{eq:rho ren class}) yielding  the same result,  ${\rho_{\mathrm{ren}}\approx \rho_{\mathrm{p}}\approx \rho_{\mathrm{cl}}}$, as in some axion dark matter models. We will analyze this possibility in section~\ref{sec:Light Fields} when dealing with light fields at late times in cosmic evolution. In the meantime, table \ref{table:preview} surveys the mode ranges in which the classical field or particle production descriptions are applicable. 

Let us conclude by emphasizing that, although we have discussed the energy density of the scalar within the specific context of cosmic transitions,  many of the  results in  section \ref{sec:Energy Density} are applicable to a much wider class of scenarios almost without modification. All that is essentially needed is for a subset of the  scalar field modes  to be in a preferred state $|0_\mathrm{in}\rangle$ such that $\hat a^\mathrm{in}_{\vec{k}}|0_\mathrm{in}\rangle=0$. How the field reaches this particular state is a question that lies beyond the formalism that we have developed in this subsection, and inflation is just a possibility.

\section{Massless Fields}
\label{sec:Massless Fields}
We shall begin our illustration of the previous results with  a massless field, which is easier to treat  analytically. This is   relevant for the production of massless particles at the end of inflation and  also facilitates our analysis of light fields  later on.  The  energy density of the field $\hat\phi$ is determined by the {\it in} mode functions $\chi_k^{\mathrm{in}}$ through equations (\ref{eq:rho ren}). Qualitatively,  the evolution of the modes throughout cosmic history is relatively simple when $m=0$:  In the asymptotic past the modes find themselves in the short wavelength regime, where they oscillate with positive frequency and constant  known amplitude, as they would do in  Minkowski space. Some of these modes are eventually  pushed by inflation to superhorizon scales, where they stop oscillating and grow in proportion to the scale factor until a fraction reenters after the end of inflation. Once they do,  they oscillate again with an enhanced amplitude, but this time with positive and negative frequencies. These properties alone suffice to determine the shape of the spectral density, which we analyze below. In particular, the evolution of those  modes that left during inflation and later reentered the horizon is behind the phenomenon of  cosmological particle production. Superhorizon and ultraviolet modes, on the other hand, do not admit a particle interpretation, as we clarify next, and only the zero mode $\vec{k}=0$ admits a description in terms of a classical field excitation in the massless case.  A summary of the results obtained for a massless field is presented in section \ref{sec:Overview}.

\begin{table}
\begin{center}
\begin{tabular}{c c l }
 \hline
 \hline
  modes & state   & \multicolumn{1}{c}{formalism} \\
 \hline
  $k=0$   & unknown     & particles: heavy mass (Sec.~\ref{sec:Beyond the Infrared Cutoff})  \\
  && classical field: always applies (Sec.~\ref{sec:Beyond the Infrared Cutoff}) 
  \\
  $0<k<\Lambda_{\mathrm{IR}}$ & unknown     & particles: possibly at high frequencies (Sec.~\ref{sec:Beyond the Infrared Cutoff})   \\
  &&  classical field: nonrelativistic modes (Sec.~\ref{sec:Beyond the Infrared Cutoff})
  \\
  $\Lambda_{\mathrm{IR}}<k<\Lambda$ & {\it in} vacuum      &  particles: possibly at  high frequencies  (Sec.~\ref{sec:Particle Production})  \\
  && classical field: nonrelativistic modes (Sec.~\ref{sec:Classical Field Description}) \\
 \hline
 \hline
\end{tabular}
\end{center}
\caption{Conditions under which the energy density of the different modes of a quantum scalar field admits a description in terms of particles or a classical field.  In the range ${\Lambda_{\textrm{IR}}<k<\Lambda}$ the state of the field is determined by the {\it in} vacuum, so we can predict its contribution to the energy density, which in some cases can be computed  using  the particle or classical field  formalisms, whichever is applicable. See the quoted sections for further details. }
\label{table:preview}
\end{table}

\subsection{Inflation}
\label{sec:Inflation}

An important advantage  of power-law inflation is that the mode functions that satisfy condition (\ref{eq:in})   are explicitly known in the massless case, 
\begin{equation}\label{eq:chi inf massless}
	\chi^\mathrm{in}_k(\eta)=i\frac{\sqrt{-\pi \eta}}{2}H^{(1)}_\nu (- k \eta  ),   \quad \nu\equiv\frac{1-2p}{2}.
\end{equation}
For convenience, we have chosen here the overall phase so that the mode functions are real and positive once modes cross the horizon, in the limit $k \eta\to 0$.  In de Sitter, these mode functions, which correspond to the Bunch-Davies vacuum 
\cite{Bunch:1978yq},   simplify considerably, so we consider this limit first.  Substituting (\ref{eq:chi inf massless}) into (\ref{eq:rho})   it is easy to see that when $p=-1$ the energy density is
\begin{equation}\label{eq:rho power}
	\rho_\mathrm{dS}= \frac{1}{2\pi^2 a^4}\left(\frac{\Lambda^4}{8}+\frac{\Lambda^2\mathcal{H}^2}{8}\right),
\end{equation}
which precisely matches the  cutoff-dependent subtraction terms in equation (\ref{eq:rho sub}) in the massless case, $m=0$. This is of course no coincidence, as the subtraction terms are supposed to cancel the cutoff dependence by construction. Note, in addition, that the energy density is infrared finite and that we have set the infrared cutoff to zero.  Subtracting equation (\ref{eq:rho sub}) from the unrenormalized energy density  (\ref{eq:rho power}) returns the renormalized energy density 
\begin{equation}\label{eq:rho dS}
 2\pi^2\rho_{\mathrm{ren}}^{\mathrm{dS}} =\delta\Lambda^f 
 -3H^2 (\delta M_P^2)^f -\frac{119H^4}{480},
\end{equation}
which is constant.  Up to the counterterms, this is the result  derived by Allen and Folacci \cite{Allen:1987tz}. Yet  in de Sitter space the  Allen-Folacci value  is degenerate with the counterterm contribution associated with the cosmological constant and the Planck mass, which are also constant,  so it is not really an observable quantity.

At any rate, as far as inflation is concerned,  de Sitter spacetime is not truly appropriate, particularly because in de Sitter there are no metric perturbations.  Away from de Sitter there are no exact analytical expressions for the energy density. In order to estimate the latter, we split the integration range into the ``infrared" ($k  \leq\mathcal{H}$) and the ``ultraviolet" ($k \ge \mathcal{H}$) and use  small and large momentum  expansions  to approximate the contribution of each domain to the integral.\footnote{ When we refer to the ultraviolet we  shall sometimes have the upper boundary at $k=\Lambda$ in mind, while others we may be  simply referring   to  subhorizon modes.}   In the ultraviolet, up to terms of fourth adiabatic order, the spectral density takes the form
\begin{equation}\label{eq:spectralinfUVnew}
 \frac{d\rho_{\mathrm{UV}}}{d\log k}\approx \frac{k^4}{4\pi^2a^4}\left[1+\frac{1}{2}\left(\frac{aH}{k}\right)^2+\frac{3}{8}\frac{p^2-1}{p^2}\left(\frac{aH}{k}\right)^4
 \right]. 
\end{equation}
As we have previously argued, equation (\ref{eq:spectralinfUVnew}) is only applicable up to wavenumbers $k$ of the order of the regulators' masses, but those scales are not accessible to low energy observations. At higher momenta the contribution of (\ref{eq:spectralinfUVnew}) to the energy density appears to diverge, but it is rendered finite by the subtraction terms (\ref{eq:rho sub}), which can be regarded as (minus) the energy density of the regulator fields.   Integrating (\ref{eq:spectralinfUVnew}) over the ultraviolet modes and subtracting (\ref{eq:rho sub}) we then arrive at the renormalized energy density
\begin{eqnarray}\label{eq:rho UV power law}
 2\pi^2\rho_{\mathrm{ren}}^{\mathrm{UV}} &\approx& \delta\Lambda^f -3H^2 (\delta M_P^2)^f
 + \frac{36(p^2-1)H^4}{p^2} \left[\delta c^f +\frac{1}{384}\log\frac{\mu^2}{4H^2}\right]
  \nonumber \\
 &&- \frac{122p^2-60p+57}{480p^2} H^4.
\end{eqnarray}
We should  point out that the error in our simple-minded approximations to the mode integrals  is expected to be of order $H^4$, mostly from the lower boundary of the ultraviolet at $k\sim\mathcal{H}$, where the adiabatic approximation breaks down. As a consequence,  equation (\ref{eq:rho UV power law}) is just an order of magnitude estimate of the actual energy density, though it does imply that  the renormalized coupling constant of the dimension four curvature invariants effectively runs logarithmically with time, as  we could have  guessed directly from equations (\ref{eq:rho sub}). A question we shall not address here is whether perturbation theory is still reliable when the logarithm is large, that is, when $H$ is far from the renormalization scale $\mu$. 

In the infrared limit, the spectral density (\ref{eq:rho})  scales like
\begin{equation}\label{eq:inf massless IR}
	\frac{d\rho_{\mathrm{IR}}}{d\log k}\approx \frac{\Gamma^2(\nu)}{2^{4-2\nu}\pi^3}(-p)^{1-2\nu}H^4
 \left(\frac{k}{aH}\right)^{5-2\nu},
\end{equation}
since the leading terms in the infrared expansion cancel in the massless case (more on this below.)  Therefore, the energy density is infrared divergent for $\nu\geq 5/2$ (${p\leq-2}$) and infrared finite otherwise.  Indeed, as we move away from $p=-1$ the long wavelength scale invariant spectrum of field fluctuations becomes redder and redder, until an infrared divergence appears in the spectral density at equations of state ${-2/3\leq w<-1/3}$ ($p\leq-2$).  This range, however, is phenomenologically less relevant, because the spectral index of the  primordial perturbations suggests that $p$ is close to $-1$ (in the simplest inflationary models.) Infrared divergences are typical of massless theories, although we shall see that in some cases they persist when fields are sufficiently light, even close to de Sitter.   In any case, as we pointed out in section \ref{sec:Energy Density}, only modes with $k > \Lambda_{\mathrm{IR}}$ can be assumed to be in the {\it in} vacuum, and in practice the comoving Hubble constant at the beginning of inflation regulates the infrared divergences.

Performing the integral of the spectral density (\ref{eq:inf massless IR}) over the infrared, ${\Lambda_{\mathrm{IR}} \le k \le \mathcal{H}}$, we obtain
\begin{equation}\label{eq:rho IR inf}
	 \rho_{\mathrm{IR}} \approx \frac{\Gamma^2(\nu)}{2^{4-2\nu}\pi^3} \frac{(-p)^{1-2\nu}}{5-2\nu}H^4\left[1-\left(\frac{\Lambda_{\mathrm{IR}}}{aH}\right)^{5-2\nu} \right],
\end{equation}
which is always positive, since it is the integral of a manifestly positive quantity. If  $-2<p<-1$ it grows from zero at the beginning of inflation and soon decays like $H^4$. On the other hand, if $p<-2$ it soon reaches  a value of order  $H^4$ shortly after the onset of inflation and subsequently decays in proportion to $1/a^2$.  In either case the vacuum energy  decays faster than the energy density of the background universe, so unless inflation begins at trans-Planckian energy densities it always remains subdominant.
Note that the cutoff has no practical impact on the energy density when $\nu<5/2$  ($p>-2$), since in that case the energy density is dominated by the shorter modes (first term in the square brackets.) In this  range, however, for the same reason as in (\ref{eq:rho UV power law}),  we can only rely on (\ref{eq:rho IR inf}) as an order of magnitude estimate. For instance, in de Sitter, where an exact analytical solution is available, $\rho_\mathrm{IR}+\rho_\mathrm{ren}^\mathrm{UV}$ overestimates the actual energy density by  about 50\%.

\begin{figure}
\begin{center}
	\includegraphics[width=12cm]{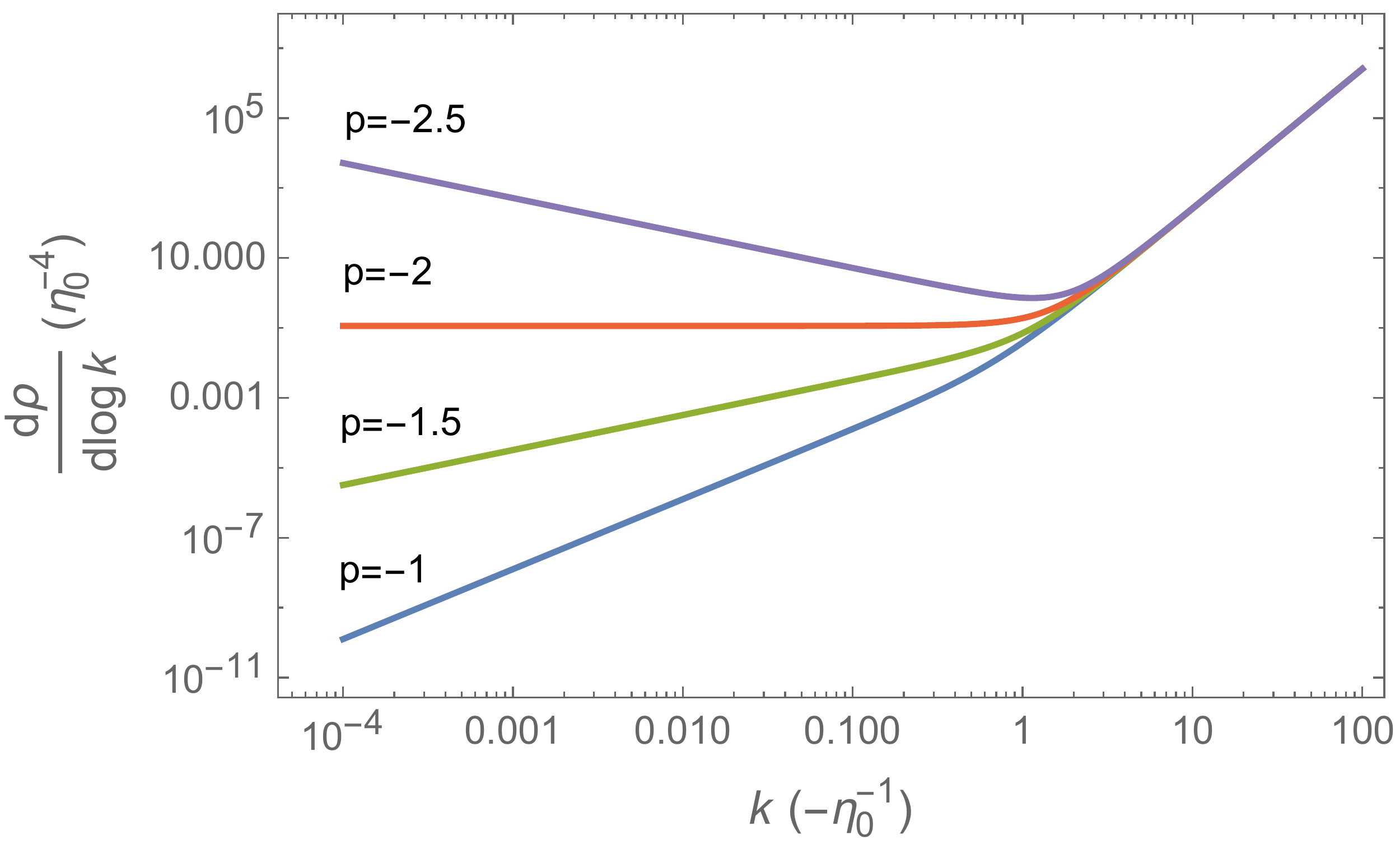}
	\caption{Spectral density during inflation for different values of the parameter $p$ in equation (\ref{eq:a inflation}). The time $\eta_0$ is arbitrary. The power-law behavior in the infrared is well approximated by equation (\ref{eq:inf massless IR}), and that in the ultraviolet  by equation (\ref{eq:spectralinfUVnew}). At $p\leq-2$ the energy density is infrared divergent.}
\label{fig:spectralInf}
\end{center}
\end{figure}

\begin{figure}
\begin{center}
	\includegraphics[width=12cm]{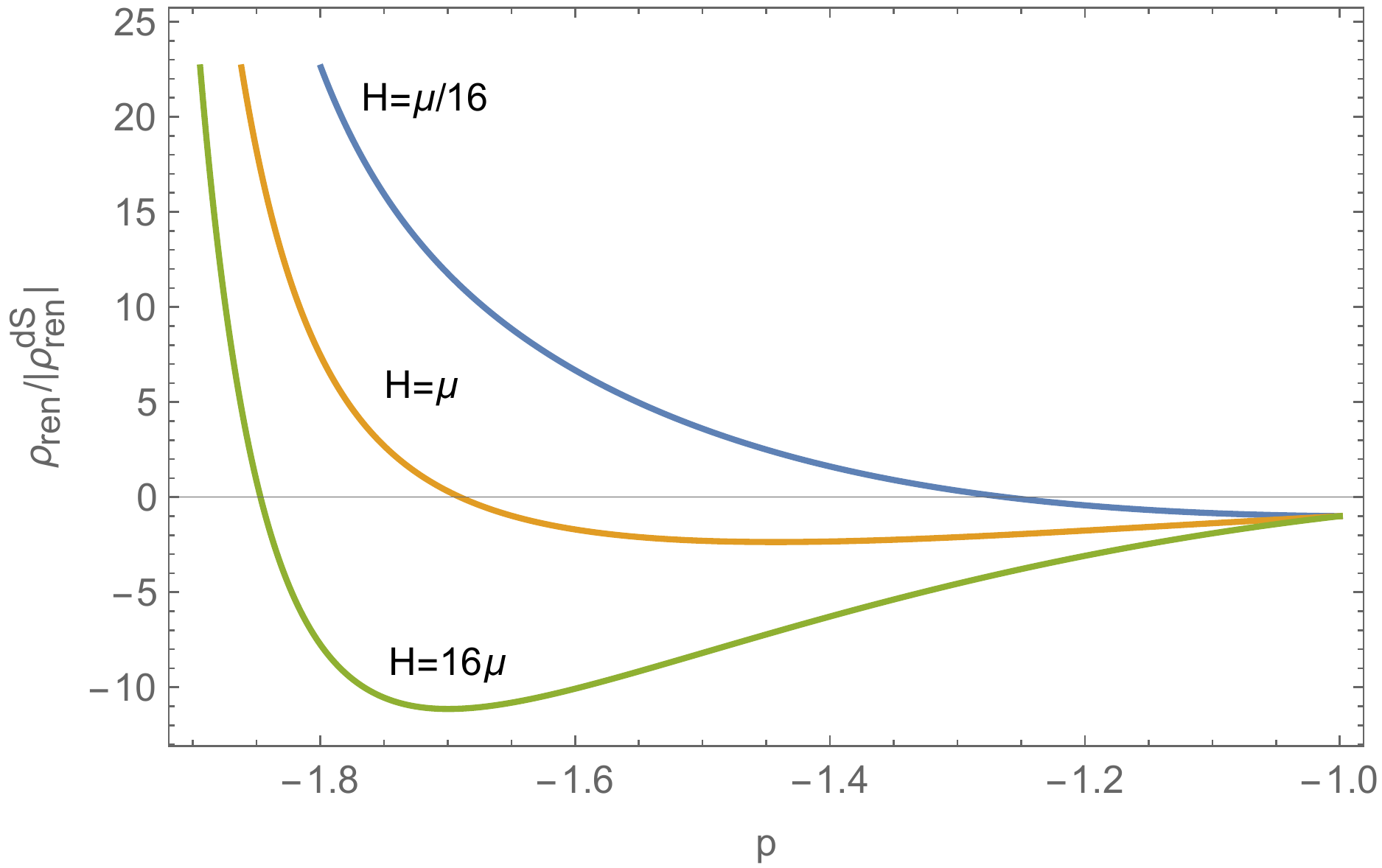}
	\caption{Renormalized energy density  in the range ${-2<p\leq -1}$ for different ratios of the renormalization scale $\mu$ to the Hubble constant $H$.  The energy density is quoted in units of the magnitude of the  Allen-Folacci value (\ref{eq:rho dS}) at the corresponding time during inflation.   We can  think of this figure as one in which time has been fixed and $\mu$ is different in each curve, or one  in which $\mu$ has been fixed and each curve corresponds to a different moment of time. In the latter case, time ``proceeds" along lines of constant $p$  upwards. Note that we have set the finite pieces of the counterterms to zero. The apparent infinite growth of the energy density as  $p=-2$ is approached  is controlled by the infrared cutoff $\Lambda_{\mathrm{IR}}$, that we have set to zero  in this plot.}

\label{fig:rhoInf}
\end{center}
\end{figure}

Figure \ref{fig:spectralInf} shows the spectral density during inflation for a few values of $p$. The two distinct asymptotic regimes in the infrared and ultraviolet are clearly visible in the spectral density, as well as the dependence of the infrared spectral index  on the value of $p$.  In figure \ref{fig:rhoInf} we also  show the numerically determined renormalized energy density  for various values of $p>-2$ and the renormalization scale $\mu$. Our order of magnitude estimate of the total energy density,  namely, the sum of ultraviolet and infrared contributions (\ref{eq:rho UV power law}) and (\ref{eq:rho IR inf}), is in good agreement with the numerical results.  In particular, the  energy density remains of order $H^4$ modulo a running that originates from the decreasing ratio $H/\mu$.  As long as $\mu$ is not too far from $H$, the renormalized energy density near  de Sitter inflation is close to the Allen-Folacci result (\ref{eq:rho dS}), but  when $-2<p<-1$ we expect the energy density to eventually become positive,  though this involves a large logarithmic correction. 

The behavior of  the energy density of a massless scalar field during inflation  is summarized in table \ref{table:rho massless}. Although we have not introduced any  cosmic transition yet,  notice that none of the results that we have obtained in this section can be recovered within the particle production formalism. In the ultraviolet modes are adiabatic,  but the corresponding Bogolubov coefficients $\beta_k$ are small, since these modes never  leave the adiabatic regime.  In the infrared  the adiabatic mode functions are not approximate solutions, and the mode frequencies $\omega_k$ are in any case small.  In either regime, at least one of the three conditions needed to justify (\ref{eq:rho ppf}) cannot be satisfied. On the other hand, as we clarified in section \ref{sec:Classical Field Description}, the formalism of classical fields is not suitable for the non-zero modes of massless scalars.

\subsection{General Infrared Evolution After Inflation}

Just as  the adiabatic expansion allows us to analyze the ultraviolet regime on general grounds, we can also describe the contribution from the infrared modes to the energy density after inflation in a  relatively model-independent way. The key here is that inflation directly determines the properties of the  superhorizon modes, whose  evolution is insensitive to the specific details of the subsequent cosmic expansion and, in particular, to the transtion to radiation domination, in a sense that we make explicit below.

 We shall study the infrared by inspecting the behavior of the $in$ mode functions around $\vec{k}=0$. As we argued in section \ref{sec:Low Frequencies}, the mode functions (\ref{eq:chi0 correction}) are   approximate solutions of the mode equation at low frequencies, $\omega_k=k\ll\mathcal{H}$, irrespectively of the scale factor evolution.   Using the latter and its complex conjugate  as basis, and the small argument expansion of (\ref{eq:chi inf massless}), we can determine the Bogolubov coefficients in the limit where $-k\eta \ll1$   by evaluating (\ref{eq:Bogolubov coeff}) at a time when both expressions are valid, e.g.\ at  an arbitrary time $\eta_0$ during inflation when the modes are larger than the horizon,
 \begin{equation}\label{eq:B light IR}
	 \beta^\mathrm{low}_k \approx -i \frac{\Gamma(\nu)}{2\sqrt{\pi}}\sqrt{\frac{M}{k}}\left(-\frac{k\eta_0}{2}\right)^{1/2-\nu}\frac{1}{a_0}
 ,\quad \alpha^\mathrm{low}_k\approx \beta^\mathrm{low}_k{}^*.
 \end{equation}
In analogy with other expressions,  the superscript `low'  in the Bogolubov coefficients indicates the basis of low-frequency approximations $\chi_k^\mathrm{low}\approx\chi^{m=0}_0$  that we have used to express $\chi_k^{\mathrm{in}}$.

In order to arrive at equations (\ref{eq:B light IR}) we have neglected the term proportional to $(-k \eta_0)^\nu$ in the expansion of the Hankel function (\ref{eq:chi inf massless}), since that term is  suppressed at long wavelengths. Such term is also proportional to the second solution of the mode equation in the limit $k\to0$, $a b$ in equation (\ref{eq:chi bar infrared}), so our approximated $in$ mode functions are purely imaginary. As a consequence, the relation between $\alpha^\mathrm{low}_k$ and $\beta^\mathrm{low}_k$ in (\ref{eq:B light IR})  directly follows from the complex conjugate of one of the equations in (\ref{eq:Bogolubov coeff}).  Furthermore, because our long wavelength approximations of the $in$ mode functions do not satisfy the Wronskian condition (\ref{eq:Wronskian}), the Bogolubov coefficients (\ref{eq:B light IR}) do not obey the normalization condition (\ref{eq:constraint}) either.  Since we used (\ref{eq:constraint}) to arrive at (\ref{eq:rho bar}), care must be taken when the approximation (\ref{eq:B light IR}) is substituted into the former, which only yields a valid approximation  when $|\beta^\mathrm{low}_k|^2$ itself is large. It is therefore necessary to choose a value of $M$ in equation (\ref{eq:B light IR}) that enforces this property. A choice that avoids introducing a new scale in the problem is $M=-\eta^{-1}_0$, which clearly results in large $|\beta_k^\mathrm{low}|$ at long wavelengths, as desired. By construction, our results do not depend on the particular choice of the arbitrary $M$; it just so happens that   (\ref{eq:B light IR})  is a valid approximation only when $M$ is chosen accordingly.

 As long as the scale factor and its first derivative are continuous, the evolution of the $in$  mode functions  on superhorizon scales does not depend on the nature of the transition.  To see this, note that the mode functions of the long wavelength modes during inflation  (\ref{eq:chi inf massless}) are essentially proportional to the scale factor, and because their value and first time  derivative are continuous,  after the transition they  must be still proportional to  $a$, which remains a particular solution of the mode equation when $\omega_k=0$, no matter how the universe expands, see equation (\ref{eq:chi bar infrared}). This is exactly what follows from equations (\ref{eq:B light IR}), which when combined with (\ref{eq:B transform}) and  (\ref{eq:chi bar infrared}) yield a mode  function  that in the infrared equals
 \begin{equation}\label{eq:chi 0 universal}
 	 \chi_k^{\textrm{in}}(\eta) \approx \frac{\Gamma(\nu)}{\sqrt{2k\pi}}\left(\frac{-k\eta_0}{2}\right)^{1/2-\nu}\frac{a}{a_0}. 
	 \end{equation}
Since  equation (\ref{eq:chi bar infrared}) remains a good approximation as long as $\omega_k = k \ll \mathcal{H}$, this expression is applicable for long wavelength modes at any time in cosmic history. As claimed,  the term proportional to $ab$ vanishes at leading order, so the $in$ mode functions find themselves in the growing mode proportional to $a$.  An alternative way to state the same  is to say that $\chi^\mathrm{in}_k/a$ is approximately conserved (i.e.\ the Fourier mode of the actual field $\hat\phi$ is frozen) on superhorizon scales,  no matter how the scale factor evolves.  Observe that in order to arrive at (\ref{eq:chi 0 universal}) we  only used equation (\ref{eq:chi inf massless}) and   $-k\eta_0\ll 1$.  In this context, then, $\eta_0$ may refer to any time during inflation at which the relevant mode is superhorizon-sized. In particular, when equation (\ref{eq:a inflation}) is substituted into (\ref{eq:chi 0 universal}) the dependence on $\eta_0$ drops out, as it also does in equations (\ref{eq:B light IR}).  

\begin{table}
\begin{center}
\begin{tabular}{c c l c c c c c c}
 \hline
 \hline
  & $\quad$ &\multicolumn{3}{c}{power-law inflation} & $\quad$ & \multicolumn{3}{c}{radiation domination} \\
 \hline
  $-2<p\le-1$   && $\rho_{\mathrm{ren}}\propto H^4$  && (IR\&UV)    && $\rho_{\mathrm{ren}}\propto a^{-2(3+p)}$ && (IR\&IM)\\
  $p\le -2$ && $\rho_{\mathrm{ren}}\propto a^{-2}$  && (IR)    && $\rho_{\mathrm{ren}}\propto a^{-2(3+p)}$ && (IM) \\
 \hline
 \hline
\end{tabular}
\end{center}
\caption{Evolution of the  energy density of a massless scalar field during inflation and radiation domination. 
In parenthesis we indicate the modes that mostly contribute to the value of $\rho_\mathrm{ren}$. When $p\approx -1$ the ultraviolet modes can dominate the energy density if the transition-dependent factor $\dot{a}_e/\dot{a}_r$ is greater than one, see table \ref{table:classical/particle massless}. In some cases the density contains an additional logarithmic running that we skip here for simplicity.
}
\label{table:rho massless}
\end{table}

Since at lowest order in the small $k$ expansion $\chi^\mathrm{in}_k/a$ is constant,  its time derivative must be of order $k^2$, and its  square  proportional to $k^4$.  Therefore, the leading contribution to the spectral density in the infrared stems from the nonderivative term in equation (\ref{eq:rho}),
 \begin{equation}\label{eq:rho IR general massless}
	\frac{d\rho_{\mathrm{IR}}}{d\log k}\approx \frac{\Gamma^2(\nu)}{2^{4-2\nu}\pi^3}(-p)^{1-2\nu}
	H_0^4
	\left(\frac{a_0}{a}\right)^{2}
	 \left(\frac{k}{a_0H_0}\right)^{5-2\nu},
 \end{equation}
which during inflation happens to be the same as equation (\ref{eq:inf massless IR}). This agreement underscores the transition independence of the spectral density in this mode range and also illustrates that equation (\ref{eq:rho IR general massless}) holds for any form of the scale factor.  Again, the conditions that we identified in section \ref{sec:Particle Production} are not satisfied here, so we cannot rely on the particle production formula (\ref{eq:rho ppf}). Even though  the modes may be  in the adiabatic regime (see the next section) and  the numbers $\beta_k$ might be large, we are considering superhorizon modes, and the second  of the three conditions in section \ref{sec:Particle Production} fails.
 
The time evolution  of the infrared contribution to the energy density depends on the properties of inflation and, in particular, on the value of $p$. Carrying out the momentum integral  of equation (\ref{eq:rho IR general massless}) over the infrared modes we arrive at 
\begin{equation}\label{eq:rho IR massless final}
	 \rho_{\mathrm{IR}} \approx 
	 \frac{\Gamma^2(\nu)}{2^{4-2\nu}\pi^3} \frac{(-p)^{1-2\nu}}{5-2\nu}H_e^4
	 \left(\frac{a}{a_e}\right)^{3-2\nu}\left(\frac{H}{H_e}\right)^{5-2\nu}
	 \left[1-\left(\frac{\Lambda_{\mathrm{IR}}}{aH}\right)^{5-2\nu} \right],
\end{equation}
where, as in all the subsequent integrated spectral densities,  we have chosen the arbitrary time $\eta_0$ to be the end of inflation, $\eta_e$, since this is when all the relevant modes are beyond the horizon. If  $p\approx-1$, this energy density tracks the background evolution.   Given that $\rho_\mathrm{IR}$ is of order $H_e^4$ at the end of inflation, the contribution of this component to the background density will remain negligible through cosmic history. On the other hand, if $p<-2$, the infrared component scales like curvature, as we already identified in the particular case of inflation.

Even though a description of the energy density in terms of particles is not expected to work in the infrared limit, where frequencies are low, one may think a description in terms of a homogeneous scalar to be appropriate at long wavelengths. However, it is easy to see that this is not possible for a massless field. As we argued in section \ref{sec:Beyond the Infrared Cutoff}, the energy density of a  massless homogeneous scalar field  is proportional to $a^{-6}$.  This is in direct contradiction with the actual behavior of the energy density in (\ref{eq:rho IR massless final}), which approximately decays like $H^2$ when $p\approx -1$. The origin of the disagreement is that the superhorizon modes effectively find themselves in the  mode proportional to $a$, as seen in equation (\ref{eq:chi 0 universal}), so the contribution of these modes to the energy density  does not stem from the terms with time derivatives in equation (\ref{eq:rho}), which are the ones that would lead to the $1/a^6$ scaling,  but from those that contain the gradient of the field, which vanish when the field is homogeneous. The energy density of the latter does not scale like curvature, as we found in equation (\ref{eq:rho below IR}), because the upper limit of integration at $k=\mathcal{H}$  implicit in  equation (\ref{eq:rho IR massless final}) depends on time. Then,  when $-2<p<-1$,  the integral is dominated by the upper boundary, which introduces an additional  time dependence in the energy density. However, when $p<-2$ the infrared modes at $k\sim\Lambda_{\mathrm{IR}}$ dominate, and the integral displays the  same curvature scaling as in section \ref{sec:Beyond the Infrared Cutoff}.

\subsection{Radiation Domination}
\label{sec:General Infrared Behavior}

Modes that left the horizon during inflation eventually reenter after its end. Because the superhorizon modes are not sensitive to the details of the transition between inflation and radiation domination, we can also analyze the spectral density during radiation domination  in a relatively model-independent way.  To do so we postulate that there must be  a time $\eta_r>\eta_e$  after which we can safely assume  the universe to be radiation dominated, see figure~\ref{fig:cosmic evolution}. Then, our analysis  applies in the range $k \ll \mathcal{H}_r$ (modes that reenter or remain outside the horizon during radiation domination) because the behavior of these modes during the transition is universal. At the same time, if inflation  ends at $\eta_e$, our analysis also applies in the range $\mathcal{H}_e\ll k$ (modes that were inside the horizon at the end of inflation) because these  remain in the adiabatic regime thoughout if the transition is gradual.  What happens in the interval $\mathcal{H}_r\lesssim k \lesssim \mathcal{H}_e$ depends on the details of the transition, and is thus model dependent (note that $\mathcal{H}$ decreases after the end of inflation by assumption.) Nevertheless, this is expected to be a small window in comparison with the mode intervals for which we can make definite predictions.

\begin{figure}[t!]
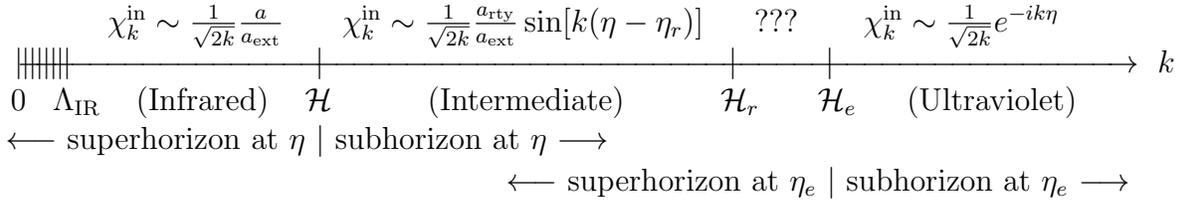


 \hspace{1.1cm} $\chi_k^{\mathrm{in}}\sim\frac{1}{\sqrt{2k}}\frac{a}{a_{\mathrm{ext}}}$ \hspace{0.5cm} $\chi_k^{\mathrm{in}}\sim\frac{1}{\sqrt{2k}}\frac{a_{\mathrm{rty}}}{a_{\mathrm{ext}}}\sin[k(\eta-\eta_r)]$ 
 \hspace{0.4cm}  ???  \hspace{0.5cm} $\chi_k^{\mathrm{in}}\sim\frac{1}{\sqrt{2k}}e^{-ik\eta}$

\vspace{-.9cm}
\[ \xrightarrow{\hspace*{14.7cm}} \; k \]

\vspace{-.935cm}
\hspace{0cm}$|$ \hspace{-.3cm} $|$ \hspace{-.3cm} $|$ \hspace{-.3cm} $|$ \hspace{-.3cm} $|$ \hspace{-.3cm} $|$ \hspace{-.3cm} $|$ \hspace{-.3cm} $|$ 
\hspace{3.1cm}$|$ \hspace{5.1cm} $|$ \hspace{0.9cm} $|\quad$

\hspace{-.05cm}0 \hspace{0.08cm} $\Lambda_{\mathrm{IR}}$ \hspace{.2cm} (Infrared) \hspace{0.2cm} $\mathcal{H}$ \hspace{1.0cm} (Intermediate) \hspace{1.0cm} $\mathcal{H}_r$ \hspace{.5cm} 
$\mathcal{H}_e$ \hspace{0.4cm} (Ultraviolet)\\ 
\vspace{.05cm}
\hspace{-.39cm} $\longleftarrow$ superhorizon at $\eta$ $|$ subhorizon  at $\eta$ $\longrightarrow$

\hspace{6.4cm} $\longleftarrow$ superhorizon at $\eta_e$ $|$ subhorizon at $\eta_e$ $\longrightarrow$

  \caption{ Modes of a massless field as function of the scale at any time $\eta>\eta_r$ after a gradual transition to  radiation domination. The scale factor subscripts `ext' and `rty' denote its values at horizon exit and reentry, respectively.  Ultraviolet (infrared) modes are shorter (larger) than the horizon at the end of inflation $\eta_e$ and also at $\eta$. On the contrary intermediate modes entered the horizon during radiation domination. The initial conditions for modes with $k<\Lambda_{\mathrm{IR}}$ are unknown, whereas those modes with $\mathcal{H}_r \lesssim k \lesssim \mathcal{H}_e$ depend on the details of reheating.  }
\label{fig:massless ranges}
\end{figure}

In order to proceed, it shall prove to be convenient to divide the different modes into three distinct ranges.   Following a similar convention as in section \ref{sec:Inflation}, we shall refer to the short wavelength regime $\mathcal{H}_e\leq  k$ as  the ``ultraviolet."  Modes that reenter during reheating, $\mathcal{H}_r \le k \le \mathcal{H}_e$, belong to the ``transition" range, while those that reenter during radiation domination, $\mathcal{H} \le k \le \mathcal{H}_r$, to the ``intermediate" one. Finally, those that remain outside the horizon during radiation domination, $k\le \mathcal{H}$, are part of the ``infrared,"
\begin{equation}\label{eq:k ranges}
\Lambda_\mathrm{IR}\leq k_\mathrm{IR}\leq \mathcal{H}\leq k_\mathrm{IM}\leq \mathcal{H}_r\leq  k_{\mathrm{T}}\leq \mathcal{H}_e\leq k_\mathrm{UV}.
\end{equation}
These ranges, along with the behavior of the mode functions in each of them after a gradual transition, are visually represented in figure \ref{fig:massless ranges}. In the infrared, in particular,  the spectal density is given by the ``universal" equation (\ref{eq:rho IR general massless}).

\paragraph{Infrared and Intermediate Regimes.}
\label{sec:Intermediate Regime}

The prescribed evolution of the scale factor during the radiation era also fixes the mode functions of those modes that enter the horizon during that time. To  see this, note that, once the transition to radiation domination has been completed, equation (\ref{eq:chi 0 universal}) remains a solution of the mode equation at long wavelengths. In that regime, the mode functions $\chi_k^\mathrm{in}$  in that equation can be cast as the Bogolubov transformation in equation (\ref{eq:B transform}), with mode functions $\chi_k$ given by
\begin{equation}\label{eq:wk0}
	\chi^\mathrm{rad}_k=\frac{e^{-i k (\eta-\eta_r)}}{\sqrt{2k}},
\end{equation}
which  happen to solve the mode equation (\ref{eq:mode equation}) during radiation domination at any $k$.   Also note that (\ref{eq:wk0}) is an adiabatic solution of the form (\ref{eq:out adiabatic}), even when  the frequency of the superhorizon modes is much smaller than the comoving Hubble constant, ${\omega_k=k\ll \mathcal{H}}$. This is a consequence of the linear dependence of the scale factor on conformal time during radiation domination, when $\ddot{a}/a=0$. In such a case, the mode equation (\ref{eq:mode equation}) reduces to that in Minkowski spacetime, where the adiabatic approximation is exact.

To express  the $in$ mode functions (\ref{eq:chi 0 universal}) as a linear combination of   (\ref{eq:wk0}) and its complex conjugate,  we evaluate first the Bogolubov coefficients (\ref{eq:Bogolubov coeff}) at a time and mode range when both are valid solutions of the mode equation,  say, at $\eta=\eta_r$ and $k\ll \mathcal{H}_r$, 
\begin{equation}\label{eq:B IR general}
	\beta_k^\mathrm{rad}\approx -i\frac{\Gamma(\nu)}{2\sqrt{\pi}}
	\left(-\frac{k \eta_0}{2}\right)^{1/2-\nu}
	\frac{a_r}{a_0}
	\left(\frac{\mathcal{H}_r}{k}+i\right),
	\quad
	\alpha^\mathrm{rad}_k\approx \beta_k^\mathrm{rad}{}^*,
\end{equation}
where the notation is analogous to that in equation (\ref{eq:B light IR}). Modes outside the horizon at $\eta_r$ satisfy $\mathcal{H}_r/k\gg 1$, which fixes the dominant term inside the last parenthesis in equation (\ref{eq:B IR general}), although it is necessary to keep the subdominant one too to reproduce the correct  infrared behavior in equation (\ref{eq:chi 0 universal}): At long wavelengths the leading term in  the parenthesis of (\ref{eq:B IR general}) picks up the mode proportional to  $ab$ in equation (\ref{eq:chi bar infrared}), whereas the subdominant one excites the desired mode proportional to $a$.  Note that the Bogolubov coefficients are large, and it would be necessary to include the contribution of the subdominant mode in equation (\ref{eq:chi 0 universal}) in  order to recover the normalization condition ${|\alpha^\mathrm{rad}_k|^2-|\beta^\mathrm{rad}_k|^2=1}$. Inserting the adiabatic $out$ mode functions (\ref{eq:wk0}) and the Bogolubov coefficients (\ref{eq:B IR general}) into equation (\ref{eq:B transform}) we recover the $in$ mode functions in the $out$ region,
\begin{equation} \label{eq:chi in rad}
 \chi_k^{\mathrm{in}} \approx \frac{\Gamma(\nu)}{\sqrt{2k\pi}}\left(-\frac{k\eta_0}{2}\right)^{1/2-\nu}\frac{a_r}{a_0}\left[\frac{\mathcal{H}_r}{k}\sin\left[k(\eta-\eta_r)\right]+
 \cos\left[k(\eta-\eta_r)\right]\right],
\end{equation}
which holds for modes in the infrared and intermediate regimes. In the infrared the amplitude of  (\ref{eq:chi in rad})  is proportional to the change in the scale factor since horizon exit, whereas in the intermedite regime the amplitude of the dominant oscillatory factor  matches the  ratio  of scale factors  at   reentry to horizon exit, as asserted in figure \ref{fig:massless ranges}.

 Substitution of  (\ref{eq:chi in rad})  into equation (\ref{eq:rho}) readily reproduces equation (\ref{eq:rho IR general massless}) in the infrared; then $a_r$ drops out of the equation and $a_0$ can be taken to be  the scale factor at an arbitrary time during inflation.  In the intermediate regime, on the other hand,  the spectral energy density becomes
\begin{equation}\label{eq:rho IM general massless}
	   \frac{d\rho_{\mathrm{IM}}}{d\log k}\approx \frac{\Gamma^2(\nu)}{2^{4-2\nu}\pi^3}(-p)^{1-2\nu}\,
	   H_0^2 H^2
 \left(\frac{k}{a_0H_0}\right)^{3-2\nu},
\end{equation}
which is nearly scale invariant if inflation is  de Sitter-like.  Yet this expression does not capture some of the subleading corrections that are relatively important for modes not too far inside the horizon. Going one order higher in the small wavelength expansion results in 
\begin{equation}\label{eq:spectral osc corrections}
\frac{d\Delta \rho_\mathrm{IM}}{d\log k} 
	\approx -
 \frac{d\rho_{\mathrm{IM}}}{d\log k}\times \frac{\mathcal{H}}{k}\sin[2k(\eta-\eta_r)],
%
\end{equation}
which we could have also recovered from the subleading term in equation (\ref{eq:rho adiabatic}) (up to the relative sign, due to the arbitrary phase in that equation.) Note that we have only used that the corresponding modes enter the horizon during radiation domination.   It is also worth stressing, as should be clear from the way in which they were derived,  that  $\eta_0$ and $\eta_r$ in  equations like (\ref{eq:chi in rad}) and (\ref{eq:rho IM general massless}) can be  taken to be arbitrary times during inflation and radiation domination, respectively, as long as the mode  is superhorizon-sized at both times. 

At this point, it is worthwhile to stress that we could have  used the particle production formalism to arrive at the spectral density (\ref{eq:rho IM general massless}), since in the intermediate regime the three conditions we quote in section \ref{sec:Particle Production} are satisfied. In particular, to arrive at  (\ref{eq:rho IM general massless}) it suffices to substitute the Bogolubov coefficients (\ref{eq:B IR general}) into the particle production formula (\ref{eq:rho ppf}). On the other hand, the subleading correction in the small wavelength expansion (\ref{eq:spectral osc corrections})   goes beyond the particle production formalism, and cannot be recovered from (\ref{eq:rho ppf}).

Finally, in order to determine the total energy density of the intermediate modes, we simply need to integrate (\ref{eq:rho IM general massless}) over the corresponding range $\mathcal{H}\leq k\leq \mathcal{H}_r$. Ignoring the oscillatory corrections  we find
\begin{equation}\label{eq:rho massless IM}
	 \rho_{\mathrm{IM}} \approx \frac{\Gamma^2(\nu)}{2^{4-2\nu}\pi^3} (-p)^{1-2\nu}H_e^4\left(\frac{a}{a_e}\right)^{3-2\nu}\left(\frac{H}{H_e}\right)^{5-2\nu}\log\frac{a}{a_r},
 \end{equation}
with the logarithm  replaced by the $(2\nu-3)^{-1}$ at late times, when $\log (a/a_r)\gtrsim (2\nu-3)^{-1}$.  Up to the former logarithmic dependence, for a universe that underwent near de Sitter inflation, this scales like radiation and is of order $H_e^4$ when extrapolated to the end of inflation, as it also happens for the infrared modes.   Remarkably, at late times this contribution to the density decays slower than curvature when ${p<-2}$, and  in fact effectively violates the weak energy condition when $p<-3$. It is well-known that quantum corrections can lead to violations of the standard energy conditions, see for instance \cite{Epstein:1965zza} for an early reference.

\paragraph{Transition and Ultraviolet Regime.}
\label{sec:massless ultraviolet}
 
Both ranges  encompass subhorizon modes at the onset of radiation domination.   In order to estimate the spectral density here  we plug  the mode functions (\ref{eq:wk0}) into the general expression (\ref{eq:rho bar}), and obtain
\begin{equation}\label{eq:spectral UV massless}
	 \frac{d\rho_{\mathrm{T}+\mathrm{UV}}}{d\log k}\approx \frac{k^4}{2\pi^2a^4}
	 \left[\frac{1}{2}+\frac{1}{4}\left(\frac{aH}{k}\right)^2
+|\beta^\mathrm{rad}_k|^2 +|\alpha^\mathrm{rad}_k\beta^\mathrm{rad}_k|\left(\frac{aH}{k}\right)\sin\left(2k\eta+\varphi\right)\right].
\end{equation}
The first two terms  in (\ref{eq:spectral UV massless}) are characteristic of the $out$ adiabatic vacuum, whereas the details of the transition are encoded in the parameters $\alpha^\mathrm{rad}_k$ and $\beta^\mathrm{rad}_k$. 

Even though in the transition range we cannot accurately estimate the magnitude of the spectral density in a model independent way, we can predict how it scales in time. Indeed, integrating equation (\ref{eq:spectral UV massless}) over the interval $\mathcal{H}_r\leq k\leq\mathcal{H}_e$ gives the energy density 
\begin{equation} \label{eq:rho T massless}
\rho_\mathrm{T}\approx\frac{1}{2\pi^2 a^4} \int_{\mathcal{H}_r}^{\mathcal{H}_e} \frac{dk}{k} k^4\left(|\beta^\mathrm{rad}_k|^2+\frac{1}{2}\right)= \rho_{\mathrm{rad}}^{\mathrm{T}}\left(\frac{a_r}{a}\right)^4,
\end{equation}
where we have restricted ourselves to the leading contribution and $\rho_{\mathrm{rad}}^{\mathrm{T}}$ is a transition dependent factor.  Thus, the total energy density in this range scales like radiation, as expected again from massless particles. In fact,  in this mode range, the mode functions are adiabatic. Hence, we expect the particle production formalism to work, at least around modes with $\mathcal{H}_r\lesssim k$, where $\beta^\mathrm{rad}_k$ ought to be relatively large,  even though we cannot precisely calculate how many particles were produced.  In any case, we do not anticipate the interval $\mathcal{H}_r \le k \le \mathcal{H}_e$ to be broad enough to  result in a substantial contribution to $\rho_{\mathrm{ren}}$, and for that reason we expect the details of the transition to be fairly inconsequential.

In the ultraviolet the  coefficients $\beta^\mathrm{rad}_k$  strongly depend on the abruptness of the transition, though the actual energy density, at least qualitatively, does not.   Consider for example a gradual transition. In this case,  the Bogolubov coefficients are highly suppressed at  $k\gg\mathcal{H}_e$, as we  argued in \ref{sec:Transitions} and illustrate in more detail  in the next section.\footnote{These coefficients are the ones we described in the paragraph below equation (\ref{eq:rho UV}). Since the mode functions (\ref{eq:wk0}) are also adiabatic we can use the superscripts `rad' and `ad' interchangeably.}  Thus, neglecting the contributions proportional to the Bogolubov coefficients,  integrating over the ultraviolet modes and subtracting (\ref{eq:rho sub})  we arrive at 
\begin{equation}\label{eq:rho massless UV}
2\pi^2 \rho_\mathrm{ren}^{\mathrm{UV}} \approx \delta\Lambda^f 
-3H^2 (\delta M_P^2)^f 
+\frac{H^4}{480} + \frac{H^2H_e^2}{8}\left(\frac{a_e}{a}\right)^2 -\frac{H_e^4}{8}\left(\frac{a_e}{a}\right)^4,
\end{equation}
which just corresponds to the renormalized energy density of the $out$ adiabatic vacuum in the corresponding mode range and, leaving the counterterms aside,  is  negligible during  radiation domination.  Note that in a radiation dominated universe, $p=1$, the counterterm proportional to $\delta c^f$ vanishes.  Then, beyond the renormalization of the cosmological and Newton's constants, the term that dominates the ultraviolet energy density at late times and can be regarded as a prediction of the quantum theory is the last one. As it turns out, though, as in the case of other integral approximations, equation (\ref{eq:rho massless UV}) should be regarded as a rough  estimate of the energy density in the ultraviolet, with an error of order $H_e^4 (a_e/a)^4$ that is comparable to our leading  prediction. Since in an abrupt transition we also expect departures from adiabaticity in the range $\mathcal{H}_e\ll k$, and because the spectral density ought to decay with increasing $k$, the lower boundary at $k=\mathcal{H}_e$ gives the dominant contribution to the integrated energy density in the ultraviolet, which is again of order 
$H_e^4 (a_e/a)^4$. 

Note that the energy densities in the infrared (\ref{eq:rho IR massless final}) and intermediate regimes  (\ref{eq:rho massless IM}) are barely sensitive to the ratio $a_e/a$ if inflation is close to de Sitter.  This allows us to predict the energy densities in those ranges at any moment of cosmic history, regardless of the transition,  provided the scale at which inflation ended, $H_e$, is known. On the other hand, in the ultraviolet the energy density (\ref{eq:rho massless UV}) explicitly contains non-zero powers of $a_e/a$. We can partially hide this ratio  by invoking the identity 
\begin{equation}\label{eq:Hubble identity}
	H=H_e \left(\frac{\dot{a}_r}{\dot{a}_e}\right) \left(\frac{a_e}{a}\right)^2,
\end{equation}
where the ensuing transition-dependent factor $\dot{a}_r/\dot{a}_e$ manifestly reflects the dependence on the unknown properties of the transition.

\subsection{Examples} 
\label{sec:Examples}

As an illustration of our general analysis and conclusions, we shall proceed to study the spectral density after the three kinds of transitions described in section \ref{sec:Transitions}. We  find that the energy densities in the  infrared and the intermediate regimes agree with the model-independent predictions we discussed earlier. On the other hand, the nature of the transition  does impact  the ultraviolet, where the spectral index of the spectral density depends on the abruptness of the transition, even though  the total energy density in this range is relatively model-independent.

\paragraph{Sharp Transition.}

With the scale factor given by (\ref{eq:disc transition}), the appropriately normalized positive-frequency solutions of the mode equation (\ref{eq:mode equation}) at $\eta\geq \eta_e$ are (\ref{eq:wk0}), where we choose $\eta_r=\eta_e$.  The actual solution during radiation domination can be cast as a linear combination of the form  (\ref{eq:B transform}), with the coefficients determined by (\ref{eq:Bogolubov coeff}). In the latter the $in$ mode functions are those of equation (\ref{eq:chi inf massless}) and the time $\eta$ is $\eta=\eta_e$.  We shall not write down the values of $\alpha_k$ and $\beta_k$ explicitly, but  merely study their asymptotic behavior.

The Bogolubov coefficients are dimensionless, so they depend on $k$ and $\eta_e$ only through the combination $k \eta_e$. Both in the infrared  and  intermediate regimes the wavenumbers obey $-k\eta_e\ll1$. In such a limit, the Bogolubov coefficients can be readily seen to match the general result in equation (\ref{eq:B IR general}), with $
\eta_0=\eta_r=\eta_e$, as pertains to  a sharp transition.  In the infrared, by combining equations (\ref{eq:B transform}),  (\ref{eq:wk0}) and (\ref{eq:B IR general}) we obtain precisely the mode function in equation (\ref{eq:chi 0 universal}), with the scale factor given by equation (\ref{eq:disc transition}), as expected from our general analysis.  Similarly, inserting the coefficients  (\ref{eq:B IR general}) and mode functions (\ref{eq:wk0}) into equation (\ref{eq:rho bar}) we obtain the leading infrared contribution to the spectral density, which exactly matches that of equation (\ref{eq:rho IR general massless}) and thus validates our model-independent analysis. There is a similar agreement between our general analysis and the results of a sharp transition in the intermediate regime. In this case the transition regime is empty, since $\eta_r=\eta_e$, so $\rho_\mathrm{T}$  vanishes.

The main differences between a gradual  and an abrupt transition, however, appear in the modes that are smaller than the horizon at the end of inflation. In the ultraviolet, up to a phase, we find that the Bogolubov coefficients approach
\begin{eqnarray}\label{eq:B UV sharp}
	\beta_k\approx \frac{p(p-1)}{4(-k\eta_e)^ 2}, \quad \alpha_k\approx 1+i\frac{p(p-1)}{2(-k\eta_e)}-\frac{p^ 2(p-1)^ 2}{8(-k\eta_e)^ 2}.
\end{eqnarray}
Observe that care must be exercised when comparing the Bogolubov coefficients defined using different sets of mode functions:
The coefficients (\ref{eq:B UV sharp})  do not agree exactly with the coefficients one gets from equations (\ref{eq:B UV}) because the corresponding mode functions $\chi^\mathrm{in}_k$ do not match either.   In any case, as noted earlier, the $1/k^2$ dependence of $|\beta_k|$ implies that the energy density is not renormalizable, in the sense that the renormalized energy density is itself logarithmically divergent.   An analogous logarithmic divergence was noted by Ford for nearly conformally coupled massless scalars \cite{Ford:1986sy}.  In the de Sitter limit, the coefficients  in (\ref{eq:B UV sharp}) are exact (the omitted terms vanish), resulting in  $|\beta_k|^2=(k \eta_e)^{-4}/4$, as found, for instance, in \cite{Yajnik:1990un}. The nonrenormalizability of the energy density implies that such a  transition is unphysical: Presumably, the (infinite) backreaction due to the quantum field would prevent the sharp transition from actually happen. The expectation of the energy-momentum tensor after a sharp transition has been also discussed in references \cite{Glavan:2013mra} and \cite{Aoki:2014ita}.

It is also instructive to determine the  energy density within the particle production formalism.  In this case we expect the total energy density to be the sum of the energy of  each mode times the number density of particles in that mode, as in equation (\ref{eq:rho ppf}). However, in order to follow previous literature \cite{Yajnik:1990un,Damour:1995pd}, we shall rely instead on equation (\ref{eq:rho pp}), which as we anticipated yields similar results. We shall  simply ignore the ill-behaved ultraviolet  and focus on the infrared and intermediate regimes.  Substituting the Bogolubov coefficients (\ref{eq:B IR general}) and the associated mode functions (\ref{eq:wk0}) into (\ref{eq:rho pp}),  for any value of $p$ at late times, the dominant contribution to the energy density becomes
\begin{equation}\label{eq:rho inf PP} 					
	\frac{d\rho_\mathrm{p}}{d\log k}\approx \frac{\Gamma^2(\nu)}{2^{4-2\nu}\pi^3}(-p)^{1-2\nu}
	H_e^4
	\left(\frac{a_e}{a}\right)^{4}
	\left[1+\frac{\mathcal{H}^2}{2k^2}\right]
 	\left(\frac{k}{a_eH_e}\right)^{3-2\nu}.
\end{equation}
This expression is correct in the intermediate regime, $\mathcal{H}\ll k\ll \mathcal{H}_r$, since it reduces to the spectral density (\ref{eq:rho IM general massless}) in that range.  In fact, as we mentioned earlier, in this regime the particle production formalism applies, and we could have simply  used the particle production formula  (\ref{eq:rho ppf}) to find the correct answer. But the approximation utterly fails in the infrared, where the leading term in equation (\ref{eq:rho inf PP})  drastically differs from equation (\ref{eq:rho IR general massless}).   The reason behind the disagreement is that equation (\ref{eq:rho inf PP}) misses the cross terms proportional to $\alpha_k\beta_k^*$ and their conjugates in equation (\ref{eq:rho bar}), which are not subdominant in the infrared, because the condition $\omega_k\gg \mathcal{H}$ is not satisfied.  In particular,   in the limit $\omega_k=k  \ll \mathcal{H}$, the latter cancel the leading terms proportional to $|\beta_k|^2$. For similar  reasons, it is also quite clear at this point that, in general, equation (\ref{eq:rho pp}) and the particle production formula (\ref{eq:rho ppf}) fail to  estimate the contribution of the long wavelength modes of a massless field. Equation (\ref{eq:rho inf PP})  is essentially the result derived in reference \cite{Yajnik:1990un}, and its  surviving contribution in the limit $\eta\to \infty$ is the one derived in  \cite{Damour:1995pd}.

\paragraph{Smooth Transition.}

The ultraviolet divergence that we have encountered in the sharp transition prevents us from making meaningful statements about the ultimately  observable  renormalized energy density.  Let us hence explore now the smooth transition to radiation domination described by (\ref{eq:cont transition}). This transition is only smooth in the sense that $\ddot{a}$ remains continuous at  $\eta=\eta_e$. The third and higher derivatives of $a$  do still experience a jump, although this has no  impact on the renormalizability of the energy density, as we explain next.  

With the scale factor given by equation (\ref{eq:cont transition})  a properly normalized solution of the mode equation (\ref{eq:mode equation}) is
\begin{equation}\label{eq:w smooth}
	\chi_k=2^{i\mu} e^{i k\eta_r}\frac{\Gamma\left(1+i\mu\right)}{\sqrt{2k}} I_{i\mu}\left(e^{-r\eta}\right) , \quad
	\mu\equiv\frac{k}{r},
\end{equation}
where $I_{i\mu}$ is the modified Bessel function of imaginary order $i\mu$, and we have introduced a time $\eta_r$ deep in the radiation era for later convenience.   Using the power series of the $I_{i\mu}$ it is readily seen that in the limit of large $\eta$  the mode functions approach the positive frequency solution (\ref{eq:wk0}). 

\begin{figure}
\begin{center}
	\includegraphics[width=12cm]{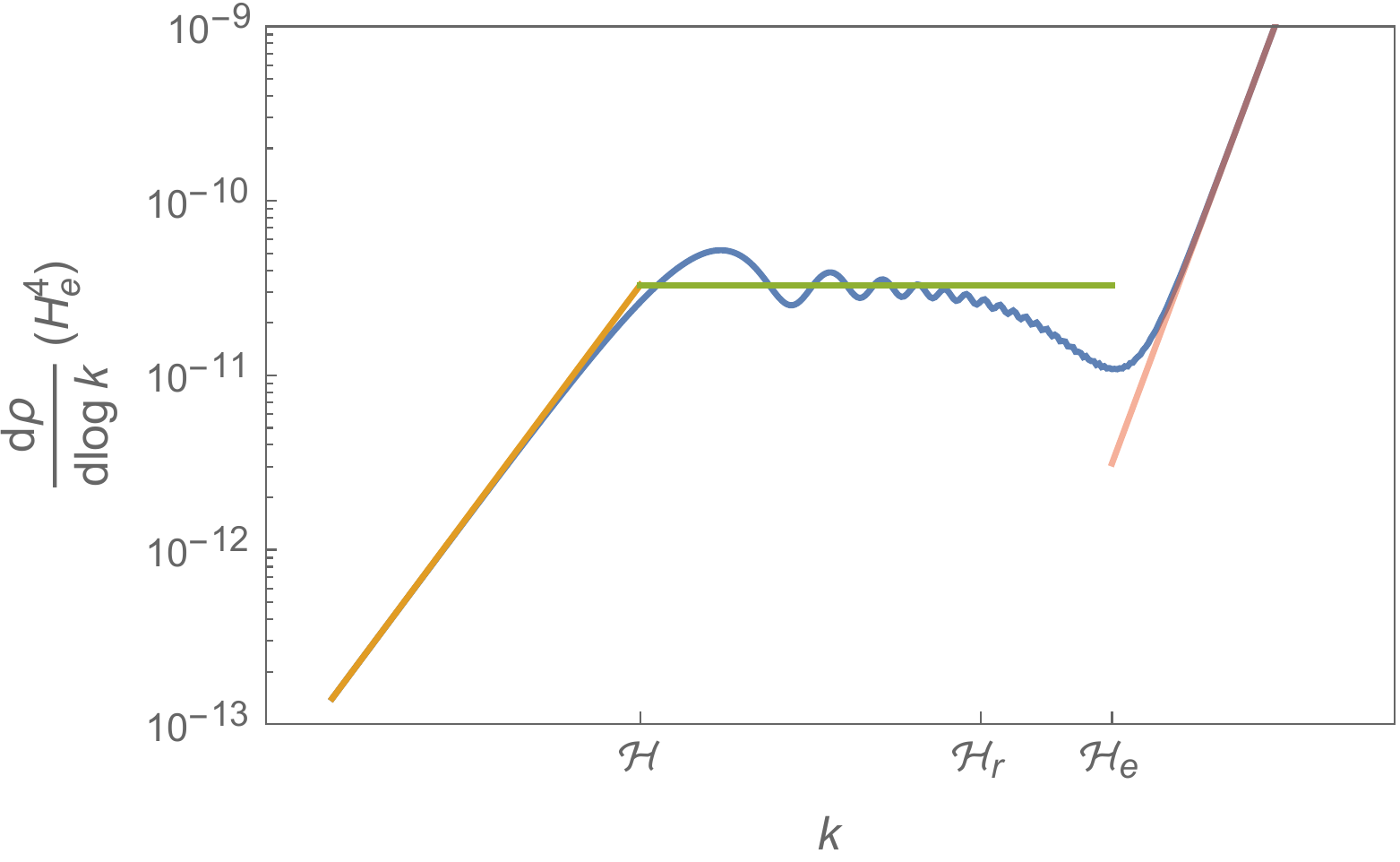}
	\caption{Spectral density  of a massless scalar field after  a smooth transition from de Sitter inflation to radiation domination (in blue).  The superimposed  power-law spectra (orange, green and red) are the corresponding analytical approximations in the infrared (\ref{eq:rho IR general massless}), intermediate (\ref{eq:rho IM general massless})  and ultraviolet regimes (\ref{eq:spectral UV massless}).  Note that the oscillations  in the intermediate regime arise from the corrections in equation (\ref{eq:spectral osc corrections}), and decrease in amplitude as $k$ increases, as expected.  The oscillations in the far end of the ultraviolet that  we expect from equation (\ref{eq:spectral smooth UV}) are not perceptible in this figure. Note that the particle production formalism reproduces the correct behavior only in the range $\mathcal{H}\le k\le \mathcal{H}_e$, although our estimates do not apply to the small transition band $\mathcal{H}_r\le k\le \mathcal{H}_e$.  In this example ${a/a_e=3\cdot 10^2}$.}
	\label{fig:smooth spectral density}
\end{center}
\end{figure}

In order to analyze the infrared and intermediate regimes we note that for modes that satisfy $k\eta_e\ll1$, and provided that $p$ is of order one, equation (\ref{eq:t0}) implies that $\mu\ll 1$. Therefore, expanding the mode function (\ref{eq:w smooth}) to first order in $\mu$  using the derivative of the Bessel function with respect to its order we arrive at the approximation
\begin{equation}\label{eq:chi bar smooth IR}
\chi_k\approx \frac{I_0(e^{-r\eta})}{\sqrt{2k}}
	+\frac{i}{\sqrt{2k}}\left[I_0(e^{-r\eta})\left(\log 2+r\eta_r-\gamma\right)-K_0(e^{-r \eta})\right]\frac{k}{r}.
\end{equation}
This is a linear combination of  two linearly independent solutions of the mode equation at $k=0$, just as in equation (\ref{eq:chi bar infrared}). In fact, with the scale factor given by (\ref{eq:cont transition}),  by changing integration variables and using the Wronskian of the Bessel functions it is seen that $b$ in equation (\ref{eq:chi bar infrared}) is also a linear combination of $I_0(e^{-r\eta})$ and $K_0(e^{-r\eta}$).  Hence, by construction, to first order in $k$ the mode function $\chi^\mathrm{in}_k$   in equation (\ref{eq:B transform}) is  proportional to the scale factor, as we established previously in equation (\ref{eq:chi 0 universal}). 

Matching the  small argument  expansion  of  (\ref{eq:chi inf massless})  to the approximate mode functions in equation (\ref{eq:chi bar smooth IR}) we arrive at the Bogolubov coefficients
\begin{equation}\label{eq:B smooth IR}
	\beta_k \approx -\frac{i}{2}\frac{\Gamma(\nu)}{\sqrt{\pi}}\left(-\frac{k\eta_e}{2}\right)^{1/2-\nu}\frac{1}{a_e}\left[\frac{cr}{k}+ic\left(r\eta_r+d+\log 2-\gamma\right)\right] ,
 \quad \alpha_k \approx \beta_k^*.
\end{equation}
We obtained these expressions by evaluating  (\ref{eq:Bogolubov coeff})  at the end of inflation $\eta=\eta_e$, and they are only valid in the superhorizon limit $k\ll \mathcal{H}_e$. Since the solutions (\ref{eq:w smooth}) also  approach equation (\ref{eq:wk0}) at late times, we can directly compare the coefficients (\ref{eq:B smooth IR})  with those in (\ref{eq:B IR general}). Inspection of the asymptotic behavior of the scale factor using equation (\ref{eq:a smooth limit})  establishes the values of $a_r$ and $\dot{a}_r$. Then,   equation~(\ref{eq:B smooth IR}) reduces to (\ref{eq:B IR general}) and we can reproduce the same results as in section~\ref{sec:Intermediate Regime}. It follows, in particular, that the spectral density in the infrared limit is given by equation (\ref{eq:rho IR general massless}), whereas in the intermediate regime we recover equations (\ref{eq:rho IM general massless}) and (\ref{eq:spectral osc corrections}).

In this particular case, we could determine the energy density in the transition range by numerical integration. It is however more instructive to derive  an upper bound on $\rho_\mathrm{T}$ instead, which happens to agree well with the numerical results. To do so,  note that the spectral density monotonically decreases in the interval $\mathcal{H}_r\leq k\leq \mathcal{H}_e$, as seen in figure \ref{fig:smooth spectral density}.  Since equation (\ref{eq:rho IM general massless}) is still a reasonable approximation for the spectral density at $k=\mathcal{H}_r$, the integral of $d\rho/d\log k$ over the transition modes is hence bounded by 
\begin{equation}\label{eq:rho T smooth}
	\rho_{\textrm{T}} \lesssim \frac{\Gamma^2(\nu)}{2^{4-2\nu}\pi^3}(-p)^{1-2\nu}H_e^4\left(\frac{a}{a_e}\right)^{3-2\nu}
	\left(\frac{H}{H_e}\right)^{5-2\nu}
	\left(\frac{a}{a_r}\right)^{3-2\nu}\log\frac{\mathcal{H}_e}{\mathcal{H}_r},
\end{equation}
where the logarithm stems from the mode interval over which we integrated.  Comparison of (\ref{eq:rho T smooth}) with (\ref{eq:rho massless IM}) thus shows that $\rho_\mathrm{T}$ is typically smaller than $\rho_\mathrm{IM}$, though, because of the slow decay of $(a/a_r)^{3-2\nu}$ and the logarithmic dependence on the mode interval size,  not parametrically so. 

In the ultraviolet regime,  applying  Hankel's expansion for large arguments to equation (\ref{eq:chi inf massless}),  and using the power series expansion for the modified Bessel function in equation (\ref{eq:w smooth}), we find, to leading order in departures from the $out$ adiabatic vacuum ($\beta_k=0$, $\alpha_k=1$), that 
\begin{equation}\label{eq:B smooth UV}
 \beta_k\approx i \,\frac{p(p-1)(1-r\eta_e)}{4 (-k\eta_e)^3},
\quad
\alpha_k \approx 1 +  i\frac{p(p-1)}{2(-k\eta_e)}\left(1-\frac{1}{2r\eta_e}\right).
\end{equation}
The  term in $\beta_k$ proportional to $1/(k \eta_e)^2$   vanishes on account of equation (\ref{eq:t0}), which follows from demanding that the second derivative of the scale factor be continuous at the transition, as we anticipated in equations (\ref{eq:B UV}). Then, with $|\beta_k|^2\propto k^{-6}$, the energy density diverges in the ultraviolet just like for the vacuum state, and one can carry out its regularization and renormalization  as described in section~\ref{sec:massless ultraviolet}. In this particular case there is an additional finite ultraviolet contribution from the nonvanishing Bogolubov coefficients that could be attributed to particle production.  In order to determine its magnitude, it just suffices to focus on the leading terms in equation (\ref{eq:rho UV}). Bearing in mind that the Bogolubov coefficients are those in (\ref{eq:B smooth UV}), we arrive at

\begin{equation}\label{eq:spectral smooth UV}
	\frac{d\rho_\mathrm{p}^{\mathrm{UV}}}{d\log k}  \!\approx \!
	\frac{(p-1)^2(1-r\eta_e)^2}{32\pi^2p^4}H_e^4 \!
	\left(\frac{a_e}{a}\right)^4 \! \! \left(\frac{a_eH_e}{k}\right)^2 
	\! \!
	- \frac{(p-1)(1-r\eta_e)}{8\pi^2p^2} H_e^3H \!\left(\frac{a_e}{a}\right)^3
	\!\!\sin[2k\eta+\varphi].
\end{equation}
The contribution from the first factor is what we would have obtained using the particle production approximation (\ref{eq:rho ppf}), but notice that, even though it oscillates, the second term has an amplitude  that dominates throughout the ultraviolet. Therefore, on top of equation (\ref{eq:rho massless UV}), there are two additional ultraviolet contributions  to the energy density  after the end of inflation, both of order $H_e^4$ around the end: One  that scales like radiation, and another one that oscillates in time with frequency $\sim 2\mathcal{H}_e $ and a decaying amplitude proportional to $a^{-6}$.  A comparison of our general analytic predictions with the actual spectral density after a smooth transition is shown in figure \ref{fig:smooth spectral density}.
 
\begin{figure}
\begin{center}
	\includegraphics[width=12cm]{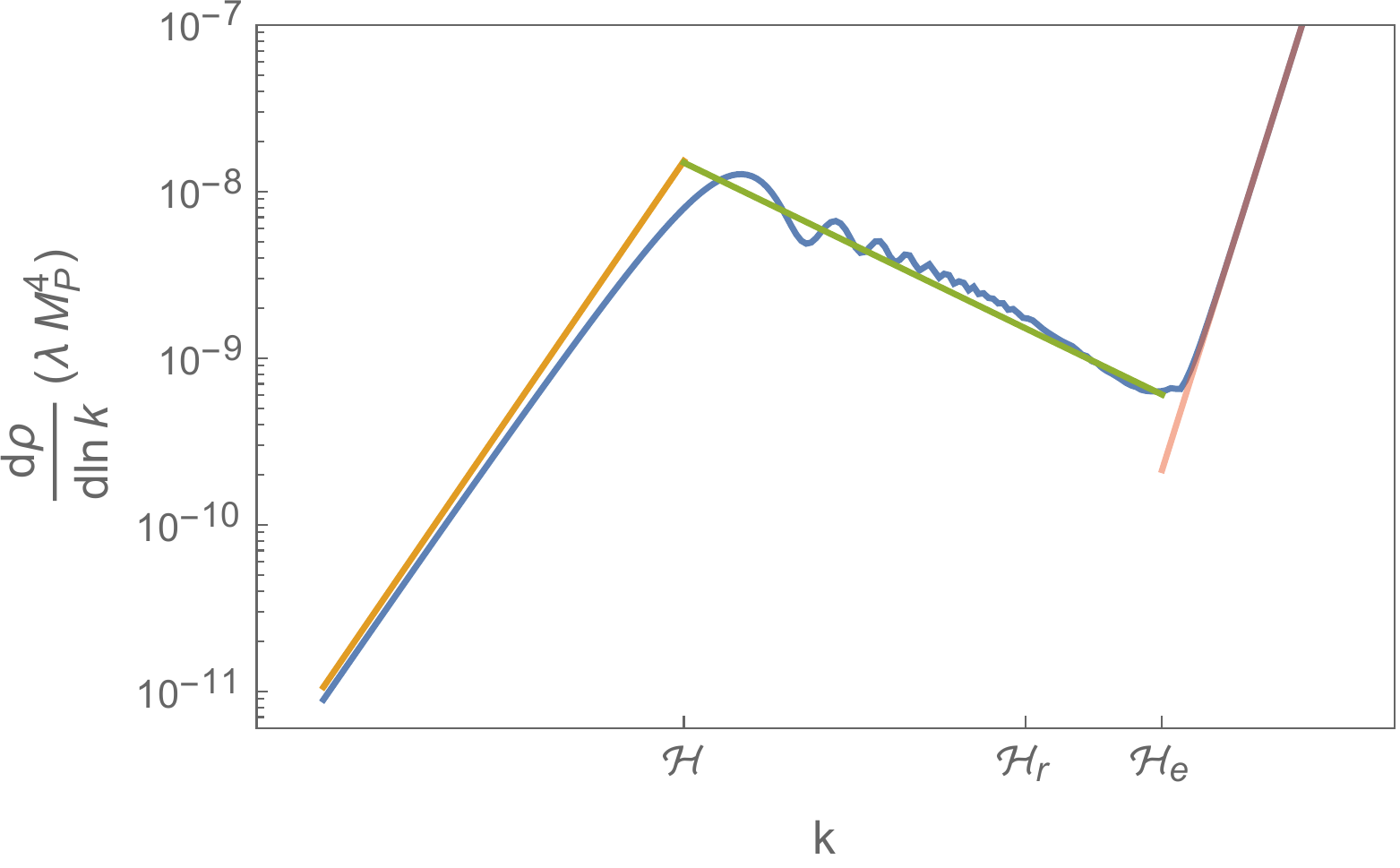}
	\caption{As in figure~\ref{fig:smooth spectral density}, but for the chaotic transition resulting from the inflationary potential (\ref{eq:chaotic}).  In this example  $a/a_e=10^2$. }
	\label{fig:spectral density}
\end{center}
\end{figure}

The smooth transition is also helpful to illustrate the behavior of transitions that deviate from our assumptions. If $p$ is large, equation (\ref{eq:t0}) implies that $r\gg \eta^{-1}_e$. In that case, the transition contains two parametrically distinct scales, $r$ and $\eta_e^{-1}$. In the range $\eta_e^{-1}\ll k\ll r$, the mode functions (\ref{eq:w smooth}) approach  those in equation (\ref{eq:wk0}), so the Bogolubov coefficients approximately equal those in a sharp transition, equation (\ref{eq:B UV sharp}). On the other hand, in the range $\eta_e^{-1}\ll r\ll k$ the assumptions of our   analysis  in the ultraviolet still hold, and the Bogolubov coefficients tend to those in (\ref{eq:B smooth UV}). The parameter $r$ here  hence plays the role of  $s$ we introduced in section \ref{sec:Transitions}. In a sense, by sending $r\to\infty$ we remove this additional scale from the problem by rendering the smooth transition sharp. One conclusion of this analysis is that in a realistic transition in  which the equation of state changes abruptly,  the spectral density would be dominated by the contribution of the lower boundary, resulting in an ultraviolet energy density of order $H_e^4 (a_e/a)^4$, just as we found in the smooth transition here.  More precisely, if the equation of state experienced a sudden change at the end of inflation, the energy density of modes in the range $\mathcal{H}_e\leq k \leq s$ would be, from (\ref{eq:B UV sharp}), of order $H_e^4 (a_e/a)^4 \log(s/\mathcal{H}_e)$.

 \paragraph{Chaotic Transition.}
 \label{sec:Differentiable Transition}

The shape of the spectral density  that we have described in our general analysis, and verified in the case of a smooth transition, survives in the gradual transition to radiation domination discussed in section \ref{sec:Transitions}. Figure \ref{fig:spectral density} shows the numerically computed  spectral density after the end of inflation. The spectral density in the infrared and intermediate regimes are well approximated by our analytical estimates, and the sharp increase  in the ultraviolet is what we would expect from the leading $k^4$ behavior in the adiabatic approximation. In order to determine the value of $p$ that appears in the analytical estimates, we have used equation (\ref{eq:p slow roll}) evaluated at a time a  mode in the corresponding  range left the horizon.  Note that the main differences  between the infrared and intermediate modes in   figures \ref{fig:smooth spectral density} and \ref{fig:spectral density} stem from inflation not being exactly de Sitter in our realization of a gradual transition. In particular, this introduces a red tilt in the spectral density of the intermediate modes. 


\begin{table}
\begin{center}
MASSLESS FIELD\\
\vspace{.2cm}
\begin{tabular}{ c c c c c}
 \hline
 \hline
   $\Lambda_\mathrm{IR}\leq k\leq \mathcal{H}$ & $\mathcal{H}\leq k\leq \mathcal{H}_r$ & $\mathcal{H}_r\le k\le \mathcal{H}_e$ & $\mathcal{H}_e\leq k$  \\ \hline \\[-2ex]
   $\displaystyle{\frac{H_e^2H^2}{16\pi^2}}$ & $\displaystyle{\frac{H_e^2H^2}{8\pi^2}\log\frac{a}{a_r}}$ & $\displaystyle{\rho_{\mathrm{rad}}^{\mathrm{T}}\left(\frac{a_r}{a}\right)^4}$ & $\displaystyle{\sim H_e^2H^2 \left(\frac{\dot{a}_e}{\dot{a}_r}\right)^2 }$ \\[1.5ex]   
   & {\footnotesize particle production} & {\footnotesize particle production} & \\
 \hline
 \hline 
\end{tabular}
\end{center}
\caption{Contributions to the renormalized energy density of a massless scalar during the radiation era following near de Sitter inflation.  All modes contribute as radiation, although only the intermediate range admits a particle interpretation.  In the latter  we just quote the early time limit, away of which the logarithm is replaced by $(2\nu-3)^{-1}$. Note that the transition dependent factor $\dot{a}_e/\dot{a}_r$  only  appears in the ultraviolet range, and that  $\rho_\mathrm{rad}^\mathrm{T}$ further encapsulates our ignorance about the transition.   The effective coupling constants are frozen. The behavior of  modes with $k<\Lambda_\mathrm{IR}$ is reviewed in section \ref{sec:Beyond the Infrared Cutoff}.}
\label{table:classical/particle massless}
\end{table}


\subsection{Overview}
\label{sec:Overview}

The energy density of a massless scalar after a transition from  inflation to radiation domination behaves as the superposition of four components, one for each of the mode ranges in equation (\ref{eq:k ranges}).  When inflation is de Sitter-like ($p \approx -1$), all scale essentially like radiation,  as shown in table \ref{table:classical/particle massless}. The contribution of infrared and intermediate modes to the energy density is sensitive to the scale at which inflation ends, $H_e$, but not to the details of the transition. On the contrary, the contribution of the ultraviolet modes does depend on reheating through an overall factor $\dot{a}_e/\dot{a}_r$, but is only mildly sensitive to the abruptness of the transition.  Note in particular that $\ddot{a}\propto \rho-3p$, so whether $\dot{a}$ increases or decreases after the end of inflation  depends on the equation of state of the universe after that time. The amplitude of modes in the small band $\mathcal{H}_r \le k \le \mathcal{H}_e$  depends on the details of the transition, although we expect the latter to be  subdominant, as we found in the  examples that we analyze in section \ref{sec:Examples}. These results imply that the total energy density of the scalar remains subdominant during radiation domination, unless the scale of inflation $H_e$ was close to the Planck mass, which clashes with cosmic microwave background constraints on the  amplitude of the primordial tensor modes. 
 
On the other hand, when inflation is far from de Sitter ($p<-2$),  during radiation domination all modes are affected by an overall transition dependent factor $\dot{a}_e/\dot{a}_r$. Speficically, the infrared component scales like  curvature,  the intermediate one like dark energy (possibly violating the energy conditions), and the  ultraviolet again like radiation.  Hence, we expect the intermediate modes to dominate in that case and possibly overcome the background energy density.  

The results of this section are further summarized in table \ref{table:rho massless}. Let us emphasize again that among the four mode ranges only the contribution of the intermediate and transition modes can be obtained within the particle production formalism, and that none of them can be cast as the energy density  of a classical homogeneous scalar.

In addition to the previous contributions to the energy density we also need to include those of the modes with $k=0$ and $0<k<\Lambda_{\mathrm{IR}}$, whose state remains unknown. As we described in section \ref{sec:Beyond the Infrared Cutoff}, if the field is in a macroscopic state the zero mode gives a non-zero contribution to the energy density, which can be cast as that of a classical, not necessarily real,  scalar.  On the other hand, modes with $0<k<\Lambda_{\mathrm{IR}}$ contribute to the energy density as a curvature component, stemming from the field gradients,  and hence do not admit an interpretation in terms of a  homogeneous classical scalar.  

We also need to consider how the different coupling constants of the theory are affected by  renormalization. In the case of a massless field  the value of the cosmological constant, the Planck mass and the dimension four curvature scalars are renormalized by constant values. During radiation domination only the first two survive. Given that during the radiation era all modes of a massless field find themselves in the adiabatic regime, see equation~(\ref{eq:wk0}), it is possible to compute  an exact and unambiguous expression of the renormalized {\it out} vacuum energy  density by extending the integral that led to (\ref{eq:rho massless UV}) to all modes, including those under the infrared cutoff,  
\begin{equation}\label{eq:rho out massless}
 2\pi^2\rho_{\mathrm{ren}}^{\mathrm{out}} = \delta\Lambda^f -3H^2(\delta M_P^2)^f +\frac{H^4}{480}.
\end{equation}
Leaving counterterms aside, we can conclude that  after inflation the $out$  adiabatic vacuum energy density of a massless field is negligible when compared with the background and scalar field energy densities.

It is also straightforward to extend our analysis to a transition to matter domination. In that case the spectral density in the infrared, transition and ultraviolet ranges is the same as during radiation domination. In the intermediate range, however, we have to distinguish between those modes that reenter the horizon during radiation domination, and those that reenter during matter domination. The former  behave in exactly the same way as they do during radiation domination; the spectral index of the  latter, however, changes from $3-2\nu$ to $1-2\nu$.

\section{Light Fields}
\label{sec:Light Fields}

Let us turn our attention to massive fields, in the limit in which their mass is much smaller than the Hubble constant during inflation. Although they share some of the properties of  massless fields, the presence of a mass term modifies the evolution of the {\it in} mode functions and introduces new features in the spectral density.  At the beginning of inflation the comoving mass of the field  is lower than the infrared cutoff, $ma<\Lambda_{\mathrm{IR}}$, and the field can be  regarded in practice as massless, as the following analysis reveals. In this regime, the results of section \ref{sec:Massless Fields} apply nearly without modification.  As time goes by, however, the comoving mass grows, and it is possible for the field to become effectively massive, $\Lambda_\mathrm{IR}< ma$.  The latter is the regime that we explore in this section, where we  describe how the initial spectral density  is increasingly distorted away  from that of a massless scalar as the universe expands. As we shall see, some time after the beginning of inflation the energy density of the  field behaves like dark energy, while once the field ceases to be light, in the $out$ region, it behaves like dark matter. The dark energy phase admits a description in terms of a classical field. The dark matter component, on the other hand, can be interpreted both as the result of the produced particles and as the classical field that used to describe dark energy early on.

\subsection{Inflation}
\label{sec:Light Fields Inflation}
When the scalar field $\hat\phi$ is massive, simple analytical solutions of the mode equation (\ref{eq:mode equation}) are known to us  only during de Sitter inflation.  Close to de Sitter, $p \approx -1$, we can substitute $m^2 a^2$ in the dispersion relation  (\ref{eq:mode equation})  by the value it would have for $p = -1$, resulting in the following  solution for the {\it in} mode functions,
 \begin{equation}\label{eq:inflation heavy chi}
 	\chi^\mathrm{in}_k(\eta)= i
	\frac{\sqrt{-\pi \eta}}{2} 
	H^{(1)}_\nu (- k \eta  ),   \quad \nu\equiv \sqrt{\frac{(1-2p)^2}{4}-p^2\frac{m^2}{H_i^2}}.
 \end{equation}
This expression differs from its massless counterpart (\ref{eq:chi inf massless}) only in the value of the index $\nu$, which agrees with that in (\ref{eq:chi inf massless}) when $m=0$.  Since $H$ is not constant away from de Sitter, we have set its value to that at the beginning of inflation, $H_i$. There is nothing particular about this choice, and indeed it does not affect our final expressions, as we show later.  By construction, the solution (\ref{eq:inflation heavy chi}) is exact when $p=-1$ (and, of course, when $m=0$), but is otherwise only an approximation to the mode functions that solve equation (\ref{eq:mode equation}) with appropriate {\it in} boundary conditions, no matter whether we are close to de Sitter  or not. 

We assess the validity of equation (\ref{eq:inflation heavy chi})   in appendix \ref{sec:Light Fields During Inflation}. In order to  do this it is convenient to split the modes into the same ultraviolet and infrared ranges as  in section \ref{sec:Inflation}.  In the ultraviolet, equation (\ref{eq:inflation heavy chi}) leads to an  approximation  of the spectral density that contains relative errors   suppressed by  factors of $(m^2/H^2)(\mathcal{H}^4/k^4)$, which are small for light fields and short wavelengths.   In the infrared, however, we need to distinguish between two limits, which depend on how close to de Sitter the universe is.  For ``ultra-light" fields, those whose mass satisfies   $m^2\ll |1+p|H^2$, equation (\ref{eq:inflation heavy chi}) is a valid approximation at zeroth order in $m^2/H^2$, and we can set $\nu\approx (1-2p)/2$ as the order of the Hankel function. This is the limit that connects  with the massless case, where the solution is exact, c.f.~equation (\ref{eq:chi inf massless}). On the other hand, for ``merely-light" fields, those whose mass satisfies  $|1+p|H^2 \ll m^2\ll H^2$, the mode functions (\ref{eq:inflation heavy chi}) remain valid at zeroth order in $1+p$. At this order the Hubble factor does not change in time, and we can write $\nu\approx 3/2 -m^2/(3H^2)$.  This limit connects  with de Sitter, where the solution is also exact.  The distinction between ultra- and merely-light fields becomes relevant in the infrared, and has an impact on the total amount of energy density at late times, as we clarify in section \ref{sec:General Infrared Behavior massive}.

\begin{figure}
\begin{center}
	\includegraphics[width=15cm]{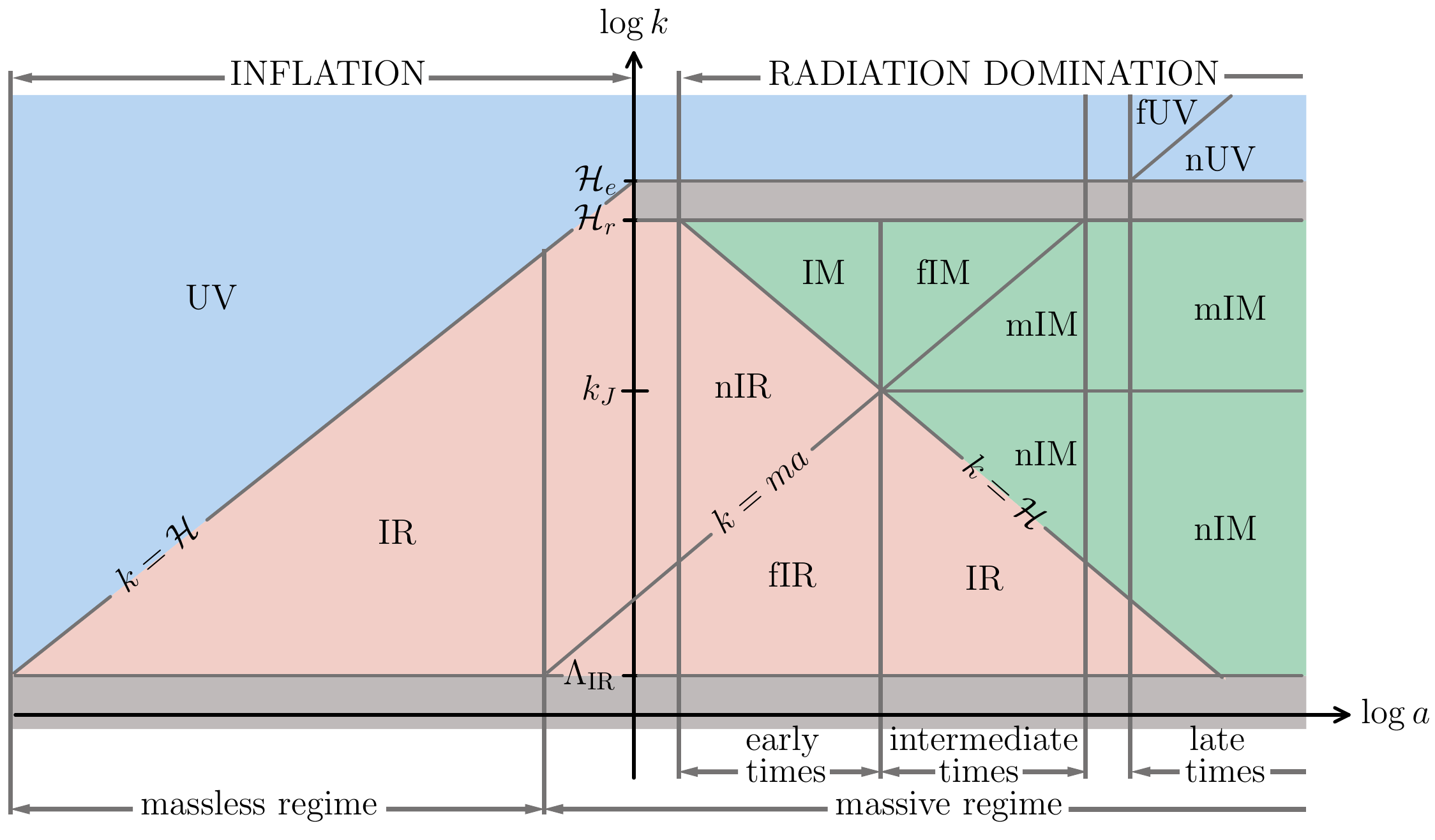}
\end{center}
  \caption{ 
  Different regimes of evolution of a massive scalar field in the light field limit, $ma_e\ll \mathcal{H}_e$.
  The scale $k_J$ remains constant  during radiation domination, whereas $ma$ grows linearly  with time and $\mathcal{H}$ decreases as the inverse of time. 
  The difference between the early, intermediate and late time regimes essentially relies on whether the scale $ma$ that sets the onset of nonrelativistic effects lies on the infrared, intermediate and ultraviolet mode ranges, respectively. In this work we restrict the $out$ region to times   such that $\mathcal{H}>\Lambda_\mathrm{IR}$, so that all modes that eventually enter the horizon originally  left during inflation.}
\label{fig:mode intervals}
\end{figure}

In the ultraviolet, and  using the adiabatic expansion (\ref{eq:out adiabatic}) to second order, it takes some work then to check that the  spectral density approaches 
\begin{eqnarray}\label{eq:massive UV inflation}
 	\frac{d\rho_{\mathrm{UV}}}{d\log k} &\approx & \frac{k^4}{4\pi^2a^4}\left[1+\frac{1}{2}\left(\frac{aH}{k}\right)^2+	\frac{1}{2}\left(\frac{ma}{k}\right)^2\right. \nonumber \\
	 &&\left.+\frac{3}{8}\frac{p^2-1}{p^2}
 	\left(\frac{aH}{k}\right)^4+\frac{1}{4}\left(\frac{ma}{k}\right)^2\left(\frac{aH}{k}\right)^2-\frac{1}{8}\left(\frac{ma}{k}\right)^4\right].
\end{eqnarray}   
Because the mode functions  (\ref{eq:inflation heavy chi}) are just an approximation, equation (\ref{eq:massive UV inflation})  does not  quite agree with the (incorrect) form of the spectral density that we would find  using  (\ref{eq:inflation heavy chi}).   
As we show in equation (\ref{eq:B2}) of the appendix, the relative difference between the two expressions remains small at short wavelengths, although the absolute error becomes large in the ultraviolet. Both approaches agree when (\ref{eq:inflation heavy chi}) is exact, namely, when $p=-1$ or $m = 0$, but away from these two particular cases it is important to use expression (\ref{eq:massive UV inflation}) in order to obtain the correct value of the renormalized energy density, as we do next. 

Integrating the spectral density (\ref{eq:massive UV inflation}) over the subhorizon modes and   subtracting equation (\ref{eq:rho sub}) we arrive at the renormalized energy density
\begin{eqnarray}\label{eq:rho ren UV inf}
 2\pi^2\rho_{\mathrm{ren}}^{\mathrm{UV}} &\approx& \left[\delta\Lambda^f -\frac{m^4}{64}\left(1+2\log\frac{\mu^2}{4H^2}\right) \right] 
 -3H^2 \left[(\delta M_P^2)^f -\frac{m^2}{72}\left(1+\frac{3}{2}\log\frac{\mu^2}{4H^2}\right) \right]\nonumber \\
&&+ \frac{36(p^2-1)H^4}{p^2} \left[\delta c^f +\frac{1}{384}\log\frac{\mu^2}{4H^2}\right] - \frac{122p^2-60p+57}{480p^2} H^4,
\end{eqnarray}
which is again finite and cutoff-independent.  The reader can compare this expression with its counterpart in the massless case, equation (\ref{eq:rho UV power law}). In the massive case, the renormalized Planck mass and cosmological constant  run  with time, since the presence of a physical quantity with the dimensions of mass allows these ``constants" to be renormalized. Cosmologies with such running constants have been studied, for instance, in  \cite{Shapiro:2004ch}.  

To find the spectral density in  the infrared we expand the mode functions (\ref{eq:inflation heavy chi}) to lowest order in $k\eta$ and substitute into equation (\ref{eq:rho}).  As we clarify in appendix \ref{sec:Light Fields During Inflation}, the former are perturbative approximations in the small parameter  $m^2/H^2$, and because the time derivative of $\chi_k^{\mathrm{in}}/a$ vanishes when $m=0$ and $k=0$, if the mass is different from zero, to lowest order in $k\eta$  the derivative must be of order $(m^2/H^2)\mathcal{H}(\chi_k^{\mathrm{in}}/a)$. Therefore, the ratio of ``kinetic'' to ``potential'' term in the spectral density is suppressed by a factor of $(m^4/H^4)(\mathcal{H}^2/\omega_k^2)$, which is small if the field is light. Then, at leading order in the long-wavelength and small mass expansions we arrive at
\begin{equation}\label{eq:dS light IR}
	  \frac{d\rho_\mathrm{IR}}{d\log k}\approx
	 \frac{\Gamma^2(\nu)}{2^{4-2\nu}\pi^3}(-p)^{1-2\nu} 
	 H^4
	 \left[\frac{m^2 a^2}{k^2}+1\right]
	 \left(\frac{k}{aH}\right)^{5-2\nu},
\end{equation}
which apart from the first  term in square brackets coincides with equation (\ref{eq:inf massless IR}). Since in the limit of light fields $ma_i \ll \mathcal{H}_i \sim \Lambda_{\mathrm{IR}}$,  such term  is negligible at the beginning of inflation,  and we can treat the field initially as effectively massless. It is not until the comoving mass $ma$ surpasses the cutoff $\Lambda_{\mathrm{IR}}$, a number of e-folds $N_{\mathrm{fIR}}\equiv \log(a_\mathrm{fIR}/a_i)\equiv \log(\Lambda_\mathrm{IR}/(ma_i))\sim \log(H_i /m)$ after the beginning of inflation, that the effects of the mass term become relevant. At this point it  proves convenient to distinguish the ``far infrared,'' $k\leq ma$, where the mass term dominates the dispersion relation, from the ``near infrared,'' $ma\leq k\leq \mathcal{H}$, where the mass term is negligible and modes are effectively relativistic. Figure \ref{fig:mode intervals} shows the limits of the  different mode ranges we consider in this section. It generalizes figure \ref{fig:massless ranges} to the case of a non-zero mass. In particular, if we set $m=0$ in figure \ref{fig:mode intervals},  lines of constant $a$ after inflation traverse the intervals shown in figure \ref{fig:massless ranges}.

In the near infrared, equation (\ref{eq:dS light IR})  replicates equation  (\ref{eq:inf massless IR}), as expected, since the mass term plays no role in this range.  Therefore,  in the near infrared the energy density is well approximated by (\ref{eq:rho IR inf}), provided that we replace the second term in the square brackets there by the appropriate contribution of the lower boundary here, $\Lambda_\mathrm{IR}\to ma$.  From now on, we shall  mainly focus our discussion on  near  de Sitter inflation, $p\approx -1$, which is phenomenologically more relevant, although equations expressed in terms of $\nu$ are still valid in general.  In this case, the lower boundary  contribution soon becomes negligible after the onset of inflation, and it remains so when the near infrared appears.  Hence, during near de Sitter inflation the near infrared  energy density is approximately that  of a massless field  quoted in (\ref{eq:rho IR inf}), with $\nu=3/2$ and the last term in the  square brackets replaced by zero. 

\begin{figure}
\begin{center}
	\includegraphics[width=12cm]{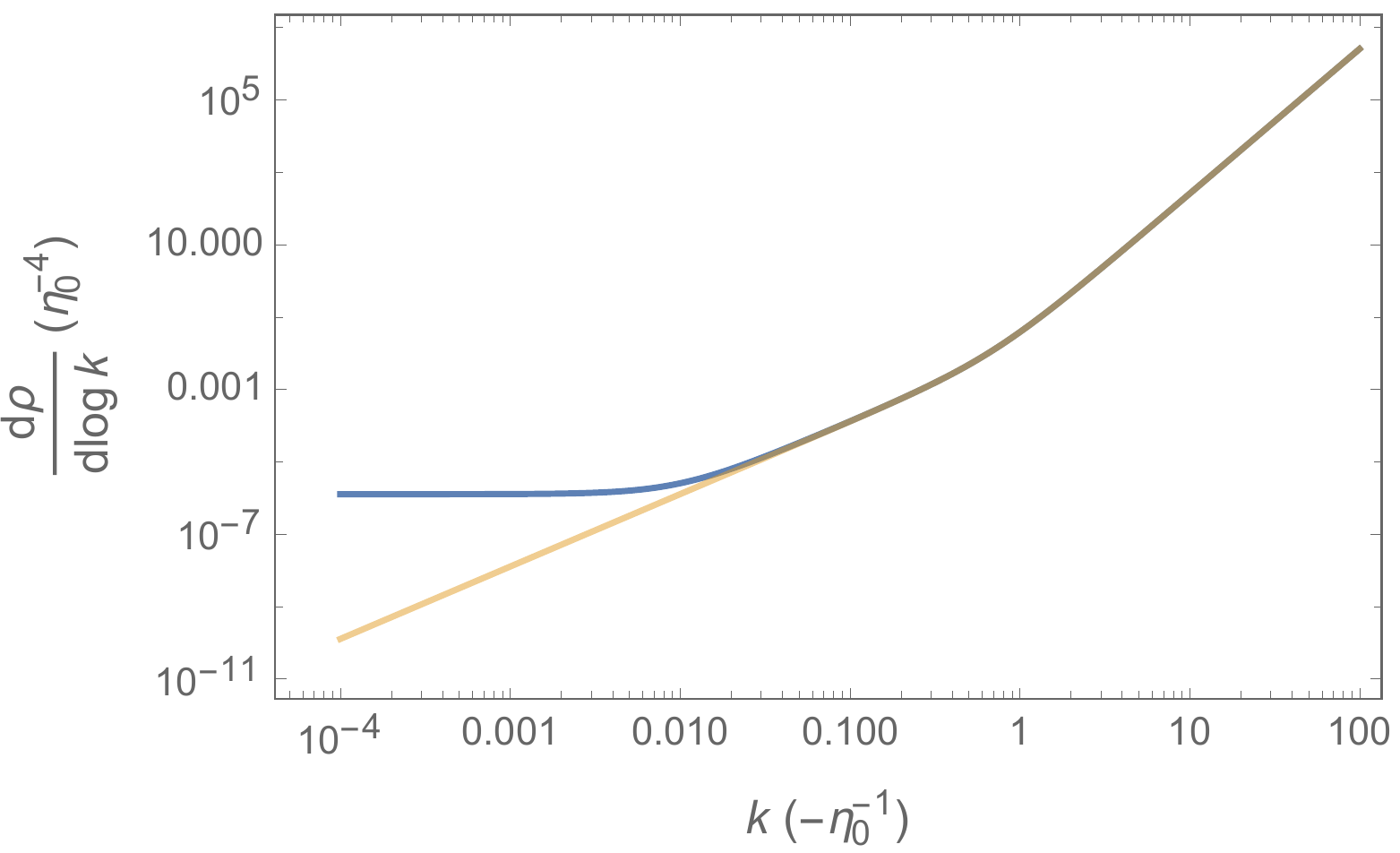}
	\caption{Spectral density during de Sitter inflation for a light field with $m=10^{-2}H_i$ (blue). For comparison we also show the spectral density in the massless case (light orange.) The  behavior in the infrared is well captured by equation (\ref{eq:dS light IR}). In particular, in the far infrared the spectral density is nearly flat, with a small blue tilt due to the mass term, and in the near infrared it matches that of a massless field. In the ultraviolet, the spectral density  is seen to approach the leading term in equation (\ref{eq:massive UV inflation}), as expected.}
	\label{fig:spectral density light inflation}
\end{center}
\end{figure}

  In the far infrared the spectral density  (\ref{eq:dS light IR}) is redder (closer to being  infrared singular) than that in the near infrared. This is perhaps unanticipated, because one would expect a mass term to regulate any existing infrared singularity in the massless theory. Nevertheless, the structure of  (\ref{eq:dS light IR})  is what our approximations would suggest. Both in the infrared and the light field limits the spectral density  is proportional to ${(k^2+m^2 a^2) |\chi^\mathrm{in}_k/a|^2}$, because on superhorizon scales the time derivatives give subdominant contributions, since $\chi_k^\mathrm{in}/a$ is essentially frozen on those scales.  In the massless case the leading infrared contribution is thus proportional to $k^2   |\chi^\mathrm{in}_k/a|^2$, as in equation (\ref{eq:inf massless IR}). But when $k^2\ll m^2 a^2$ the leading term is then proportional to $m^2 a^2  |\chi^\mathrm{in}_k/a|^2$, which explains the softer infrared behavior and the additional factor of $m^2a^2$ in equation (\ref{eq:dS light IR}). We plot the spectral density during de Sitter inflation in figure \ref{fig:spectral density light inflation}.

Integrating the spectral density (\ref{eq:dS light IR}) over the far infrared, provided it is not empty, we  arrive at
 \begin{equation}\label{eq:rho fIR inf}
 \rho_{\mathrm{fIR}} \approx \frac{\Gamma^2(\nu)}{2^{4-2\nu}\pi^3}\frac{(-p)^{1-2\nu}}{3-2\nu}m^2H^2
 \left[\left(\frac{m}{H}\right)^{3-2\nu}-
 \left(\frac{\Lambda_\mathrm{IR}}{a H}\right)^{3-2\nu}
 \right].
 \end{equation}
Close to de Sitter the behavior of this positive energy density  is quite sensitive to the parameters of the model, and, particularly, to the departures of the spectral index from scale invariance.  There, because  
\begin{equation}
	\nu\approx \frac{3}{2}+|1+p|-\frac{1}{3}\frac{m^2}{H_i^2},
\end{equation}
 the spectral index $3-2\nu$ vanishes at zeroth order, and this time it is necessary to keep the $|1+p|$ and/or $m/H_i$ terms in this expression. Which of the two dominates depends on whether the field is  ultra-light or  merely-light.   

In the limit of ultra-light fields $\nu\approx 3/2+|1+p|$, and in this case the energy density takes the form 
\begin{equation}\label{eq:rho fIR ultra}
\rho_{\mathrm{fIR}} \approx \frac{m^2H_i^2}{16\pi^2|1+p|}\left[1-\left(\frac{\Lambda_\mathrm{IR}}{ma}\right)^{2|1+p|}\right].
\end{equation} 
Since the spectral index in this range is negative, the energy density is dominated by the lower boundary at $k=\mathcal{H}_i$, which is why we have opted to express $\rho_{\mathrm{fIR}}$ in terms of $H_i$. Once the far infrared window opens,  $\rho_{\mathrm{fIR}}$
initially grows like ${m^2H_i^2/(8\pi^2)\log (ma/\Lambda_{\mathrm{IR}})}$, until it eventually reaches   $m^2H_i^2/(16\pi^2|1+p|)\approx m^2/(|1+p|H^2)(H_i/H)^2\rho_{\mathrm{nIR}}$, which is constant. The transition between these two regimes occurs after $N\sim 1/(2|1+p|)$ e-folds since the opening of the far infrared window, $N \equiv \log(a/a_{\mathrm{fIR}}) \equiv \log(ma/\Lambda_{\mathrm{IR}})$. In models where the slow-roll parameter $\epsilon_1\approx |1+p|$ is not too small ($\epsilon_1\sim 0.01$ in realistic cosmological scenarios) this happens relatively early. Because $H$ is not expected to  change significantly during inflation and $m^2/(|1+p|H^2)$ is a small number in the ultra-light case,  typically $\rho_{\mathrm{fIR}} \ll \rho_{\mathrm{nIR}}$,  and in practice, during inflation the infrared behavior is expected to be the  same as in the massless case.

In the  opposite limit of merely-light fields  $\nu\approx 3/2-(m^2/H_i^2)/3$, and under the assumption of nearly constant $H$ discussed in appendix \ref{sec:Light Fields During Inflation}, $H \approx H_i$, the spectral  density reduces to
\begin{equation}\label{eq.edfIR2dS}
 \rho_{\mathrm{fIR}} \approx \frac{3H_i^4}{16\pi^2}\left[1-\left(\frac{\Lambda_\mathrm{IR}}{ma}\right)^{\frac{2}{3}\frac{m^2}{H_i^2}}\right].
\end{equation}
The spectral index is positive here, so the energy density is dominated by the upper limit of the integral. Although it would be natural to express $\rho_{\mathrm{fIR}}$ in terms of $H_e$, we are working under the approximation that $H_e\approx H_i$, so the two quantities are interchangeable. Again, the energy density (\ref{eq.edfIR2dS}) initially approaches ${\rho_\mathrm{fIR}\approx  m^2H_i^2/(8\pi^2) \log (ma/\Lambda_\mathrm{IR})}$, until it eventually reaches  the constant value $\rho_\mathrm{fIR}\approx 3H_i^4/(16\pi^2)\approx 3\rho_{\mathrm{nIR}}$.  This time the transition happens about $N\sim 3H_i^2/(2m^2)$ e-folds since the opening of the far infrared window, which can be a very large number in models where the scalar field mass is much lower than the scale of inflation. It is the presence of this additional contribution, of order $\rho_{\mathrm{nIR}}$,  what causes the well-known discontinuity of the renormalized energy density in de Sitter spacetime as the massless limit  is approached. As discussed in \cite{Kirsten:1993ug}, the disagreement with the massless case can be alternatively traced back to the absence of an appropriate de Sitter invariant \emph{Fock} vacuum state for a massless scalar. However,  an exponentially large period of inflation is necessary for the discontinuity to appear, since otherwise the contribution of the far infrared during inflation is negligible, as it  happens in the ultra-light case too. 

Note that both for ultra-light and merely-light fields the far infrared energy density initially grows in proportion to the number of e-folds since the opening of the far infrared window, $N$. This  growth can be interpreted as the result of a random walk experienced  by a light field $\hat\phi$ in a de Sitter background, during which the field amplitude changes by $H_i/(2\pi)$ each Hubble time,   $\langle \hat\phi^2\rangle\sim N (H_i/2\pi)^2 $ \cite{Vilenkin:1982wt}. Near de Sitter, however, such an interpretation breaks down after the two different number of e-folds $N$ quoted above, when the far infrared energy density $\rho_{\mathrm{fIR}}$ settles down to a constant value. We further comment on this random walk in section \ref{sec:The Particle and Classical Field Approximations}.

\subsection{General  Infrared Evolution After Inflation}
\label{sec:General  Evolution After Inflation}

Just as we did in the case of a massless field, it is possible to explore the behavior of the spectral density in the infrared  in relatively full generality. Again, the key here is to utilize the approximate solution in equation (\ref{eq:chi bar infrared}), which  generally remains valid at low frequencies, that is, when modes are long and the field is light, as  we argue in section \ref{sec:Validity of the Low Frequency Approximation} of  appendix \ref{sec:Validity of the Low and High  Frequency Approximations}. Hence, this approximation only works while  ${ma\ll \mathcal{H}}$, which we shall refer to as the limit of early times. 

During inflation the {\it in} mode functions of a massless field find themselves in the growing solution, and they remain so if the field is light, as we discuss in appendix \ref{sec:Light Fields During Inflation}. On long wavelengths this solution is well approximated by the first line in equation (\ref{eq.exp.pl}), as during inflation, where we have included the next to leading order in the low frequency expansion. It is possible then to estimate the contribution of the ``kinetic" term to the spectral density. In the growing mode such a contribution is suppressed with respect to that of the ``potential" term by a  factor of order $(m^4/H^4)(\mathcal{H}^2/\omega_k^2)$, which again renders the kinetic term negligible at early times. In that limit the mode functions are still given by (\ref{eq:chi 0 universal}) at leading order, and substitution into (\ref{eq:rho}) returns
\begin{equation}\label{eq:spectral massive IR} 
	\frac{d\rho_{\mathrm{IR}}}{d\log k}\approx \frac{\Gamma^2(\nu)}{2^{4-2\nu}\pi^3}(-p)^{1-2\nu}H_0^4\left(\frac{a_0}{a}\right)^2\left[\frac{m^2a^2}{k^2}+1\right]
 \left(\frac{k}{a_0H_0}\right)^{5-2\nu},
\end{equation}
which  apart from the first term in  square brackets coincides with equation (\ref{eq:rho IR general massless}). The former  also  agrees approximately with the spectral density  that we derived in equation (\ref{eq:dS light IR}) during inflation by different means. The origin of the difference, a factor of order $(a_0/a)^{2m^2/(3H_i)^2}$, stems from the somewhat different time evolution of the mode functions (\ref{eq:inflation heavy chi}), which we used to arrive at (\ref{eq:dS light IR}), and the mode functions (\ref{eq:chi 0 universal}), which we used to arrive at (\ref{eq:spectral massive IR}).  In any case, as we discuss in appendix \ref{sec:Light Fields During Inflation}, the difference  between the two is small for ultra-light fields, and also remains negligible  for merely-light fields when the condition under which we can trust (\ref{eq:inflation heavy chi}), equation (\ref{eq:merely light validity}), is satisfied.    Otherwise equation (\ref{eq:spectral massive IR}) is universal, in the sense that it applies at early times, $ma\leq \mathcal{H}$, to all superhorizon modes, just as their massless counterpart (\ref{eq:rho IR general massless}) did. Recall, in particular, that  $\eta_0$ is an arbitrary time during inflation, which actually drops out of equation (\ref{eq:spectral massive IR}) to zeroth order in $m^2/H_i^2$.  It ought to be clear from our derivation that at early times the particle production formalism does not apply in the infrared, where frequencies are low. 

As we noted in section \ref{sec:Light Fields Inflation}, at the beginning of inflation, when $ma < \Lambda_{\mathrm{IR}}$, the first term in square brackets in equation (\ref{eq:spectral massive IR}) is negligible and we can treat the field as effectively massless. The effects of the mass term become relevant at $ma\geq \Lambda_{\mathrm{IR}}$, when it is convenient to distinguish between the far and near infrared. In the near infrared interval $ma\leq k \leq \mathcal{H}$ the modes are relativistic and outside the horizon at the time of interest, and the spectral density (\ref{eq:spectral massive IR}) reduces to that  of the massless case in the infrared, equation  (\ref{eq:rho IR general massless}). Although the lower integration limit is  different here, we can also borrow the expression for the energy density  in the massless case (\ref{eq:rho IR massless final})   by replacing $\Lambda_\mathrm{IR}\to ma$. Then, close to de Sitter,  the lower boundary gives a negligible contribution, and  the near infrared energy density tracks the background evolution, as it did when $m=0$. The tracking proceeds until the comoving mass approaches the size of the horizon towards the end of early times, when the near infrared range disappears and its corresponding energy density vanishes, see figure \ref{fig:mode intervals}. We can hence conclude that after the end of inflation there must be a period during which the far infrared modes come to dominate the infrared energy density. 

As we also described in section \ref{sec:Light Fields Inflation}, if $p\approx -1$  the precise contribution of the far infrared $k\leq ma$ to the energy density delicately depends on the values of the different parameters of the model,
\begin{equation}\label{eq:fIR general light}
 \rho_{\mathrm{fIR}} \approx  \frac{\Gamma^2(\nu)}{2^{4-2\nu}\pi^3}\frac{(-p)^{1-2\nu}}{3-2\nu}m^2H_e^2
 \, \left[\left(\frac{m a}{a_e H_e}\right)^{3-2\nu}
 \!\! -\left(\frac{\Lambda_\mathrm{IR}}{a_e H_e}\right)^{3-2\nu}
 \right],
\end{equation}
where we have chosen again, as in equation (\ref{eq:rho IR massless final}), $\eta_0=\eta_e$. In the ultra-light field limit this expression reduces to equation (\ref{eq:rho fIR ultra}), whereas in the merely-light limit to (\ref{eq.edfIR2dS}). We can therefore conclude that (\ref{eq:rho fIR ultra}) and (\ref{eq.edfIR2dS})  are generally valid and not only restricted to the inflationary period.   We return to the question whether (\ref{eq:fIR general light})  can be cast as the energy density of an appropriate homogeneous classical scalar field in section \ref{sec:The Particle and Classical Field Approximations}.

Equations (\ref{eq:spectral massive IR}), as well as (\ref{eq:rho IR massless final}) and (\ref{eq:fIR general light}), apply  as long as the field remains light, $ma\ll\mathcal{H}$, no matter during which cosmological epoch. The approximately constant  (and positive)  $\rho_\mathrm{fIR}$ in this regime  is what motivated reference \cite{Aoki:2014dqa} to identify the light scalar with a late time dark energy candidate. Although there is an  additional nearly constant contribution to the vacuum energy stemming from the subtraction terms, the far infrared modes start to behave like nonrelativistic matter later on, so it is in principle possible to disentangle (\ref{eq:fIR general light}) from other   contributions to an effective cosmological term.  It is in fact conceivable, though beyond the scope of our present analysis,  that an analogous  form of  dark energy, just with a heavier mass $m$, may resolve the much debated ``Hubble tension'' \cite{Planck:2018vyg}, as discussed in \cite{Poulin:2018cxd} (see \cite{Jedamzik:2020zmd}, however.) To the extent that our quantized field shares some of the phenomenology of an oscillating classical scalar (see below for further details),  this would agree with the analysis in \cite{Smith:2019ihp}.  A somewhat related explanation of the Hubble tension  that relies on the  field fluctuations of a light scalar during inflation has been  proposed in \cite{Belgacem:2021ieb}.

\subsection{Radiation Domination}
\label{sec:General Infrared Behavior massive}

As in the massless case, in order to compute the spectral density during radiation domination we shall simply assume that  there must be  a (smallest) time $\eta_r$  after which we can safely assume  the universe to be radiation dominated. Here, the evolution of the modes depends on the location of the comoving mass $ma$ relative to the comoving horizon, so it shall prove convenient to differentiate  among early times, $m a \leq \mathcal{H}$, intermediate times, $\mathcal{H}\leq m a \leq \mathcal{H}_r$, and late times, $\mathcal{H}_e\leq ma$, although not all these periods necessarily occur during radiation domination.\footnote{In late dark energy models with  $m\sim 10^{-33}\mathrm{ eV} \ll H_\mathrm{eq}\sim10^{-28}\,\mathrm{eV}$ the limit of intermediate times extends into the matter era.}  These splits leave out a  time window during which $\mathcal{H}_r\leq ma \leq \mathcal{H}_e$, that  we expect to be small and shall not discuss explicitly.

During each of the previous epochs the location of a given mode within the different  scales in the problem changes, so, as in the massless case,  it is useful to split the different modes into distinct, nonoverlapping, ranges.  In the following we proceed to discuss the spectral density  during these epochs at all  mode ranges.  We recommend that the reader  consult figure \ref{fig:mode intervals} for a visual reference of the different mode intervals at the three different epochs.

\paragraph{Early Times ($m\, a \leq \mathcal{H}$).} 
 
We argued in section \ref{sec:General  Evolution After Inflation} that, as long as the field remains light, the behavior of the spectral density in the infrared, $k\leq \mathcal{H}$, is universal. Hence, at early times during radiation domination the infrared spectral density  is given by equation (\ref{eq:spectral massive IR}).  In the intermediate interval $\mathcal{H}\leq k \leq \mathcal{H}_r$, on the other hand,  the modes are relativistic at the end of inflation and remain that way at early times, so the spectral energy density  behaves like that of a massless field, and thus agrees with the one quoted  in equation (\ref{eq:rho IM general massless}).  It follows then that in the far infrared the energy density is  given by equation (\ref{eq:fIR general light}), and that in the near infrared and intermediate regime it is given by the corresponding expressions in table \ref{table:classical/particle massless}, where we have assumed that the infrared is dominated by the near modes and the universe is close to de Sitter during inflation.

In general, we cannot predict  the amplitude of the (relativistic) modes in the transition range $\mathcal{H}_r\leq k\leq \mathcal{H}_e$. However, as in the massless case, we expect their contribution to the total energy density to remain subdominant; see equation (\ref{eq:rho T massless}) and the paragraph below.  In the  ultraviolet range $\mathcal{H}_e\leq k$, given that  the field mass  is irrelevant, at least at leading order,  the same relation between the Bogolubov coefficients and the smoothness of the transition that we discussed in section \ref{sec:Examples} applies. If the transition is gradual, then, we expect a sharp fall-off of the Bogolubov coefficients, and   the  spectral density  approaches  that in equation (\ref{eq:massive UV inflation}), provided we simply set $p=1$, as appropriate during radiation domination. The  contribution of the modes with $\mathcal{H}_e\leq k\sim \Lambda$  to the  energy density is subtracted out in equation (\ref{eq:rho ren def}), so the  renormalized energy density in this range  stems from the  $out$ adiabatic vacuum, 
\begin{eqnarray}\label{eq:rho UV EA}
	2\pi^2 \rho_\mathrm{ren}^{\mathrm{UV}} &\approx&
	 \left[\delta\Lambda^f -\frac{m^4}{64}\left(1+2\log\frac{\mu^2}{4H_e^2}\frac{a^2}{a_e^2}\right) \right] 
-3H^2 \left[(\delta M_P^2)^f -\frac{m^2}{18}\left(1+\frac{1}{2}\log\frac{\mu^2}{4H_e^2}\frac{a^2}{a_e^2}\right) \right]\nonumber\\
&&+\frac{H^4}{480} + \frac{H^2H_e^2}{8}\left(\frac{a_e}{a}\right)^2 -\frac{H_e^4}{8}\left(\frac{a_e}{a}\right)^4.
\end{eqnarray}
Setting $m=0$ in (\ref{eq:rho UV EA})  returns  the energy density of a massless field in equation (\ref{eq:rho massless UV}). The main difference between the two cases is that here the cosmological and Newton's ``constants" run. Terms proportional to the curvature counterterm $\delta c^f$ vanish during radiation domination, but the energy density  nevertheless contains a contribution of fourth adiabatic order, proportional to $H^4$, that is of the same order as the one generally expected from the curvature invariants. Again, the adiabatic approximation breaks down around ${k\sim \mathcal{H}_e}$, so we expect an error in $\rho_\mathrm{ren}^{\mathrm{UV}}$ of same order as the leading prediction of the quantum theory, namely, the term proportional of order  $H_e^4 (a_e/a)^4$. As in the massless cases, we also expect this to be the magnitude of the ultraviolet energy density after an abrupt transition. 

\begin{figure}
\begin{center}
	\includegraphics[width=12cm]{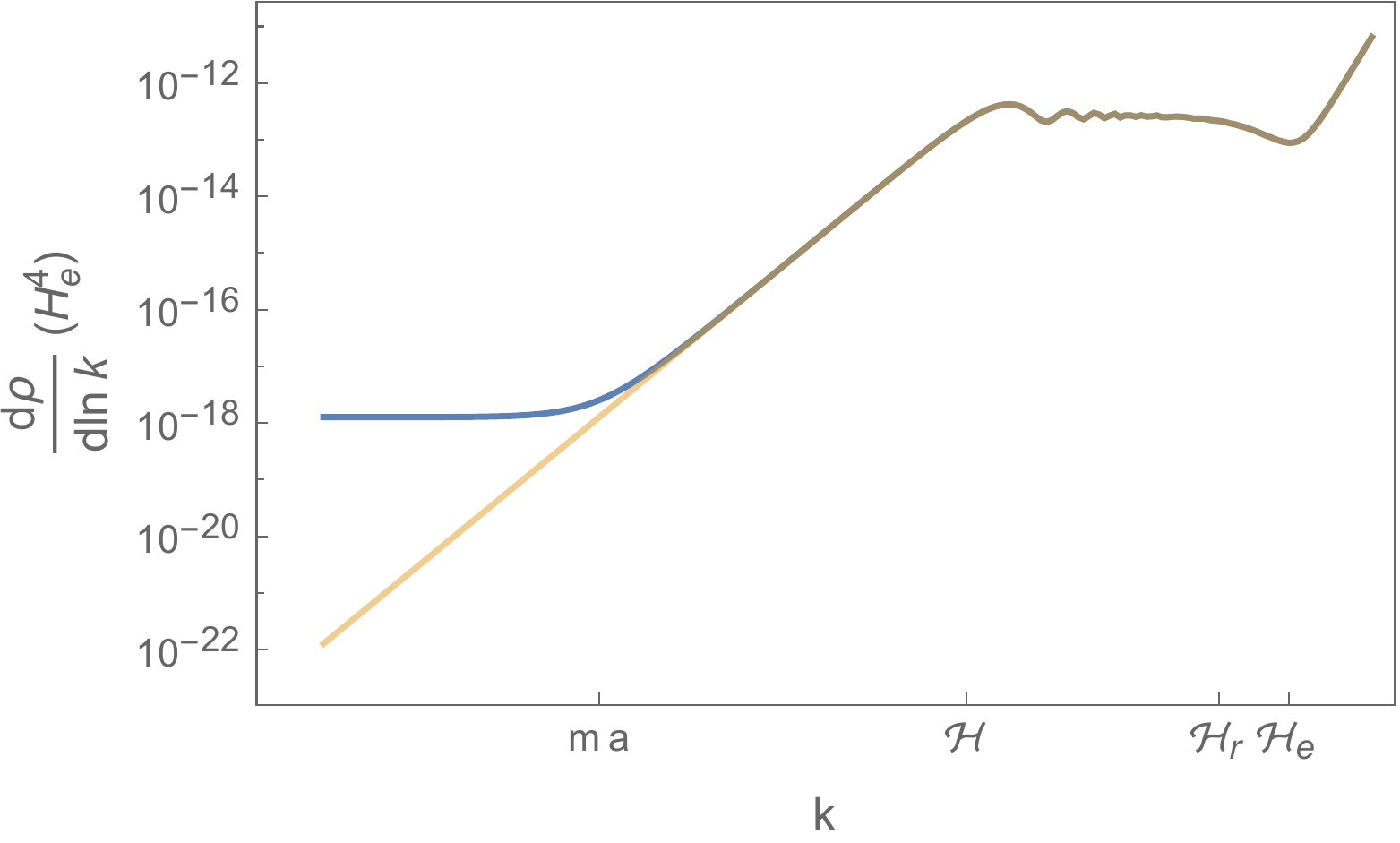}
	\caption{Spectral density at early times, $ma\leq \mathcal{H}$, after a smooth transition from de Sitter.  In this example $m=10^{-8}\,H_e$ and $a/a_e=10^3$ (blue). For comparison we also plot the spectral density in the massless case (light orange). The spectral density is well approximated by  (\ref{eq:spectral massive IR}) when  $k \leq \mathcal{H}$, (\ref{eq:rho IM general massless}) when $\mathcal{H}\leq k \leq \mathcal{H}_r$ and (\ref{eq:massive UV inflation}) when $\mathcal{H}_e\leq k$. The only significant difference with the massless case  is the presence  of a flat tail in the far infrared, whose energy density behaves like a dark energy component. Its relevance depends on the parameters of the model. The particle production formalism only applies in the intermediate range  $\mathcal{H}\leq k \leq \mathcal{H}_r$, though it does not capture the oscillations seen in the spectral density. }
	\label{fig:spectral light early}
\end{center}
\end{figure}

Figure \ref{fig:spectral light early} shows a plot of the  spectral density at early times after a smooth transition from de Sitter inflation, which  neatly displays  the expected behavior in the four relevant mode ranges.  As during inflation, see figure \ref{fig:spectral density light inflation}, at early times the spectral density differs from that of the massless case only in the far infrared. Away from this range we recover the same result as in the massless case; see section \ref{sec:Overview} for a discussion. But as the universe approaches intermediate times the far modes begin to dominate the infrared, mimicking a dark energy component with equation of state $w \approx -1$ that is absent in the massless case. Like in the massless case, only intermediate modes admit an interpretation in terms of particles. In contrast, the far infrared admits an interpretation in terms of a classical field, in addition to that  of the modes with $k<\Lambda_{\mathrm{IR}}$, as we discuss in more detail in section \ref{sec:The Particle and Classical Field Approximations}.

\paragraph{Intermediate Times ($\mathcal{H}\leq m\, a \leq \mathcal{H}_r$).} 

When $\mathcal{H}\leq m\,a $, the low frequency approximation (\ref{eq:chi bar infrared}) we relied on at early times and long wavelengths breaks down.   Still, the exact solutions of the mode equation during radiation domination are known to be the parabolic cylinder functions \cite{Birrell:1982ix,Aoki:2014dqa}, 
\begin{equation}\label{eq:exact massive rad}
	\chi^\mathrm{rad}_k= \frac{1}{(2ma\mathcal{H})^{1/4}}
	\exp\left(\frac{\pi}{8}\frac{k^2}{ma\mathcal{H}}\right)
	 D_\alpha\left[(1+i) \sqrt{\frac{m}{H}}\right], 
	\quad
	\alpha\equiv-\frac{1}{2}\left(1+i\frac{k^2}{ma\mathcal{H}}\right),
\end{equation}
which we have normalized by demanding that they approach the adiabatic limit (\ref{eq:out adiabatic}) in the asymptotic future,  when all modes are nonrelativistic. These solutions are the massive counterparts of equation (\ref{eq:wk0}), even though it is nontrivial to take the limit $m\to0$ in the former. 
Inspection of the index $\alpha$ indicates the appearance of a new momentum scale in the problem, 
\begin{equation}\label{eq:kJ}
	k_J\equiv \sqrt{m a\mathcal{H}},
\end{equation}
which is constant during radiation domination. 
As we discuss in appendix \ref{sec:Adiabaticity During Radiation Domination}, modes with  $k\leq k_J$  are not adiabatic at early times, but modes with $k_J\leq k$  remain adiabatic throughout radiation domination. It is then simpler to work in this range with the adiabatic approximation (\ref{eq:out adiabatic}), as we shall do later.   Note that this new scale is the inverse of the Jeans length of a self-gravitating scalar field identified in \cite{Hu:2000ke}.

Since we already determined the universal  solution of the mode equation that remains valid when both $k$ and $m$ are negligible,  equation (\ref{eq:chi 0 universal}),  we can simply find the appropriate infrared and intermediate mode functions $\chi^\mathrm{in}_k$  throughout radiation domination  by matching this expression to a linear combination of (\ref{eq:exact massive rad}) at $\eta_r$.  In the interval $k\leq k_J$ (which includes both the infrared and near intermediate modes, see figure \ref{fig:mode intervals}) the Bogolubov coefficients are 
\begin{equation}\label{eq:BC light IR}
	\beta^\mathrm{rad}_k \approx -i\frac{\Gamma(\nu)\Gamma(\frac{1}{4})}{4\pi}(-p)^{1/2}\left(-\frac{k\eta_0}{2}\right)^{-\nu}\left(\frac{\dot{a}_r}{\dot{a}_0}\right)^{3/4}\left(\frac{H_0}{m}\right)^{1/4},
	\quad 
 \alpha^\mathrm{rad}_k \approx \beta^\mathrm{rad}_k{}^*,
\end{equation}
where we have used the asymptotic form of the parabolic cylinder functions at early times, when ${m/H\leq 1}$. By the nature of the derivation, this analysis only applies if the field is still light at $\eta_r$, which is natural in this context,  unless  the mass of the field was close to the Hubble scale at the end of inflation. Plugging these Bogolubov coefficients into equation (\ref{eq:B transform}) and  exploiting the form of the parabolic cylinder functions at intermediate and late times (large arguments) we get 
\begin{equation}\label{eq:chi in IM massive}
 \chi_k^{\mathrm{in}}\approx  \frac{\Gamma(\nu)\Gamma(\frac{1}{4})}{2\pi}(-p)^{1/2}\left(-\frac{k\eta_0}{2}\right)^{-\nu}\left(\frac{\dot{a}_r}{\dot{a}_0}\right)^{3/4}\left(\frac{H_0}{m}\right)^{1/4} 
\frac{1}{\sqrt{2ma}}\sin\left(\frac{m}{2H}+\frac{\pi}{8}\right),
 \end{equation}
which happens to oscillate with  decaying amplitude in time, as opposed to being proportional to the scale factor as during early times. Finally,  substituting (\ref{eq:chi in IM massive}) into the spectral density (\ref{eq:rho}) we arrive at 
\begin{equation}\label{eq:spectral IR IM}
 	\frac{d\rho_{\mathrm{IR+nIM}}}{d\log k} \approx
	 \frac{\Gamma^2(\nu)\Gamma^2(\frac{1}{4})}{2^{5-2\nu}\pi^4}(-p)^{1-2\nu} 
	 m^2H_0^2 \left(\frac{a_\mathrm{osc}}{a}\right)^3 \left(\frac{k}{a_0H_0}\right)^{3-2\nu}.
 \end{equation}
Here, $a_\mathrm{osc}$ denotes the value of the scale factor at the threshold of intermediate times,   namely,   the time  at which  superhorizon modes start to oscillate,
\begin{equation}\label{eq:aosc}
 	ma_\mathrm{osc}= \mathcal{H}_\mathrm{osc}.
\end{equation}

In order to obtain the energy density of the modes in the infrared and near intermediate ranges we need to integrate (\ref{eq:spectral IR IM}). Since we assume that inflation lasted long enough to predict the amplitude of all subhorizon modes, $\Lambda_\mathrm{IR}<\mathcal{H}$, the Jeans scale $k_J$ is also above the infrared cutoff. In that case the mode interval is finite and its boundaries are fixed, so the energy density equals
\begin{equation}\label{eq:massive IM IRIM1} 
 	\rho_{\mathrm{IR+nIM}}\approx 
	\frac{\Gamma^2(\nu)\Gamma^2(\frac{1}{4})}{2^{5-2\nu}\pi^4}\frac{(-p)^{1-2\nu}}{3-2\nu}
 m^2H_e^2 
 \left[\left(\frac{k_J}{a_e H_e}\right)^{3-2\nu}
- \left(\frac{\Lambda_\mathrm{IR}}{a_e H_e}\right)^{3-2\nu} \right]
\left(\frac{a_\mathrm{osc}}{a}\right)^3,
\end{equation}
which  scales like nonrelativistic matter, and whose magnitude again depends on the precise balance of two unrelated quantities. Note that the dark energy component we identified at early times in equation (\ref{eq:fIR general light})  morphs into cold dark matter  at intermediate times, essentially because the nonevolving spectral density in the early time far infrared starts to decay like a presureless fluid at intermediate times. Indeed, at the intermediate threshold $a=a_\mathrm{osc}$, where $m a=k_J$, the far infrared energy density (\ref{eq:fIR general light}) we calculated at early times is of the same order as the energy density in (\ref{eq:massive IM IRIM1}), as one could have also guessed by looking at figure \ref{fig:mode intervals}.

In  the mode range $k_J \leq k$ it is somewhat cumbersome to work with the parabolic cylinder functions in equation (\ref{eq:exact massive rad}).   As we discuss in appendix \ref{sec:Adiabaticity During Radiation Domination},  modes in this interval remain adiabatic throughout radiation domination. Therefore, it is simpler to rely on the adiabatic mode functions (\ref{eq:out adiabatic}) here. When their overall phase is appropriately chosen, the latter reduce to equation (\ref{eq:wk0}) around the radiation time $\eta_r$, when all modes with $k_J \leq k$ remain still relativistic, so that in the interval $k_J \leq k \leq \mathcal{H}_r$ we can then lean on the analysis that led to the Bogolubov coefficients (\ref{eq:B IR general}) in the massless case. To proceed, it is convenient to differentiate between the mid and far intermediate modes of figure \ref{fig:mode intervals}.  In the mid intermediate range $k_J \leq k \leq ma$,  at intermediate times the adiabatic mode functions do not reduce to (\ref{eq:wk0}), but to the nonrelativistic counrerparts ${\chi_k\approx \exp[-i (m/2)(1/H-1/H_r)]/\sqrt{2ma}}$ instead. Therefore, by plugging (\ref{eq:B IR general}) into (\ref{eq:B transform}) the $in$ mode functions  at intermediate times become
\begin{equation}\label{eq:chi in IM}
 \chi_k^{\mathrm{in}}\approx -\frac{\Gamma(\nu)}{\sqrt{\pi}}\left(-\frac{k\eta_0}{2}\right)^{1/2-\nu}\frac{a_r}{a_0}\frac{\mathcal{H}_r}{k}\frac{1}{\sqrt{2ma}}
 \sin\left[\frac{m}{2}\left(\frac{1}{H}-\frac{1}{H_r}\right)\right],
\end{equation}
which substituted into (\ref{eq:rho}) finally leads to
\begin{equation}\label{eq:light sharp massive}
 	\frac{d\rho_{\mathrm{mIM}}}{d\log k}\approx \frac{\Gamma^2(\nu)}{2^{4-2\nu}\pi^3}(-p)^{1-2\nu}
	m^2H_0^2\left(\frac{a_{\mathrm{osc}}}{a}\right)^{3}\left(\frac{k_J}{a_0H_0}\right)\left(\frac{k}{a_0H_0}\right)^{2-2\nu}.
\end{equation}
The origin of the transition-dependent  $k_J=\sqrt{m\dot{a}_r}$ in equations like (\ref{eq:light sharp massive}) can be traced back to the amplitude of the mode  at the time it enters the horizon during radiation domination; compare (\ref{eq:chi in IM}) with (\ref{eq:chi in rad}). 

Integrating over the modes in the mid intermediate interval, then, we find that the energy density is dominated by the lower boundary of the integral, and, again, scales like nonrelativistic matter, as expected,
\begin{equation}\label{eq:rho IM2 IM}
 \rho_{\mathrm{mIM}}\approx \frac{\Gamma^2(\nu)}{2^{4-2\nu}\pi^3}\frac{(-p)^{1-2\nu}}{2\nu-2}m^2H_e^2\left(\frac{k_J}{a_eH_e}\right)^{3-2\nu}\left[1-\left(\frac{ma}{k_J}\right)^{2-2\nu}\right]\left(\frac{a_{\mathrm{osc}}}{a}\right)^{3}.
\end{equation}
 In particular, note that the  square brackets can be approximated by one soon after the comoving mass surpasses $k_J$ at the beginning of the intermediate time period.   Even though we have derived the spectral densities in (\ref{eq:spectral IR IM}) and (\ref{eq:light sharp massive}) by rather different means, note that they agree where they overlap, at  $k\sim k_J$. Since the spectral density is nearly flat in the interval $k\leq k_J$,  the infrared and near intermediate modes typically dominate over those in the mid intermediate range.

Modes in the far intermediate range $m \, a\leq k\leq \mathcal{H}_r$ are adiabatic and relativistic both at the transition and intermediate times, so  here we can directly  borrow the result quoted  in equation (\ref{eq:rho IM general massless}). But since the interval boundaries in the far intermediate range are different from those in the massless case, the energy density here is instead
\begin{equation}\label{eq:rho fIM}
  \rho_{\mathrm{fIM}} \approx \frac{\Gamma^2(\nu)}{2^{4-2\nu}\pi^3}\frac{(-p)^{1-2\nu}}{3-2\nu} \left(\frac{m}{H_e}\right)^{3-2\nu} H_e^2H^2 \left(\frac{a_e}{a}\right)^{2\nu-1}
 \left[1-\left(\frac{H}{m}\right)^{3/2-\nu}\right].
\end{equation}
which approximately behaves like the energy density of a fluid with equation of state $w=1$ and remains subdominant in comparison to (\ref{eq:massive IM IRIM1}) and (\ref{eq:rho IM2 IM}).  In the transition and ultraviolet ranges  the spectral density has the same form  as during early times, so the corresponding energy densities are again given by equations (\ref{eq:rho T massless}) and  (\ref{eq:rho UV EA}), respectively. 

\begin{figure}
\begin{center}
	\includegraphics[width=12cm]{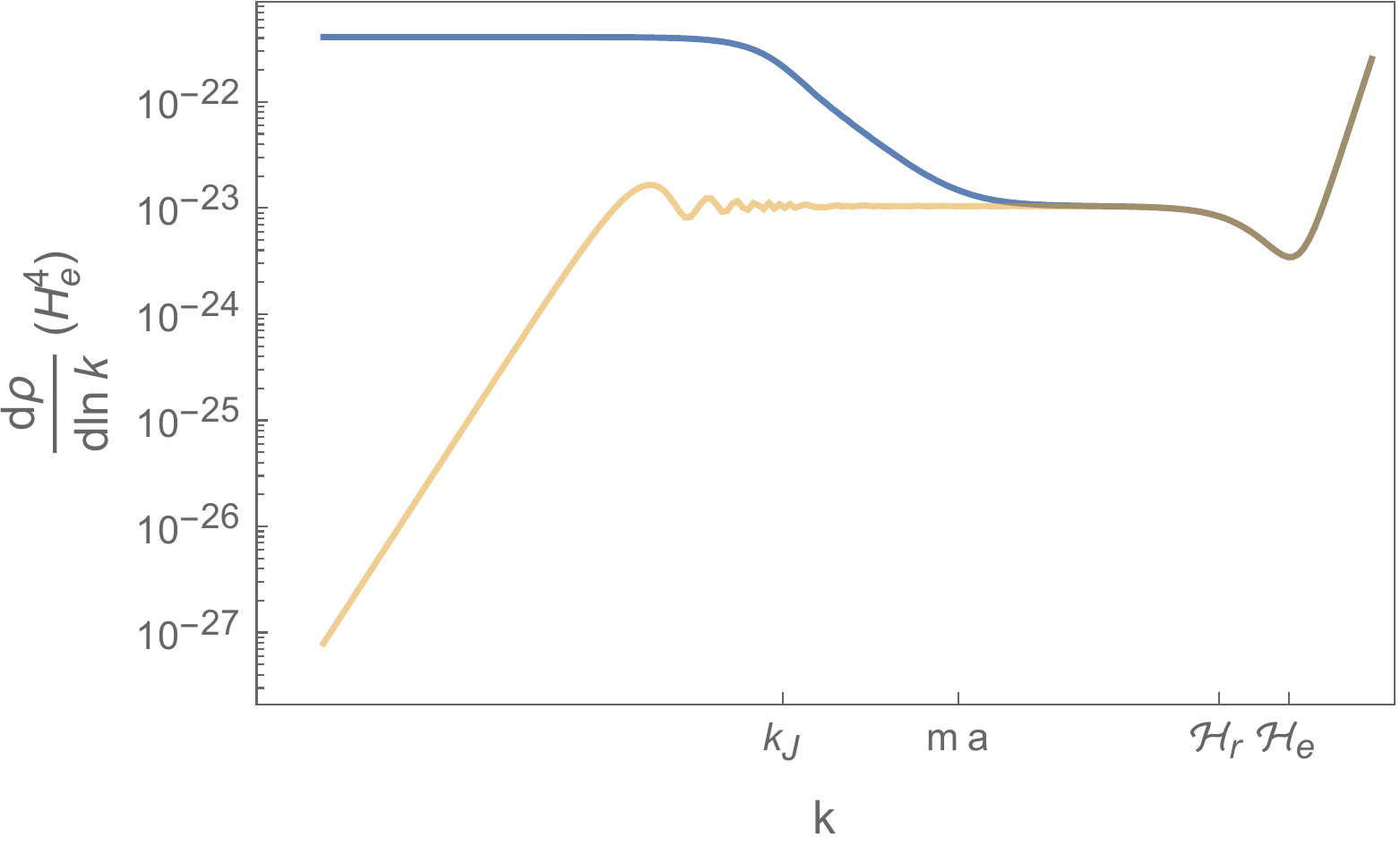}
	\caption{As in figure \ref{fig:spectral light early}, but  evaluated at an intermediate time  $\mathcal{H}\ll ma\ll \mathcal{H}_e$, with  $a/a_e=4\cdot 10^5$. The spectral density equals (\ref{eq:spectral IR IM}) when ${k\leq k_J}$, (\ref{eq:light sharp massive}) when $k_J\leq k \leq m a$, (\ref{eq:rho IM general massless}) when $m \, a\leq k \leq \mathcal{H}_r$ and (\ref{eq:massive UV inflation})  when $\mathcal{H}_e\leq k$.  At intermediate times, the particle production formalism reproduces the correct spectral density in the infrared and intermediate mode ranges, $k\leq\mathcal{H}_r$.}
	\label{fig:spectral light intermediate}
\end{center}
\end{figure}

Although the spectral density of a massless field  looks very similar to that of a light field at early times, as shown in figure \ref{fig:spectral light early}, as time goes by the presence of the growing mass term  deforms and alters the spectral density of a light field.  This is seen in figure \ref{fig:spectral light intermediate}, which  shows a plot of the  spectral density at intermediate times.   Again, the distinct asymptotic behaviors in the four different mode ranges are clearly recognizable.  In conclusion, at intermediate times the energy density is dominated by the infrared and near intermediate modes. They encompass the mode range that behaves like a cosmological constant at early times, and starts to redshifts as a cold dark matter component  at the onset of the intermediate times. The  latter can be interpreted as due to  particle creation, or the result of an homogeneous classical scalar displaced from equilibrium, as we explain in section \ref{sec:The Particle and Classical Field Approximations}.

 \begin{figure}
\begin{center}
	\includegraphics[width=12cm]{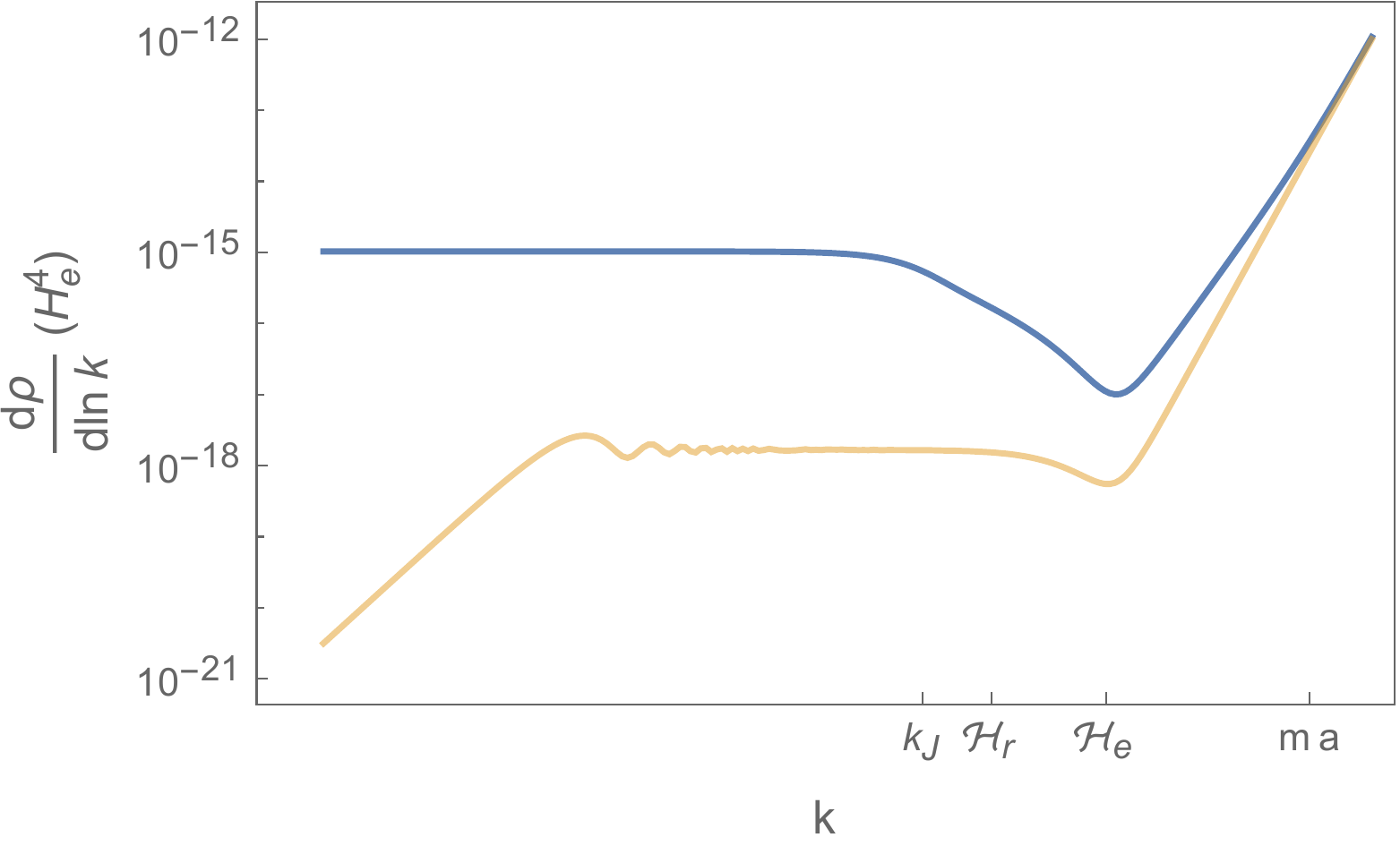}
	\caption{As in figures \ref{fig:spectral light early} and \ref{fig:spectral light intermediate}, but  evaluated at a late time  $ \mathcal{H}_e\ll ma$. For numerical reasons in this example we have chosen $m = 10^{-3} H_e$ and $a/a_e = 2\cdot 10^4$. The spectral density equals (\ref{eq:spectral IR IM}) when ${k\leq k_J}$, (\ref{eq:light sharp massive}) when $k_J\leq k \leq \mathcal{H}_r$, (\ref{eq:spectral light late}) when $\mathcal{H}_e\leq k \leq m\,a$, and (\ref{eq:massive UV inflation})  when $m\,a\leq k$. As during intermediate times, the particle production   formalism only  applies when $k\leq \mathcal{H}_r$.}
	\label{fig:spectral light late}
\end{center}
\end{figure} 

\paragraph{Late Times ($\mathcal{H}_e\leq ma$).}

The far intermediate range  present at intermediate times disappears at late times, by definition, see figure \ref{fig:mode intervals}. But, otherwise, all the expressions that we have derived for the  infrared and the two intermediate ranges at intermediate times effectively apply here too. The only difference is that in the equation for the energy density in the mid intermediate range,  (\ref{eq:rho IM2 IM}), $ma$ should be replaced by $\mathcal{H}_r$ to take into account the appropriate upper limit of the mode integral.   

In the transition range ${\mathcal{H}_r\leq k \leq  \mathcal{H}_e}$ modes are nonrelativistic and adiabatic. Therefore, although we ignore the magnitude of the Bogolubov coefficients in this range, say, equation (\ref{eq:rho bar}) immediately implies that the energy density here scales like non-relativistic matter 
\begin{equation} \label{eq:rho T massive}
	\rho_\mathrm{T}\approx \frac{m}{2\pi^2 a^3} \int_{\mathcal{H}_r}^{\mathcal{H}_e} \frac{dk}{k} k^3\left(|\beta^\mathrm{rad}_k|^2+\frac{1}{2}\right)\equiv \rho_{\mathrm{nr}}^{\mathrm{T}}\left(\frac{a_r}{a}\right)^3.
\end{equation}
For arbitrary $|\beta^\mathrm{rad}_k|^2$ the energy densities (\ref{eq:rho T massless}) and (\ref{eq:rho T massive}) are therefore unrelated. Yet, in ``reasonable" models we would expect both integrals to be dominated by the lower boundary at $k=\mathcal{H}_r$, since these modes left the horizon before the end of inflation (see figure \ref{fig:spectral light late}.) Under that assumption, it is then easy to see that $\rho_\mathrm{nr}^\mathrm{T}= \rho_\mathrm{rad}^\mathrm{T} (m/H_r)$. This result  precisely reproduces what we would find if the energy density (\ref{eq:rho T massless}) were dominated by particles of comoving momenta $k=\mathcal{H}_r$, which would  therefore become non-relativistic  at a time when $\mathcal{H}_r= m a$, and subsequently scale like (\ref{eq:rho T massive}).

In the  ultraviolet, $\mathcal{H}_e\leq k$, provided the transition is gradual,  it is safe to assume that modes remain adiabatic, so the spectral density reduces to that of the $out$ adiabatic vacuum, which we quote in equation (\ref{eq:spectral sub}) at fourth order. In the near ultraviolet the mass terms dominate and the spectral density becomes 
\begin{equation}\label{eq:spectral light late}
	\frac{d\rho_\mathrm{nUV}}{d\log k}\approx  \frac{1}{4\pi^2} \frac{k^3 m}{a^3},
\end{equation}
while in the far ultraviolet we recover again the relativistic result (\ref{eq:massive UV inflation}).  Although the relevant modes are in the adiabatic regime, these expressions cannot be recovered from the particle production formalism because the coefficients $\beta^\mathrm{ad}_k$ are small when $\mathcal{H}_e\leq k$. By integrating  equation (\ref{eq:spectral sub}) over the ultraviolet modes, and subtracting equation (\ref{eq:rho sub}), we obtain the renormalized energy density
\begin{eqnarray}
 2\pi^2 \rho_{\mathrm{ren}}^{\mathrm{UV}} &\approx& \left[\delta\Lambda^f+\frac{m^4}{32}\log \frac{m^2}{\mu^2}\right]-
 3H^2\left[(\delta M_P^2)^f+\frac{m^2}{48}\log \frac{m^2}{\mu^2}\right] \nonumber\\
 &&- \frac{1}{6}\left(\frac{\dot{a}_r}{\dot{a}_e}\right)^{-3/2}\left(\frac{m}{H_e}\right)^{1/2}m^2H_e^2\left(\frac{a_\mathrm{osc}}{a}\right)^3,
\end{eqnarray}
which again agrees with that of the $out$ adiabatic vacuum.  Leaving renormalization aside, at late times the leading contribution behaves like nonrelativistic matter, with a density that is suppressed in comparison with those in equations (\ref{eq:massive IM IRIM1})  and (\ref{eq:rho IM2 IM}). Observe that the renormalized cosmological constant and Planck mass, which used to run at early times, equation (\ref{eq:rho UV EA}), approach constants at late times. 

\begin{figure}
\begin{center}
	\includegraphics[width=12cm]{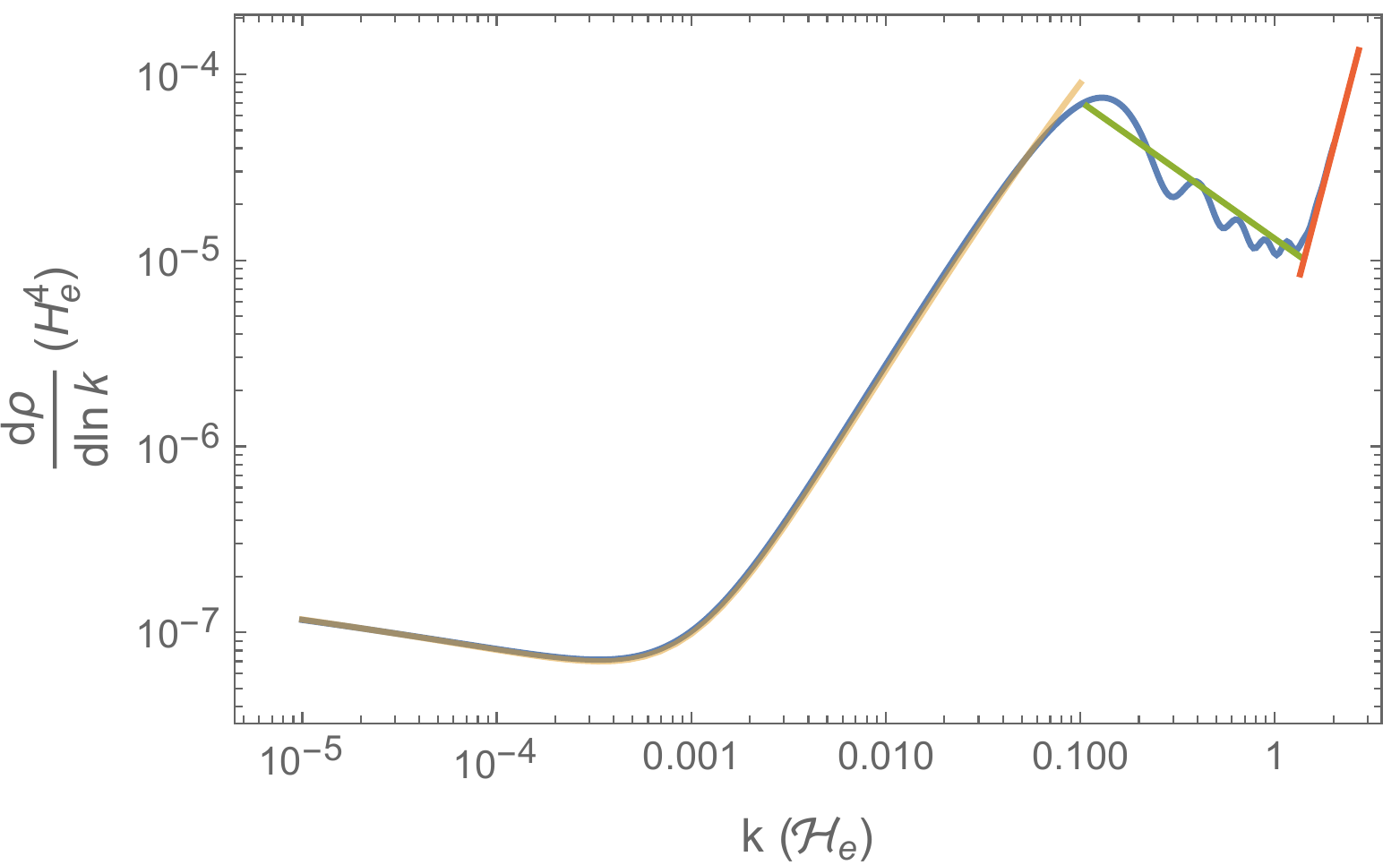}
	\caption{Spectral density in an inflationary model with chaotic potential (\ref{eq:chaotic}) at early times. In this example $m=10^{-4}H_e$   and $a/a_e=10$. The superimposed  curves  follow equations (\ref{eq:spectral massive IR}), (\ref{eq:rho IM general massless}) and (\ref{eq:massive UV inflation}) in the corresponding mode ranges. With $ma=10^{-3} \mathcal{H}_e$ and $\mathcal{H}\approx 7\cdot 10^{-2} \mathcal{H}_e$  the transition between the different approximations  happens where expected.  In  equation (\ref{eq:spectral massive IR}) we have  taken $\eta_0$ to be the time at which the given mode crosses the horizon, with $p$ determine by equation (\ref{eq:p slow roll}) at the same time. The approximation works so well that it overlaps with the actual spectral density. \label{fig:chaotic massive}}
\end{center}
\end{figure}

Figure \ref{fig:spectral light late} shows the spectral density at late times after a smooth transition to radiation domination. As opposed to what happened at early times, the spectral density here is quite different from that of the massless case.  As further illustration of our general analysis in a realistic case, we  consider again the evolution of a scalar field in the quartic potential (\ref{eq:chaotic}) that we analyzed in section \ref{sec:Differentiable Transition}. Figure \ref{fig:chaotic massive} compares the  numerically determined spectral density with our analytical estimates, and confirms again their validity.

\subsection{The Particle and Classical Field Approximations}
\label{sec:The Particle and Classical Field Approximations}

We may also return to the question whether part of the previous results can be reinterpreted in the light of the  homogeneous classical  scalar field we introduced in section \ref{sec:Classical Field Description}. Indeed, at early times, in the far infrared,  substitution of the Bogolubov coefficients  (\ref{eq:B light IR}) into equations (\ref{eq.identification0mode}) reveals that, up to a phase,
\begin{equation}\label{eq:hom scalar}
	B_0\approx A^*_0\approx -i\sqrt{\frac{1}{2\pi^2} \int_{\Lambda_\mathrm{IR}}^{ma} \frac{dk}{k} k^3\, |\beta_k|^2}
\end{equation}
solves the equations required for a classical description, (\ref{eq:identification}). This identification applies whenever the Bogolubov coefficients satisfy $\alpha_k\approx \beta_k^*$ and $|\beta_k|\gg 1$, as  happens  with  (\ref{eq:B light IR}) and (\ref{eq:BC light IR}). In the light field limit we are concerned here with,  the mode functions $\chi_0$  in (\ref{eq:phi cl}) can be approximated by the massless expression   (\ref{eq:chi bar infrared}). The latter then implies that such a classical scalar would be real, which agrees with the character of the $in$ mode functions at long wavelengths that we identified in (\ref{eq:chi 0 universal}). The field amplitude of the scalar is  nearly frozen,
\begin{equation}\label{eq:classical amplitude}
	\phi_{\mathrm{cl}}\approx \sqrt{\frac{2}{M}}| B^\mathrm{low}_0|,
\end{equation} 
where the superscript in $B_0^\mathrm{low}$ indicates that this is the value of $B_0$ obtained by substitution of  (\ref{eq:B light IR}) into (\ref{eq:hom scalar}).  Despite appearances, the field magnitude does not depend on  $M$ because $|B_0^\mathrm{low}|$ is linear in  $M^{1/2}$. Its energy density is $\rho_\mathrm{cl}\approx m^2 |\phi_\mathrm{cl}|^2/2$, which precisely matches the energy density in  equation (\ref{eq:fIR general light}). 
 It is important to stress, though, that  the field amplitude (\ref{eq:classical amplitude}) does not evolve exactly like that of a classical scalar, at least initially, since its magnitude also changes  as $ma$ grows and additional nonrelativistic modes contribute to its mean square. That is because this classical field does not arise from the  expectation of the scalar in a coherent state, but captures its quantum fluctuations in the  {\it in} vacuum over scales $\Lambda_{\mathrm{IR}}\leq k \leq ma$, $|\phi_{\mathrm{cl}}|^2=\left\langle\phi^2(x)\right\rangle_{\Lambda_{\mathrm{IR}}\leq k\leq ma}$, which is why we can think of such a classical  field  as performing a random walk during inflation.  In particular, prior to the opening of the far infrared window, this classical amplitude vanishes. Soon afterwards,  $|\phi_\mathrm{cl}|^2$  grows rather slowly, since $3-2\nu$ is a small number. Its actual asymptotic value, in particular, crucially depends on whether the field is ultra-light or merely-light, as we emphasized following equation (\ref{eq:rho fIR inf}).

At intermediate and late times the energy density is dominated by the infrared and near intermediate mode ranges, equation (\ref{eq:massive IM IRIM1}), whose contribution is also readily explained in terms of the  homogeneous classical scalar field we introduced earlier. This is so because in the  range $\Lambda_{\mathrm{IR}}\leq k\leq k_J$   modes are nonrelativistic  and the boundary that dominates the energy density remains effectively fixed.  The amplitude of the classical scalar field can be inferred by substituting  the Bogolubov coefficients (\ref{eq:BC light IR})  into equation (\ref{eq:hom scalar}), and then using  the  form of the corresponding mode functions (\ref{eq:exact massive rad})  at $k=0$ in the asymptotic future, 
\begin{equation}\label{eq.classical.intermediate}
 \phi_{\mathrm{cl}}\approx \sqrt{\frac{2}{Vm}}|B_0^\mathrm{rad}|\frac{1}{a^{3/2}}\sin\left(\frac{m}{2H}+\frac{\pi}{8}\right).
\end{equation}
Therefore, as expected, the scalar field starts to oscillate once its mass surpasses the Hubble parameter.  The energy density of such an oscillating field scales like that of cold dark matter, so  at intemediate and late  times it is of order ${\rho_\mathrm{cl}=\rho_\mathrm{osc} (a_\mathrm{osc}/a)^3}$, as we found in equation (\ref{eq:massive IM IRIM1}). The amplitude of the originally frozen scalar is an initial condition in the classical theory, whereas in the quantum theory it is predicted  by equation (\ref{eq:classical amplitude}),  up to the contribution of modes below the infrared cutoff that we described in section \ref{sec:Beyond the Infrared Cutoff}. 

Remarkably, we could have also arrived at  the energy density in (\ref{eq:massive IM IRIM1})  using the particle production formalism of section \ref{sec:Particle Production}.  We argue in appendix \ref{sec:Adiabaticity During Radiation Domination} that at intermediate and late  times all modes of a massive field are adiabatic, including those that are nonrelativistic. Therefore, since we are dealing with  frequencies $\omega_k\approx ma\gg \mathcal{H}$ and   the coefficients $\beta^\mathrm{rad}_k$ in equation (\ref{eq:BC light IR}) are large,  the three conditions in section \ref{sec:Particle Production} are met.  Indeed, if we substitute the Bogolubov coefficients (\ref{eq:BC light IR}) into the particle production formula (\ref{eq:rho ppf}), with $\omega_k\approx ma$, we  precisely recover equation (\ref{eq:spectral IR IM}), while substitution of (\ref{eq:BC light IR}) into (\ref{eq:rho ren ppf}) leads to (\ref{eq:massive IM IRIM1}). In other words, we can also think of the scalar as an ensemble of nonrelativistic particles of number density $n_\mathrm{p}= \rho_\mathrm{ren}/m$.

\subsection{Overview}
\label{sec:Light Fields Overview}

A  scalar field that is light during inflation displays a rich variety of behaviors that cannot be entirely captured by the particle production formalism or simply by a classical homogeneous field.  This can be appreciated in table \ref{table:classical/particle},  which summarizes the evolution of the energy density of a light scalar during radiation domination in all mode ranges.

 At early times, when the scalar field mass obeys $ma\leq \mathcal{H}$, the spectral density agrees with that of the massless case, up to the presence of a  flat tail in the far infrared, where $k\leq ma$, which appears once $\Lambda_{\mathrm{IR}}<ma$.  The  contribution of this tail behaves like a  dark energy component that is typically subdominant at the end of inflation and comes to dominate the scalar energy density before the end of the early time regime, when the field mass surpasses the Hubble constant.  Once in the intermediate time regime, $\mathcal{H}\leq ma\leq \mathcal{H}_r$, the infrared modes that were previously frozen begin to oscillate and the former dark energy component morphs then into nonrelativistic matter, a behavior that persists into the late time regime, when ${\mathcal{H}\leq ma}$.    In both regimes the dominant contribution of the  scalar to the energy density is quite sensitive to the particular parameters of the theory; it can be expressed in the form $\rho \sim \rho_{\mathrm{osc}}(a/a_{\mathrm{osc}})^3$, with $\rho_{\mathrm{osc}}\sim m^2H_i^2/|1+p|$ if the field is ultra-light and $\rho_{\mathrm{osc}}\sim H_i^4$ if it is merely-light (these estimates are upper limits that are not always saturated.)  Both the dark energy- and cold dark matter-like behaviors follow from general solutions of the mode equation, and therefore remain valid regardless of the nature of cosmic expansion. This implies that they will persist even after the radiation era has ended.   The scalar field also renormalizes  the cosmological constant and the Planck mass. As opposed to what happens in the massless case,  these ``constants" experience a characteristic running with the logarithm of the scale factor that stops at late times, when $ma$ surpasses  $\mathcal{H}_e$. The corresponding change in these parameters as the mass crosses such threshold is a unique signature of this scenario that does not depend on the unknown values of the counterterms. We  determine the renormalized energy density of the $out$ adiabatic vacuum at intermediate and late times in the next section, equation (\ref{eq:rho dS massive}). Up to the counterterms, its contribution  relative to that of the dark matter- and dark energy-like components we just mentioned is negligible. 


\begin{table}
\begin{center}
LIGHT FIELD DURING INFLATION\\
\vspace{.2cm}
Early times ($ma\leq \mathcal{H}$) \quad Effective coupling constants run\\
\begin{tabular}{ c c c c c}
 \hline
 \hline
  $\Lambda_{\mathrm{IR}} \leq k\leq ma$   & $ma\leq k\leq \mathcal{H}$ & $\hspace{-.4cm}$ $\mathcal{H}\leq k\leq \mathcal{H}_r$ &  $\hspace{-.4cm}$ $\mathcal{H}_r\le k\le \mathcal{H}_e$ &  $\hspace{-.4cm}$ $\mathcal{H}_e\leq k$  \\ \hline \\[-2ex]
  {\color{red}$\displaystyle{\rho_{\mathrm{osc}}}$}   & ${\color{red}\displaystyle{\frac{H_e^2 H^2}{16\pi^2}}}$ & $\hspace{-.4cm}$ {\color{red}$\displaystyle{\frac{H_e^2 H^2}{8\pi^2}\log\frac{a}{a_r}}$} & $\hspace{-.4cm}$ $\displaystyle{\rho_{\mathrm{rad}}^{\mathrm{T}}\left(\frac{a_r}{a}\right)^4}$  & $\hspace{-.4cm}$ {\color{red}$\displaystyle{\sim H_e^2 H^2 \left(\frac{\dot{a}_e}{\dot{a}_r}\right)^2  }$} \\[1ex]
   {\footnotesize {\color{red} classical field}} & & $\hspace{-.4cm}$ {\footnotesize {\color{red}particle production}} &  {\footnotesize particle production} & \\
 \hline
 \hline 
\end{tabular}
\vspace{.5cm}

Intermediate times ($\mathcal{H}\leq ma \leq \mathcal{H}_r$)
\quad Effective coupling constants run
\begin{tabular}{ c  c c c c }
 \hline
 \hline
  $\Lambda_{\mathrm{IR}} \leq k\leq k_J$   & $k_J\leq k\leq ma$ & $\hspace{-.2cm}$ $ma\leq k\leq \mathcal{H}_r$ & $\hspace{-.3cm}$ $\mathcal{H}_r\le k\le \mathcal{H}_e$ &  $\hspace{-.5cm}$ $\mathcal{H}_e\leq k$    \\
 \hline \\[-2ex]
  {\color{red}$\displaystyle{\rho_{\mathrm{osc}}\left(\frac{a_{\mathrm{osc}}}{a}\right)^3}$}  & $\displaystyle{\rho_{\mathrm{osc}}^{(2)}\left(\frac{a_{\mathrm{osc}}}{a}\right)^3}$ & $\hspace{-.2cm}$ $\displaystyle{\rho_{\mathrm{stiff}}\left(\frac{a}{a}\right)^6}$ & $\hspace{-.3cm}$ $\displaystyle{\rho_{\mathrm{rad}}^{\mathrm{T}}\left(\frac{a_r}{a}\right)^4}$  & $\hspace{-.5cm}$ $\displaystyle{\sim H_e^2 H^2  \left(\frac{\dot{a}_e}{\dot{a}_r}\right)^2  }$ \\[1ex]
 {\footnotesize {\color{red}classical field}} & {\footnotesize classical field}& $\hspace{-.2cm}$ {\footnotesize particle } & {\footnotesize particle } & \\[-1ex]
 {\footnotesize {\color{red}particle production}} & {\footnotesize particle production}&   {\footnotesize  production} &  {\footnotesize  production}& \\
 \hline
 \hline 
\end{tabular}
\vspace{.5cm}

Late times ($\mathcal{H}_e\leq ma$) \quad Effective coupling constants frozen
\begin{tabular}{ c c c c}
 \hline
 \hline
  $\Lambda_{\mathrm{IR}} \leq k\leq k_J$   & $ma\leq k\leq \mathcal{H}_r$ & $\mathcal{H}_r\le k\le \mathcal{H}_e$ &  $\mathcal{H}_e\leq k $  \\
 \hline \\[-2ex]
  {\color{red}$\displaystyle{\rho_{\mathrm{osc}}\left(\frac{a_{\mathrm{osc}}}{a}\right)^3}$}     & $\displaystyle{\rho_{\mathrm{osc}}^{(2)}\left(\frac{a_{\mathrm{osc}}}{a}\right)^3}$  & $\displaystyle{\rho_{\mathrm{nr}}^{\mathrm{T}}\left(\frac{a_{\mathrm{nr}}}{a}\right)^3}$ &  $\!\!\!\!\displaystyle{\sim\left(\frac{\dot{a}_e}{\dot{a}_r}\right)^{3/2} \!\! m^{5/2}H_e^{3/2}\left(\frac{a_\mathrm{osc}}{a}\right)^3}$ \\[1ex]
 {\footnotesize {\color{red}classical field}} & {\footnotesize classical field}& {\footnotesize classical field} & \\[-1ex]
 {\footnotesize {\color{red}particle production}} & {\footnotesize particle production}& {\footnotesize particle production} & \\
 \hline
 \hline 
\end{tabular}
\end{center}
\caption{Contributions to the renormalized energy density of a light scalar during the three epochs of radiation domination following near de Sitter inflation.  Equations in red denote the potentially dominant contributions. Apart from the far infrared, the early times regime is identical to that of a massless field, see table~\ref{table:classical/particle massless}. 
The differences with the massless case  manifest themselves at intermediate and late times. The values of $\rho_{\mathrm{osc}}^{(2)}$,  $\rho_{\mathrm{stiff}}$, $\rho^\mathrm{T}_{\mathrm{rad}}$ and $\rho^\mathrm{T}_{\mathrm{nr}}$ are not particularly relevant, but can be inferred from equations~(\ref{eq:rho IM2 IM}), (\ref{eq:rho fIM}),  (\ref{eq:rho T massless}) and (\ref{eq:rho T massive}), respectively.  On the other hand, $\rho_{\mathrm{osc}}\sim m^2H_i^2/|1+p|$ if the field is ultra-light and $\rho_{\mathrm{osc}}\sim H_i^4$ if the field is merely-light; see equations~(\ref{eq:rho fIR ultra}) and~(\ref{eq.edfIR2dS}) and the discussion below. Finally, $k_J$ is defined in equation (\ref{eq:kJ}) and  $a_\mathrm{osc}$ in equation (\ref{eq:aosc}). Note that at intermediate and late times the dominant contributions to the renormalized energy density (in red) depend on the details of inflation, but are insensitive to those of reheating. The behavior of  modes with $k<\Lambda_\mathrm{IR}$ is studied in section \ref{sec:Beyond the Infrared Cutoff}.}
\label{table:classical/particle}
\end{table}

As indicated on table \ref{table:classical/particle}, some of the previous results can be readily derived within the particle production or the classical field approximations. The simplicity of these approximations underscores the convenience of these approaches in the cases in which  do apply.  Along these lines, it is also worth emphasizing that, if the scalar field is light during inflation, the mode ranges that dominate the energy density at intermediate and late times simultaneously admit  an interpretation  in terms of  a classical field and an ensemble of particles. Furthermore, in this case the contribution of the $out$ adiabatic vacuum is negligible and only affects the values of the effective coupling constants of the theory. 

\section{Heavy Fields}
\label{sec:Heavy Fields}
 
We shall end our analysis with massive fields,  in the limit in which their mass is much larger than the Hubble constant during inflation.  In this limit,  we can reap the benefits of the general discussion in section \ref{sec:Cosmological Transitions}. Because the only dimensionless parameter that controls the size of the Bogolubov coefficients $\beta_k$ is $H/m$, these are small by assumption.  Hence, the downside of this regime is that it does not leave much room for particle production or the emergence of a sizable classical field.

 By construction,   the renormalized energy density of the $out$ vacuum is expected to be small, since it is of sixth adiabatic order. To estimate its magnitude, we expand $d\rho_\mathrm{out}/d\log k$ in equation (\ref{eq.density.out}) to sixth adiabatic order, and subtract equation (\ref{eq:spectral sub}).  Integrating over all modes below the ultraviolet cutoff and restoring the contribution of the counterterms we obtain the renormalized energy density of the $out$ adiabatic vacuum
\begin{eqnarray}\label{eq:rho dS massive}
 	2\pi^2 \rho^\mathrm{out}_{\mathrm{ren}} &\approx& \left[\delta\Lambda^f+\frac{m^4}{32}\log \frac{m^2}{\mu^2}\right]
	-3H^2\left[(\delta M_P^2)^f+\frac{m^2}{48}\log \frac{m^2}{\mu^2}\right]
	 \\
	 &&+48\left[\delta c^f-\frac{1}{384}\log\frac{m^2}{\mu^2}\right]\left(\frac{\mathcal{H}^2 \ddot{a}}{a^5}+\frac{\ddot{a}^2}{4a^6}-\frac{\mathcal{H}\dddot{a}}{2a^5} \right)
	 \nonumber\\
 	&&+\frac{1}{m^2}
	\Bigg[\frac{211 H^6}{5040}-\frac{173}{280}\frac{H^4\ddot{a}}{a^3}
	+\frac{1237}{2240} \frac{H^2 \ddot{a}^2}{a^6}
	+\frac{673}{10080}\frac{\ddot{a}^3}{a^9} 
	+\frac{683}{1680}\frac{H^3 \dddot{a}}{a^4}
	\nonumber \\
	&&-\frac{673}{3360}\frac{H \ddot{a} \dddot{a}}{a^7}
	+\frac{17}{2240} \frac{\dddot{a}^2}{a^8}
	-\frac{17}{140}\frac{H^2 a^{(4)}}{a^5}
	-\frac{17}{1120}\frac{\ddot{a}\, a^{(4)}}{a^8}
	+\frac{17}{1120}\frac{H a^{(5)}}{a^6}
	\Bigg]. \nonumber
\end{eqnarray}
The terms of zeroth, second and fourth adiabatic order in this equation depend on the contribution of the counterterms, so these are not predictions of the quantum theory.  Those of fourth order renormalize the dimension four curvature invariants, which vanish in de Sitter and during radiation domination. Note that none of the coupling constants run with cosmic expansion. An expansion in powers of $m$ of the standard renormalized energy density in de Sitter \cite{Dowker:1975tf}  reveals that the first nonvanishing term is indeed proportional to $H^6/m^2$, and exactly agrees with the predicted sixth-order term in  (\ref{eq:rho dS massive}).

Although  $\rho^\mathrm{out}_\mathrm{ren}$ is small, when the field is heavy during inflation and the transition is gradual we expect it to  provide the dominant contribution to the renormalized energy density.  All we need to assume is that  the modes remain adiabatic  between inflation and radiation domination  (we analyze abrupt transitions in the next subsection.)  Then,  the  uninterrupted validity of the adiabatic approximation implies that the Bogolubov coefficients $\beta^\mathrm{ad}_k$ vanish, so the total energy density is just that of the $out$ adiabatic vacuum.   Therefore, setting $\ddot{a}= 0$ in equation (\ref{eq:rho dS massive}) yields the energy density during radiation domination,
\begin{equation}\label{eq:rho massive general}
 	2\pi^2\rho_{\mathrm{ren}} \approx \left[\delta\Lambda^f+\frac{m^4}{32}\log \frac{m^2}{\mu^2}\right]-
 3H^2\left[(\delta M_P^2)^f+\frac{m^2}{48}\log \frac{m^2}{\mu^2}\right]+\frac{211}{5040}\frac{H^6}{m^2}.
\end{equation}
As we noted above, the contribution of order $H^4$ to the energy density is absent. This main result is summarized in table \ref{table:classical/particle heavy}.

\begin{table}
\begin{center}
HEAVY FIELD DURING INFLATION\\
\vspace{.2cm}
\begin{tabular}{ c  }
 \hline
 \hline
    $0< k$  \\ \hline \\[-2ex]
    $\displaystyle{\frac{211}{10080\pi^2}\frac{H^6}{m^2}}$  \\[1.5ex]   
   {\footnotesize {\it out} adiabatic vacuum} \\
 \hline
 \hline 
\end{tabular}
\end{center}
\caption{Renormalized energy density of a heavy scalar during the radiation era following near de Sitter inflation.  
This expression does not depend on the properties of inflation nor reheating. It coincides with 
the energy density of the {\it out} adiabatic vacuum and does not admit an interpretation in terms of produced particles or a classical field. The effective coupling constants are frozen. The zero mode $k=0$ is discussed in section \ref{sec:Beyond the Infrared Cutoff}.}
\label{table:classical/particle heavy}
\end{table}


\subsection{Examples}

To arrive at equation (\ref{eq:rho massive general}) we have assumed that the transition is gradual. If the derivatives of the scale experience a sudden jump, the Bogolubov coefficients do not vanish, and the renormalized energy density receives additional   contributions from the produced particles, $\rho_\mathrm{p}$, as on the examples we discuss next. 

\paragraph{Sharp Transition.}

In a universe that expands like equation (\ref{eq:disc transition}), the solutions of the  mode equation (\ref{eq:mode equation}) are still given  by the parabolic cylinder functions (\ref{eq:exact massive rad}). Yet these are  somewhat unwieldy, particularly in the limit of large masses.  Since we are interested in the limit of heavy fields anyway, it turns out to be simpler just to consider  the adiabatic approximation, which remains valid for all modes throughout  inflation and radiation domination. From equations (\ref{eq:B UV}), at two time derivatives the corresponding Bogolubov coefficients become
\begin{equation}
	\beta^\mathrm{ad}_k \approx \frac{p(p-1)}{4}\left[\frac{\mathcal{H}_e}{p\omega_k(\eta_e)}\right]^2\left[1+\frac{m^2a_e^2}{2\omega_k^2(\eta_e)}\right]  ,\quad \alpha^\mathrm{ad}_k \approx 1.
\end{equation}
The main difference with the massless case is that in the infrared limit, the mode functions and the Bogolubov coefficients remain finite. In this limit, the coefficient $\beta^\mathrm{ad}_k$ is suppressed, as expected, by a power of $(H_e/m)^2$. The suppression extends to the ultraviolet, where $\beta^\mathrm{ad}_k$ is of order $(\mathcal{H}_e/k)^2$. As we discussed earlier, this renders the energy density of the field nonrenomalizable.  In fact,  it is readily seen that in the ultraviolet the value of $\beta^\mathrm{ad}_k$ agrees with the one we found in the massless case, equation (\ref{eq:B UV sharp}). Yet, again, since the energy density is not renormalizable, it is not strictly possible to obtain a sensible  finite value that captures the ultraviolet contribution. We shall instead evaluate the renormalized energy density  for a smooth transition, which we  take as a representative case of an abrupt transition.

\paragraph{Smooth Transition.}

With the scale factor given by (\ref{eq:cont transition}), it is not possible to find analytical solutions of the mode equation (\ref{eq:mode equation}). We shall hence  focus again on the  adiabatic approximation, which remains  valid  for all modes  as long as the field is sufficiently heavy. 

In order to find the Bogolubov coefficients in the adiabatic approximation, we shall use equations (\ref{eq:B UV}) one more time. Because the scale factor is continuous up to its second derivative, it suffices to focus on those terms that contain three time derivatives. Doing so we arrive at
\begin{equation}\label{eq:B smooth huge}
	\beta^\mathrm{ad}_k \approx i\frac{p(p-1)(1-r\eta_e)}{4}\left[\frac{\mathcal{H}_e}{p\omega_k(\eta_e)}\right]^3\left[1+\frac{m^2a_e^2}{2\omega_k^2(\eta_e)}\right]  ,\quad 
 \alpha^\mathrm{ad}_k \approx 1 +\beta^\mathrm{ad}_k{}^*.
\end{equation}
In the infrared limit, the coefficient $\beta^\mathrm{ad}_k$ is suppressed by $(H_e/m)^3$, one additional power than in the case of a sharp transition. This again illustrates that particle production is related to departures from adiabaticity: The smoother the transition, the higher the suppression. In the ultraviolet regime we find $\beta^\mathrm{ad}_k\sim (\mathcal{H}_e/k)^3$, which yields a renormalizable energy density and agrees with equation (\ref{eq:B smooth UV}).

Because the Bogolubov coefficients $\beta^\mathrm{ad}_k$ are only mildly  suppressed here, the renormalized energy density of the field in an abrupt transition may stray away from that of the field in the $out$ vacuum in equation (\ref{eq:rho massive general}). In order to determine the contribution due to the nonzero $\beta^\mathrm{ad}_k$, we simply substitute equation (\ref{eq:B smooth huge}) into equation (\ref{eq:rho adiabatic}). The integral of the term proportional to $|\beta^\mathrm{ad}_k|^2$ can be evaluated exactly,  but we just reproduce  here its limit when $a\gg a_e$, 
\begin{equation}\label{eq:rho ppf 1}
	\rho^{(0)}_{\mathrm{p}} \approx \frac{5p^2(p-1)^2(1-r\eta_0)^2}{2048\pi p^6}H_e^4\frac{H_e^2}{m^2}\left(\frac{a_e}{a}\right)^3.
\end{equation}
This  scales like nonrelativistic matter, as expected, and dominates over the ``vacuum" piece in equation (\ref{eq:rho massive general}). Note that equation (\ref{eq:rho ppf 1}) is what we would have obtained from the particle production formula (\ref{eq:rho ppf}). To estimate the contribution  proportional to $|\alpha^\mathrm{ad}_k\beta^\mathrm{ad}_k|$,
we restrict our attention to the nonrelativistic modes, which do not oscillate with $k$ and ought to give the dominant contribution to the mode integral. In the same  limit we find 
\begin{equation}
	\rho^{(1)}_{\mathrm{p}} \approx	\frac{3(p^2-p)(1-r\eta_e)}{4\pi^2p^3} H_e^4 \left(\frac{H_e}{m}\right)^{-1} \frac{H}{m}
	\log \left(\frac{a}{a_e}\right)
	 \left(\frac{a_e}{a}\right)^3  \sin \left(m\!\int^\eta \!\!  \tilde{a}\, d\tilde{\eta}\right).
\end{equation}
Note that  both    energy densities $\rho_\mathrm{p}^{(0)}$ and $\rho_\mathrm{p}^{(1)}$ are essentially determined by  two independent dimensionless expansion parameters, $H_e/m$ and $H/m$. The former is the one that controls the magnitude of the Bogolubov coefficients, whereas the latter is the one that controls the accuracy of the derivative expansion. Although $\rho_\mathrm{p}^{(1)}$ indeed contains one more power of $H/m$ than $\rho_\mathrm{p}^{(0)}$, the latter is suppressed by three additional powers of $H_e/m$. Hence,  $\rho_\mathrm{p}^{(1)}$ may actually dominate the energy density of the ``produced" particles if $H_e/m$ is small enough.  The time average of the energy density $\rho_\mathrm{p}^{(1)}$ over a Hubble time  further suppresses $\rho_\mathrm{p}^{(1)}$ by an additional power of $H/m$, but it does not change that its relative enhancement by a factor of $(m/H_e)^3$. As we have repeatedly emphasized, these considerations illustrate that there are cases in which the particle production formula (\ref{eq:rho ppf}) is not the correct approximation to the energy density of the field, even when all the relevant modes are in the adiabatic regime.

\section{Summary and Conclusions}

In this article we have explored the conditions under which the energy density of a minimally coupled free scalar can be computed using the particle production and classical field formalisms, particularly after a transition from inflation to radiation domination.  To face these questions, we need to deal with renormalization first.  In cosmological spacetimes, Pauli-Villars regularization proves to be extremely useful in regulating and renormalizing loop integrals that otherwise diverge. In this approach, a set of massive regulator fields cancel  the divergent vacuum energy density of the scalar in the ultraviolet and thus render it finite. This allows us to concentrate on the spectral density and to study the contribution of different mode ranges to the energy density separately. At energy scales lower than the regulator masses, the regulators can be largely ignored. Only at higher scales do the latter impact the  spectral density and  renormalize the cosmological constant, the Planck mass and the coupling constant of an appropriate combination of curvature invariants. For light fields, the running caused by this renormalization leads to changes in these ``constants" that do not depend on the counterterms themselves; these are then testable predictions of the quantum theory. 

In order for the particle production formula (\ref{eq:rho ppf}) to correctly approximate the  spectral energy density in a particular mode range, it is not only necessary for the corresponding modes to be in the adiabatic regime, but  also  that frequencies are large ($\omega_k\gg \mathcal{H}$) and particle production is significant ($|\beta^\mathrm{ad}_k|\gtrsim 1$.) In the case of a massless field,  the first condition is satisfied during radiation domination by all modes,  the second fails on superhorizon scales, and the third is violated in the ultraviolet, regardless of the field mass.  This is why work in the literature that estimated the energy density using the particle production formalism arrived at incorrect values in the infrared regime. In particular, equation (\ref{eq:rho ren ppf}) correctly  approximates the renormalized energy density only when  the range of modes that satisfy the previous three conditions gives the  dominant contribution to the total density.

The classical field formalism applies to the zero mode when the latter is in a macroscopically excited state, though, in general, the matching classical scalar is complex. Provided that the appropriately defined Bogolubov coefficients are large, the formalism also  applies to the non-relativistic modes of a massive field. In that case   the field gradients are negligible and the modes collectively act like a massive, homogeneous scalar. At early times, when $m\leq H$, such a classical field is effectively frozen and behaves like dark energy. In contrast, at late times, when $H\leq m$, the field oscillates and on average behaves like cold dark matter. It is in this regime where the particle production formalism also applies. There are cases, however, in which neither the particle nor the classical field approximations work. The spectral density of relativistic superhorizon modes, for instance, scales like  curvature, regardless of the particular quantum state of the modes. Again, when the classical field approximation applies to the modes that give the dominant contribution to the renormalized energy density, the latter can be approximated by equation (\ref{eq:rho ren class}). 

When the adiabatic approximation does apply to all the modes of the field, one can also disentangle the total energy density of the produced particles from that of the $out$  vacuum. For a massless field, during radiation domination, the latter is given by (\ref{eq:rho out massless}), whereas for a massive field at late times it is determined by (\ref{eq:rho dS massive}). These energy densities therefore set bounds above which the total energy density is dominated by the produced particles, rather than by the {\it out} adiabatic vacuum contribution. In the case of the homogeneous classical field approximation the same happens when we can disregard the renormalized energy density of the relativistic modes. 

The previous statements apply to any quantum state. If the field is heavy during inflation,  we expect all its modes to be in the $in$ vacuum. When the scalar is massless, or its mass is light during inflation, on the other hand, only modes that are subhorizon at the beginning of inflation can be assumed to be in the  {\it in} vacuum. The state of superhorizon modes at that time remains unknown to us, so the comoving Hubble constant at the beginning of inflation   acts as  an infrared cutoff.  When  modes are in the $in$ vacuum,  the  form of the spectral density critically depends on the ratio of the mass of the scalar to the Hubble scale during inflation.  We summarize the behavior of the spectral density of a massless scalar in section \ref{sec:Overview}, and that of a massive scalar that is light during inflation in section \ref{sec:Light Fields Overview}. Although in some mode ranges it is possible to reproduce the corresponding energy densities by invoking the particle formalism or a classical homogeneous scalar, there is no case in which all of the spectral density can be described by either. In particular, the spectral density typically displays a diversity of behaviors that cannot be simply captured by a single fluid. When the scalar field is heavy during inflation, neither the particle production nor the classical field approximations do strictly apply, since in this case its renormalized energy density is dominated by that of the $out$ adiabatic vacuum, equation (\ref{eq:rho massive general}).

The smoothness of the transition to radiation domination strongly influences the spectral density in the ultraviolet. If the transition is gradual, we can make model-independent predictions  because the Bogolubov coefficients are highly suppressed in the ultraviolet,  where the spectral density simply approaches that of the $out$ adiabatic vacuum. This implies that the low momentum modes typically provide the dominant contribution to the energy density, which  happens to be insensitive to the details of the transition to radiation domination when inflation was de Sitter-like.  

Models in which a derivative of the scale factor changes discontinuously can serve as idealizations of  abrupt transitions, in which there is a sudden change in the evolution of the scale factor. If the second derivative of the scale factor changes discontinuously, particle production is so copious that the renormalized energy density diverges in the ultraviolet. Such a transition is therefore unphysical and should not be taken too literally at ultrahigh frequencies. To the extent that the energy density in the ultraviolet regime of such idealizations is dominated by the  lower boundary of the ultraviolet regime, we can also make rather general model-independent estimates of the energy density within this class of transitions.

Our results illustrate that care must be taken when using the particle production or classical field approximations, which we view as a simple shortcut to the spectral density when the appropriate conditions apply. Yet, in the end, other than convenience, we do not see a strong reason to introduce particles or classical fields in a framework that ultimately deals with quantum fields. In the field theories we use to describe the universe all that ought to be relevant is the expectation value of  field operators like the energy density. Some of this work can be interpreted as the first step in the grand scheme of characterizing  all of cosmology by  appropriate expectation values \cite{Armendariz-Picon:2020tkc}. 

\acknowledgments

ADT is supported in part by CONACyT grant No.~286897 ``Materia oscura: Implicaciones de sus propiedades fundamentales en las observaciones astrof\'isicas y cosmol\'ogicas" and by DAIP.  He also acknowledges the hospitality extended to him during his stay at the Physics Department of St.~Lawrence University, where part of this work was completed.

\appendix
 
 \section{Validity of the High and Low  Frequency Approximations}
 \label{sec:Validity of the Low and High  Frequency Approximations}
 
 In this appendix we explore in more detail the regime of validity of the high  and  low frequency approximations of sections  \ref{sec:High Frequencies}  and \ref{sec:Low Frequencies}. Particular attention is paid to the  power-law expanding universes in equation (\ref{eq:a inflation}).

\subsection{High Frequency Approximation}
 \label{sec:Adiabaticity During Radiation Domination}
 
As we mention in section \ref{sec:High Frequencies}, the {\it adiabatic expansion} is an expansion of the solutions of the mode equation (\ref{eq:mode equation}) in the number of time derivatives of the scale factor.   The expansion is exact at zeroth order if the spacetime is nonexpanding, and also during radiation domination, when $\ddot{a}=0$, if the the field is massless.  To the extent that we typically keep only a finite number of terms of the expansion, we shall refer to it as the adiabatic approximation.

The validity of the adiabatic approximation demands that the effective frequencies introduced in equations (\ref{eq:W}) satisfy $W^{(0)}_k\gg {}^{(2)} W_k\gg {}^{(4)}W_k\gg \dots\,$. The first inequality translates into
\begin{equation}\label{eq.condition.ad}
 \frac{3}{8}\frac{\dot{\omega}_k^2}{\omega_k^4}-\frac{1}{4}\frac{\ddot{\omega}_k}{\omega_k^3}-\frac{1}{2}\frac{\ddot{a}}{\omega_k^2a}\ll 1.
\end{equation}
Hence, it is generally necessary to impose $\omega_k\gg \mathcal{H}$ in order for the last term on the left hand side of equation~(\ref{eq.condition.ad}) to be  smaller than one. This alone guarantees that the first two terms of this expression are also small, and in addition implies that the remaining  inequalities above are generically satisfied too.  At early times, $ma\leq \mathcal{H}$, the condition of adiabaticity imposes $k\gg \mathcal{H}$; only subhorizon modes are adiabatic. At intermediate or late times, $ma\geq \mathcal{H}$, however, all modes are adiabatic. (See section \ref{sec:General Infrared Behavior massive}  for the distinction between ``intermediate" and ``late" times.)

\begin{figure}[t!]
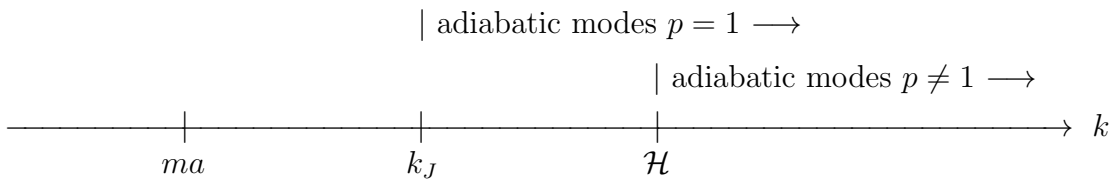


\hspace{5.88cm}$|$ adiabatic modes $p=1$ $\longrightarrow$\\

\vspace{-0.3cm}
\hspace{9cm}$|$ adiabatic modes $p\neq 1$ $\longrightarrow$\\

\vspace{-1.3cm}
\[ \xrightarrow{\hspace*{14cm}} \; k \]

\vspace{-0.95cm}
\hspace{2.59cm} $|$ \hspace{2.75cm} $|$ \hspace{2.75cm} $|\quad$\\
\vspace{-0.2cm}
\hspace{2.2cm} $ma$ \hspace{2.4cm}  $k_J$ \hspace{2.45cm} $\mathcal{H}$\\ 
  \caption{ 
  The adiabatic regime at early times, when $ma\leq \mathcal{H}$. In general,  only subhorizon modes $k\gg\mathcal{H}$ are adiabatic. However, during radiation domination this condition is relaxed, $k\gg k_J\equiv \sqrt{ma\mathcal{H}}$, and there are superhorizon modes which are also adiabatic.  In both instances  all adiabatic modes are relativistic at early times, $k\gg ma$.
  As time goes by $ma$ ($\mathcal{H}$) moves to the right (left) in the figure, whereas $k_J$ remains constant. The intersection of the three quantities at $ma\sim k_J \sim \mathcal{H}$ defines the transition between early and intermediate times. At intermediate and late times, $ma\geq \mathcal{H}$, all modes are adiabatic.}
\label{Fig:adiabatic}
\end{figure}

During radiation domination ($p=1$)  adiabaticity  is less restrictive because the mode equation does not contain an explicitly time-dependent factor when  $m=0$. In this case the last term on the left hand side of equation (\ref{eq.condition.ad}) vanishes and a  necessary and sufficient condition for the validity of (\ref{eq.condition.ad}) is instead
\begin{equation}\label{eq.condition.ad2}
 \frac{m^2a^2\mathcal{H}^2}{\omega_k^4}\ll 1.
\end{equation} 
Higher orders in the adiabatic approximation lead to similar conclusions, since $\ddot{a}= 0$ and the scale factor always appears in conjunction with $m$.  At early times, ${ma\leq \mathcal{H}}$, the inequality (\ref{eq.condition.ad2}) demands $k\gg k_J$, where $k_J$ is the scale defined in equation (\ref{eq:kJ}). Note that the scale $k_J$ is constant during radiation domination, and it is superhorizon-sized at early times as indicated in figure~\ref{Fig:adiabatic}.  Therefore, in addition to the subhorizon modes, there are superhorizon modes which are also adiabatic at early times.  At intermediate or  late times, $ma\geq \mathcal{H}$, again, all modes are adiabatic. If we set $m=0$,  all modes satisfy~(\ref{eq.condition.ad2}) at any time and the zeroth order term in the adiabatic approximation is already exact; the solution of the mode equation is (\ref{eq:wk0}).

In a power-law expanding universe, some of the properties discussed above  are manifest in the following  expansion of the mode functions to second adiabatic order,  which is valid in the limit of  short wavelengths, $\mathcal{H}\ll k$, and  light fields, $ma\ll \mathcal{H}$,  
\begin{eqnarray}\label{eq.adiabatic1}
 \chi_k^{\mathrm{ad}(2)} &\approx& \frac{1}{\sqrt{2k}}e^{-ik\eta}\left\{1-i\left[\frac{p-1}{2}+\frac{p}{2(2p+1)}\left(\frac{m}{H}\right)^2\right]\left(\frac{aH}{k}\right)\right.
 \\
 && \left. -\left[\frac{(1+p)(1-p)(2-p)}{8p}+\frac{p^2+p+1}{4(2p+1)}\left(\frac{m}{H}\right)^2+\frac{p^2}{8(2p+1)^2}\left(\frac{m}{H}\right)^4\right]\left(\frac{aH}{k}\right)^2\right\}. \nonumber
\end{eqnarray}
The corrections to the zeroth adiabatic order are suppressed by factors of $\mathcal{H}/k$, which are small inside the horizon. If the field is massless, these corrections disappear during radiation domination, when the adiabatic expansion is exact and reproduces (\ref{eq:wk0}).  The scale $k_J$ defined in equation (\ref{eq:kJ}) arises from the expansion of $\chi_k^{\mathrm{ad}(2)}$ to order $1/k^4$, and cannot be inferred just from (\ref{eq.adiabatic1}).  We shall use this expression in appendix \ref{sec:Light Fields During Inflation} in order to determine the validity of the approximation~(\ref{eq:inflation heavy chi}) at short scales.

\subsection{Low Frequency Approximation}
 \label{sec:Validity of the Low Frequency Approximation}
 
The {\it low frequency expansion} in section \ref{sec:Low Frequencies}  is an expansion of the solutions of the mode equation (\ref{eq:mode equation}) in powers of $\omega_k^2$. The leading order of this expansion provides the exact solution of the zero mode equation of a massless field, no matter the form of the scale factor, equation (\ref{eq:chi bar infrared}). To the extent that we typically keep only a finite number of terms of the expansion, we shall refer to it as the low frequency approximation.
 
 The validity of the low-frequency approximation demands that the corrections to the mode functions decrease as  the order of the series increases, ${\chi_k^{\mathrm{low}(0)} \gg {}^{\mathrm{low}(2)}\chi_k\gg \dots}$.  Here ${}^{\mathrm{low}(n+2)}\chi_k\equiv \chi_k^{\mathrm{low}(n+2)} - \chi_k^{\mathrm{low}(n)}$ denotes the correction of order $\omega_k^{n+2}$ in the expansion of the mode function $\chi_k$, which can be obtained from equation (\ref{eq:chi0 correction}) by replacing $\chi_k^{\mathrm{low}(n)}$ by $^{\mathrm{low}(n)}\chi_k$ and dropping the lowest order approximation $\chi_k^{\mathrm{low}(0)}$.   Taking into account the definition of $b$  in equation (\ref{eq:chi bar infrared})  we can roughly estimate the relative difference between two of the consecutive approximations generated  by  iteration of  (\ref{eq:chi0 correction}),  
\begin{equation}
	{}^{\mathrm{low}(n+2)}\chi_k
	\sim \eta\,  a^2 b \, \omega_k^2\, \, {}^{\mathrm{low}(n)}\chi_k \sim  \eta^2 \omega_k^2 \, \, {}^{\mathrm{low}(n)}\chi_k,
\end{equation}
implying that our approximate solution remains valid as long as $\omega_k\ll \mathcal{H}\sim 1/\eta$, that is, in the expected regime of long wavelengths and light fields.  

 It is in fact useful to evaluate the second order correction to $\chi_k^{\mathrm{low}(0)}$  during power-law expansion (\ref{eq:a inflation}). For arbitrary fixed $p$, and setting $n=0$,  the integral in equation (\ref{eq:chi0 correction}) can be explicitly evaluated, yielding
\begin{eqnarray}\label{eq.exp.pl}
	\chi_k^{\mathrm{low}(2)}(\eta) &=& -i\frac{a}{\sqrt{2M}}\left[1-\frac{p^2}{2(1+2p)}\left(\frac{k}{a H}\right)^2-\frac{p^2}{2(1+4p)}\frac{m^2 }{(1+p) H^2}\right]
	 	\nonumber\\
 	&&-ab\,\sqrt{\frac{M}{2}}\left[1-\frac{p^2}{2(3-2p)}\left(\frac{k}{aH}\right)^2-\frac{p^2}{6}\frac{m^2}{(1+p)H^2}\right],
\end{eqnarray}
which obeys the  Wronskian  condition (\ref{eq:Wronskian}) at first order in $m^2$ and $k^2$. The corrections to $\chi_k^{\mathrm{low}(0)}$ are generically small whenever $k\ll \mathcal{H}$ and $ma\ll \mathcal{H}$,  which agrees with our previous condition $\omega_k\ll \mathcal{H}$.  This hence supports the validity of (\ref{eq:chi bar infrared}) as a zeroth order approximation to the mode functions in the limit of light fields and superhorizon modes. 

It appears, however, that sufficiently close to de Sitter inflation,  when the combination ${|1+p|H^2}$ becomes of order $m^2$, the perturbative expansion breaks down, for in that case the corrections  proportional to $m^2$ in equation (\ref{eq.exp.pl}) become of the same order as the leading term. This situation merits more attention, given that it corresponds to the merely-light field limit ${|1+p|H^2\ll m^2\ll H^2}$ that we explore in section~\ref{sec:Light Fields}.
In order to address this limit, it is convenient to take advantage of the freedom one has to add multiples of the lowest order solution $\chi_k^{\mathrm{low}(0)}$ to the leading correction $\chi_k^{\mathrm{low}(2)}$ in equation (\ref{eq.exp.pl}). This amounts to choosing the lower integration limit in equations (\ref{eq:chi bar infrared}), (\ref{eq:scattering}) and (\ref{eq:chi0 correction}) appropriately. Doing so, and focusing on the problematic terms with $k=0$  we can write
\begin{eqnarray}\label{chi0dS3}
 \chi_0^{\mathrm{low}(2)}(\eta) &=& -i\frac{a}{\sqrt{2M}} \left[1-\frac{p^2}{2(1+4p)}\frac{m^2}{(1+p)H^2}+\frac{p^2}{2(1+4p)}\frac{m^2}{(1+p)H_i^2}\right]\nonumber\\
 &&-ab \sqrt{\frac{M}{2}} \left[1-\frac{p^2}{6}\frac{m^2}{(1+p)H^2}-\frac{p^2}{2(1+4p)}\frac{m^2}{(1+p)H_i^2}\right].
\end{eqnarray}
By construction, equation (\ref{chi0dS3}) solves the mode equation (\ref{eq:mode equation}) and satisfies again the 
 Wronskian condition (\ref{eq:Wronskian}) at first order in $m^2$. In this form, the limit $p\to -1$ remains finite and well-behaved. Actually, if we substitute the Hubble parameter $H$ in equation (\ref{chi0dS3}) by the expansion
\begin{equation}\label{eq:moH expansion}
 H=H_i \left(\frac{a_i}{a}\right)^{\frac{1+p}{p}} = H_i\left(1+\frac{1+p}{p}\log\frac{a_i}{a}+\ldots\right),
\end{equation}
we arrive at 
\begin{eqnarray}\label{chi0dS4}
 \chi_0^{\mathrm{low}(2)}(\eta) &=& -i\frac{a}{\sqrt{2M}} \left[1-\frac{m^2}{H_i^2}\left(\frac{p}{1+4p}\log\frac{a}{a_i}+\ldots\right)\right]\nonumber \\
 &&-ab \sqrt{\frac{M}{2}}  \left[1-\frac{m^2}{H_i^2}\left(\frac{2p^2}{3(1+4p)}+\frac{p}{3}\log\frac{a}{a_i}+\ldots\right)\right].
\end{eqnarray}
This expression is consistent with the exact result we directly obtain in de Sitter, when $p=-1$.  The ellipses refer to higher orders in $1+p$, and the subindex $i$ denotes the initial time of inflation (although it can be also replaced by an arbitrary instant of time during inflation if necessary.) The form of the leading corrections therefore implies that  near de Sitter we can trust our zeroth order solution $\chi^{m=0}_0$ in equation (\ref{eq:chi bar infrared})  as long as
\begin{equation}
\frac{m^2}{H_i^2}\log \frac{a}{a_i}\ll 1. 
\end{equation}
In the merely-light limit, this condition also guarantees the validity of the expansion (\ref{eq:moH expansion}) that we used to arrive at (\ref{chi0dS4}). As it turns out, however, in the merely-light limit it is more convenient to choose equation (\ref{eq:inflation heavy chi}) (with $p=-1$) as our zeroth order solution in $
|1+p|$, aas the approximate solution of the mode equation at long wavelengths.

Note also that, in the massless case $m=0$, during radiation domination, $p=1$, the low frequency expansion~(\ref{eq.exp.pl}) takes the simpler form 
\begin{equation}\label{eq:m=0 rad long}
 \chi^{\mathrm{low}}_k(\eta) = \frac{1}{\sqrt{2k}}\left[1-i(k\eta)-\frac{1}{2}(k\eta)^2+\frac{i}{6}(k\eta)^3+\ldots\right],
\end{equation}
where the ellipsis make reference to higher orders in $k\eta$, and where we have chosen $M\equiv a_0^2/(\eta_0^2 k)$. This series can be summed for {\it any} value of $k$, resulting in  $e^{-ik\eta}/\sqrt{2k}$, which corresponds to the exact solution of the mode equation that we identified in (\ref{eq:wk0}), and also to the leading order term  of the adiabatic  expansion~(\ref{eq.adiabatic1}). Again, this is a consequence of the absence of a characteristic time scale that differentiates between high and low frequencies in the massless case. In particular, since $\ddot{a}=0$ during radiation domination, the condition for the validity of the approximate solution (\ref{eq:chi bar infrared}), which does provide the first two terms in (\ref{eq:m=0 rad long}),  is not ${\omega_k^2\ll \ddot{a}/a}$, as one may have naively expected from (\ref{eq:mode equation}).

 \section{Light Fields During Inflation}
 \label{sec:Light Fields During Inflation}
 
 In the massive case, away from de Sitter, the mode equation (\ref{eq:mode equation}) has no exact solution known to us. In order to construct an  approximate solution, we note that the dispersion relation during power-law inflation reads
 \begin{equation}\label{frequency.mode2}
 \tilde{\omega}_k^2 =k^2+
 \left[
 	p^2\frac{m^2}{H_i^2}\left(\frac{\eta_i}{\eta}\right)^{-2(1+p)}
	-p(p-1)
 \right]\frac{1}{\eta^2},
 \end{equation}
 where for convenience we have chosen $\eta_0$ in equation (\ref{eq:a inflation}) to be the time at which inflation begins, $\eta_i$. Clearly, around the vicinity of the ``expansion point'' $\eta_i$ we can replace  the power proportional to  $m^2$  in the dispersion relation by the (constant) value it would have 
in de Sitter, which remains a valid approximation as long as $\left| (1+p)\log (\eta/\eta_i)\right|\ll 1$. Then, the solution of the mode equation is indeed equation  (\ref{eq:inflation heavy chi}). To further assess the validity of this approximate solution beyond the vicinity of the expansion point $\eta_i$, and away from the de Sitter limit $p=-1$,  we shall explore its short and long wavelength regimes.  

 \subsection{Short Wavelengths}
 
Let us  compare first the  short wavelength limit $-k\eta\gg 1$ 
of equation (\ref{eq:inflation heavy chi}) with the adiabatic expansion (\ref{eq:out adiabatic}), which we already know to be a valid approximation on subhorizon scales. For convenience, we shall focus directly on the spectral density, which is the quantity we are interested in.  Let $d\rho_\mathrm{in}/d\log k$ and  $d\rho^\mathrm{ad(4)}/d\log k$ be the spectral densities computed  by using   the expansions of (\ref{eq:inflation heavy chi})   to fourth order in $(k\eta)^{-1}$, and of equation  (\ref{eq:out adiabatic}) to fourth adiabatic order,  respectively.  Their relative difference with the zeroth order adiabatic spectral density $d\rho^\mathrm{ad(0)}/d\log k$ then is
 \begin{equation}\label{eq:B2}
	\frac{d\rho_\mathrm{in}-d\rho^\mathrm{ad(4)}}{d\rho^\mathrm{ad(0)}} =-\frac{1}{4p}\left[1+(2p+1)\frac{H^2}{H_i^2}\right]\left(\frac{ma}{k}\right)^2\left(\frac{aH}{k}\right)^2 + \frac{1}{8}\left[1-\frac{2H^2}{H_i^2}+\frac{H^4}{H_i^4}\right]\left(\frac{ma}{k}\right)^4,
\end{equation}
which gives a measure of the error committed when using the approximate solution (\ref{eq:inflation heavy chi}). Surely, equation~(\ref{eq:B2}) vanishes in de Sitter, $p=-1$ and $H=H_i$, and in the massless limit,  $m=0$, where the solution (\ref{eq:inflation heavy chi}) is exact. Away from these two particular cases, the difference (\ref{eq:B2}) tends to $\sim (m/H)^2 (\mathcal{H}/k)^4$, which is heavily suppressed for light fields at short wavelengths.   Nevertheless, although the relative difference of the two spectral densities becomes small at short distances, the absolute error  grows large as $k$ increases in the ultraviolet. Since it is crucial for renormalizability that the spectral density  display the correct absolute value there, we cannot rely on the approximation (\ref{eq:inflation heavy chi}) to determine the energy density in the ultraviolet, as we clarify in the paragraph below equation (\ref{eq:massive UV inflation}). Other than that,  the approximation in equation (\ref{eq:inflation heavy chi}) remains appropriate at short distances. 

\subsection{Long Wavelengths}

During inflation, however, the comoving Hubble factor increases in time, and some modes  eventually leave the horizon. Since at $k=aH$ the relative difference (\ref{eq:B2}) is still suppressed by powers of $(m/H)^2$, we expect the relative error in the spectral density to be small at horizon crossing when the field is light.  To determine to what extent we can further trust the approximation (\ref{eq:inflation heavy chi}) it is convenient to consider its behavior in the long wavelength limit $k=0$, 
\begin{equation}\label{eq:41 long}
\chi_0^\mathrm{in}\propto  \left(\frac{a}{a_i}\right)^{\frac{1-2\nu}{2p}}\approx \frac{a}{a_i}\, \left(\frac{a}{a_i}\right)^{\frac{p}{1-2p}\frac{m^2}{H_i^2}},
\end{equation}
which we have normalized to one at $\eta=\eta_i$, and  where we have only kept the leading term in  the $(m/H_i)^2$ expansion in the exponent. The square of this nonoscillating solution directly determines the spectral density at long wavelengths, so we shall focus on $\chi^\mathrm{in}_0$ as a proxy for this quantity. We would like to compare the behavior of equation (\ref{eq:41 long}) with the one captured by equation (\ref{eq.exp.pl}) when $k=0$, which correctly approximates the actual solution of the mode equation in the limit of long wavelengths and light fields. We could have obtained the second order correction therein by inserting the zeroth order solution (\ref{eq:inflation heavy chi}) (with $m = 0$) into equation (\ref{eq:chi0 correction}) (with $\omega_k^2=m^2 a^2$) and then taking the limit $k = 0$. This ought to be the same as first taking the limit $k= 0$ and then substituting into (\ref{eq:chi0 correction}), which is precisely what led to the growing mode  proportional to $a$  in equation (\ref{eq.exp.pl}). In particular, since the zeroth order solution (\ref{eq:inflation heavy chi}) is $\chi_k^{\mathrm{in}} \propto a$, the {\it in} mode functions at first order in $(m/H_i)^2$ must be provided by the first line of equation (\ref{eq.exp.pl}) with $k=0$. What this says is that, again, the {\it in} mode functions find themselves in what used to be the growing mode in the massless case. In the following we shall hence restrict ourselves to this growing mode. 
  
To continue, it is convenient to distinguish between the ultra-light field limit, where $m^2\ll |1+p| H^2$, and the merely-light field limit, where $|1+p| H^2\ll  m^2\ll H^2$. If the field is ultra-light, at leading order in the long wavelength expansion, $k=0$, equation (\ref{eq.exp.pl}) becomes
\begin{eqnarray}\label{eq:chi0 k=0 approx}
	\chi_0^{\mathrm{low}(2)}(\eta) \propto \frac{a}{a_i} \left[1-\frac{p^2}{2(1+4p)}\frac{m^2 }{(1+p) H^2}\right].
\end{eqnarray} 
Clearly, the time dependence of~(\ref{eq:41 long})  agrees with that of~(\ref{eq:chi0 k=0 approx})  in the massless limit. Since equation (\ref{eq:inflation heavy chi}) is exact in this case, this is what we expected. Yet away from the massless case,  the evolution of (\ref{eq:41 long})  differs from that of  (\ref{eq:chi0 k=0 approx}) when $a/a_i$ is sufficiently large. To  keep the relative error in our estimate of the spectral density  small at long wavelengths, it suffices then that the  $(m/H_i)^2$  correction in~(\ref{eq:41 long}) remain small, which happens  as long as
\begin{equation}\label{eq:41 validity}
	 \left(\frac{a}{a_i}\right)^{\frac{p}{1-2p}\frac{m^2}{H_i^2}}\approx 1.
\end{equation}
At the same time, the low frequency approximation~(\ref{eq:chi0 k=0 approx}) remains valid as long as
\begin{equation}\label{eq:ultralight}
\frac{m^2}{|1+p| H^2}=\frac{m^2}{|1+p| H_i^2} \left(\frac{a}{a_i}\right)^{\frac{2(1+p)}{p}}\ll 1,
\end{equation}
which guarantees that the field is ultra-light.  Since in the ultra-light limit this last power of $a$ is much bigger than that in equation (\ref{eq:41 validity}), in this case we can thus trust the approximate  solution (\ref{eq:inflation heavy chi}) as a zeroth order expansion in $(m/H_i)^2$. This is equivalent to setting $m=0$ in  equation (\ref{eq:inflation heavy chi}) from the beginning and working in terms of the massless solution (\ref{eq:chi inf massless}), which is otherwise exact in $1+p$.

In the opposite limit, $|1+p| H^2\ll  m^2\ll H^2$, it is more convenient to work with equation (\ref{chi0dS4}), where again we  only retain the growing mode,
\begin{equation}\label{eq:chi0 merely light}
	\chi_0^{\mathrm{low}(2)}(\eta) \propto \frac{a}{a_i} \left[1-\frac{p}{1+4p}\frac{m^2}{H_i^2}\log\frac{a}{a_i} +\cdots\right].
\end{equation}
Equation (\ref{eq:41 long}) agrees with equation   (\ref{eq:chi0 merely light}) at first order in $m^2/H_i^2$ in the de Sitter limit, when $1+p=0$.  Since equation (\ref{eq:inflation heavy chi}) is also exact in this case, this is again what we expected. Yet away from de Sitter, the evolution of   (\ref{eq:41 long})  differs once again from that of (\ref{eq:chi0 merely light}) when $a/a_i$ is sufficiently large. The $1+p$ correction denoted by the ellipsis in (\ref{eq:chi0 merely light}) remains negligible as long as $|1+p|(m^2/H_i^2)\log (a/a_i)\ll 1$. This condition is less restrictive that the one we relied on  to arrive at (\ref{chi0dS4}),   
\begin{equation}\label{eq:merely light validity}
	|1+p|\log \frac{a}{a_i}\ll 1,
\end{equation}
so we can only guarantee that (\ref{eq:inflation heavy chi}) remains a valid approximation to the exact mode equation  when (\ref{eq:merely light validity})  holds.  Recall that the latter  is necessary for the validity of the expansion~(\ref{eq:moH expansion}), and therefore  implies that the Hubble parameter does not change significantly, as it  happens in de Sitter, where it remains constant.  When  (\ref{eq:merely light validity}) is satisfied, it suffices to work at zeroth order in $1+p$, which is equivalent to setting $p=-1$ in equation~(\ref{eq:inflation heavy chi}) from the very beginning.   In spite of the  limited range of validity of this approximation, in most cases of interest, such as in some dark matter and dark energy models, the mass of the field is much smaller than the Hubble factor during inflation, and condition~(\ref{eq:merely light validity}) is satisfied during an extremely large number of e-folds $ N\lesssim 10^{56} (H_i/M_P)^2 (m/\mathrm{eV})^{-2}$.

\end{document}